\documentclass[12pt,preprint]{aastex}
\shortauthors{}
\shorttitle{}

\begin{document}

\title{A new approach to analyzing solar coronal spectra
       and updated collisional ionization equilibrium calculations.
       II. Updated ionization rate coefficients }       

\author{P. Bryans,\altaffilmark{1}
	E. Landi,\altaffilmark{2}
	and D. W. Savin\altaffilmark{1}}
\altaffiltext{1}{Columbia Astrophysics Laboratory, Columbia University,
New York, NY 10027}
\altaffiltext{2}{US Naval Research
Laboratory, Space Science Division, 4555 Overlook Avenue, SW, Code 7600A,
Washington, DC 20375}

\begin{abstract}

We have reanalyzed SUMER observations of a parcel of coronal gas
using new collisional ionization equilibrium (CIE) calculations.
These improved CIE fractional
abundances were calculated using state-of-the-art electron-ion recombination
data for K-shell, L-shell, Na-like, and Mg-like ions of all elements from H through Zn
and, additionally, Al- through Ar-like ions of Fe.
They also incorporate the latest  recommended electron impact ionization data
for all ions of H through Zn.
Improved CIE calculations based on these recombination
and ionization data are presented here.
We have also developed a new systematic method for determining the 
average emission
measure ($EM$) and electron temperature ($T_e$) 
 of an isothermal plasma.
 With our new CIE data
and our new approach for determining average $EM$ and $T_e$,
we have reanalyzed SUMER observations of the solar corona.
We have compared our results with those of previous studies
and found some significant differences for
the derived $EM$ and $T_e$.
We have also calculated the enhancement of coronal elemental abundances compared
to their photospheric abundances, using
the SUMER observations themselves to determine the
abundance enhancement factor for each of the emitting elements.
Our observationally derived first ionization potential
(FIP) factors are in reasonable agreement with the theoretical model of
\citet{Lami08a}.

\end{abstract}


\section{Introduction}
\label{sec:intro}

Investigating the dynamics of the solar corona is crucial if one is to
understand fundamental solar and heliospheric physics.   The corona also
greatly influences the Sun-Earth interaction, as it is from here that the solar
wind originates. Explosive events in the corona can deposit up to $2 \times
10^{16}$~g of ionized particles into the solar wind \citep{Hund93a}. These can
have a profound effect on the Earth's magnetosphere and ionosphere. Hence the
investigation of the corona is of obvious importance. 

Over the years there has been a significant amount of  research invested in
developing our understanding of the corona 
(reviewed by Aschwanden 2004 and Foukal 2004).
However, gaps remain in our understanding of some of the most fundamental 
processes taking place in the corona. For example, the so-called coronal heating
problem  remains unsolved \citep{Gudi05a,Klim06a} and we are still unable to
explain the onset processes that cause solar flares and coronal mass ejections
\citep{Forb00a,Prie02a}.

One of the most powerful tools for understanding the properties of the solar
corona is spectroscopy \citep{Tand88a,Fouk04a}.  Analyzing the spectral emission
of the corona can give the temperature and density of the plasma, as well as
information on the complex plasma structures common in this region of the Sun's
atmosphere. One common approach to this end is to calculate the emission measure
($EM$) of the gas \citep[e.g.,][]{Raym81a}.   

The $EM$ technique is particularly useful for studying the properties of the upper
solar atmosphere.  In this region, conditions are such that the plasma can often
be described as  low-density and in steady-state and the emitting region as constant
in density and temperature. These relatively simple conditions allow one to
neglect density effects and to assume all emission is from an isothermal plasma.  For
example, \citet{Land02a} compared off-disk spectral observations of the solar
corona with predictions from the CHIANTI version 3 atomic database
\citep{Dere97a,Dere01a}. \citet{Land02a} calculated the $EM$ of the plasma based
on the observed intensities using the atomic data assembled together in CHIANTI.
From this, they also infer the electron temperature ($T_e$) of the emitting plasma.  
However, the power of this spectroscopic
diagnostic can be limited by our understanding of the underlying atomic physics
that produce the observed spectrum.

Reliable $EM$ calculations require accurate fractional abundances for the
ionization stages of the elements present in the plasma. 
For a plasma in collisional ionization
equilibrium (CIE; sometimes also called coronal equilibrium),
the atomic data needed for such a spectral analysis  includes
rate coefficients for electron-ion recombination and electron-impact ionization.
These data directly affect the calculated ionic fractional abundances of the
gas. The fractional
abundances, in turn, are used to determine the $EM$. Hence the reliability of
the CIE calculations is critical.

The recommended CIE calculations at the time of the work by \citet{Land02a} were
those of \citet{Mazz98a}. Recently, however,  state-of-the-art electron-ion
recombination data have been published for K-shell, L-shell, and Na-like ions of
all elements from H through Zn
\citep{Badn03a,Badn06a,Badn06b,Badn06c,Gu03a,Gu03b,Gu04a}.  Based
on these new recombination data, a significant update of the recommended CIE
fractional abundances was published recently by 
\citet[][Paper~I in this series]{Brya06a}.
Since then additional recombination data have been published for Mg-like ions
of H through Zn \citep{Altu07a} and Al- through Ar-like ions of Fe \citep{Badn06d,Badn06e}.
Electron impact ionization (EII) data have also been updated recently by
\citet{Suno06a}, \citet{Dere07a}, and \citet{Matt07a}.  
Of these three, the recommended EII data of
\citet{Dere07a}, which we adopt, provide the only complete available set
of rate coefficients for all ions of H through Zn.
Here we have updated the results of \citet{Brya06a} using these new
recombination and ionization data.
One of the motivations behind this paper is to investigate the effects of  
the recent improvements in CIE calculations on solar observations.

Since the \citet{Land02a} paper there have been other improved
atomic data
(e.g., the improvement of the model for N-like ions).
These have been made available in a more recent CHIANTI release---version 
5.2 \citep{Land06a}. 
It is this version we use here.

We also investigate here the observed
relative elemental abundances and the first ionization potential (FIP) effect.
The FIP effect is the discrepancy between the coronal and photospheric
elemental abundances, possibly explained by the pondermotive
force induced by the propagation of Alfv\'{e}n waves
through the chromosphere \citep{Lami04a,Lami08a}.  
Elements with a FIP of below $\sim 10$~eV appear to have a
coronal abundance that is enhanced by a factor of a few over their photospheric abundance
\citep[see, e.g., the review by][]{Feld00a}.
Often, the FIP effect is accounted for by multiplying the abundance of the 
low-FIP elements by a single scaling factor \citep[such as 3.5 as was done in][]{Land02a}.
In the present work, we investigate the reliability of this approach by
 quantifying the FIP effect based on the 
observations themselves.
We determine the $EM$ from the high-FIP element Ar and then scale the elemental
abundances of the moderate- and low-FIP elements so that their derived  $EM$s match that of Ar.
We compare our derived abundances with those of a previous analysis of
the same observation \citep{Feld98a} as well as with theoretical
predictions \citep{Lami08a}.

An important aspect of the present paper is the development of a sound mathematical 
method of determining the average $EM$ and $T_e$ of an isothermal plasma.
Previous studies have done this in a less rigorous manner.  \citet{Land02a}, for example,
evaluate plots of $EM$ versus $T_e$
curves and give a ``by eye'' estimate of the average value of the $EM$ and $T_e$
and their associated errors.
This method allows human bias to become important when deciding which curve
crossings to include in the selection.
In addition, it is unclear to what this ``average'' actually corresponds mathematically.
The fact that the analysis is performed on graphs with logarithmic axes suggests
that by-eye average is closer to the geometric mean than the arithmetic mean.
Finally, no account is taken of the reliability of the atomic data
used to calculate fractional abundances.
\citet{Brya06a} showed that CIE results are unreliable at temperatures 
where the ionic fractional abundances are less than 1\%.
Previous studies have failed to account for this
when using the CIE data in the $EM$ analysis.

Taking the above four paragraphs into account, we have reanalyzed
the observations of \citet{Land02a}.  The rest of this
paper is organized as follows: In Sec.~\ref{sec:observations} we give a description of
the observing sequence, the observed lines and their categorizations
by \citet{Land02a}.  Section~\ref{sec:em}
defines the $EM$ and explains the method we use to determine the plasma
temperature from the observed line intensities. In Sec.~\ref{sec:cie} we review
the recent developments in the understanding of dielectronic and radiative recombination
and electron impact ionization, and
the subsequent improvement in CIE calculations.
We also present updated tables of these CIE calculations, which
supersede those of \citet{Brya06a}.
In Sec.~\ref{sec:method} we describe our new approach for
determining the  $EM$ and temperature of an isothermal 
plasma based on the observed spectral
line intensities.
Section~\ref{sec:fip} discusses our method of determining the elemental
abundance enhancement factors due to the FIP effect.
In Sec.~\ref{sec:groups} we
present the results of our $EM$ calculations for each of the categorizations
introduced by \citet{Land02a}.
Section~\ref{sec:discuss} discusses the consequences of these
results, in particular highlighting discrepancies between the results of this
paper and those of \citet{Land02a}.
In Sec.~\ref{sec:future} we propose future observations needed to
address some of the remaining issues raised by our results here.
Concluding remarks are given in Sec.~\ref{sec:conclusion}.

\section{Observations}
\label{sec:observations}

The spectrum analyzed by \citet{Land02a}, and revisited here, was detected using
the Solar Ultraviolet Measurements of Emitted Radiation spectrometer
\citep[SUMER;][]{Wilh95a}  onboard  the Solar and Heliospheric
Observatory (SOHO). 
The observation spans over 5 hours, from 21:16 UT
on 1996 November 21 to 02:28 UT November 22, and was collected in 61 
spectral sections.
The observing slit imaged at a height $h$ of
 $1.03R_\odot\lesssim h \lesssim 1.3R_\odot$ above the
western limb.
The resulting spectrum covers the
entire SUMER spectral range of 660--1500~\AA. 
\citet{Land02a} give 
a full description of the
observation sequence and data reduction.

Table~\ref{tab:linelist} lists the  coronal lines identified in the spectrum
 and their
corresponding transitions  \citep[reproduced
from][]{Land02a}.
Known typos in the line assignment labels of \citet{Land02a} have been corrected;
these do not affect their reported results.
Landi et al.\ estimate uncertainties on the extracted line intensities
of 25--30\%.
Twelve of the emission lines observed in this run are omitted from the table here due to
their being blended with other emission lines or having uncertain intensities.
The remaining spectral lines are split into three distinct groups,
labeled in the first column of Table~\ref{tab:linelist} as:
\begin{itemize}
\item[I.] Forbidden transitions within the ground configuration.
\begin{itemize}
\item[Ia.] Non--N-like transitions.
\item[Ib.] N-like transitions.
\end{itemize}
\item[II.] Transitions between the ground and the first excited configuration:
\begin{itemize}
\item[IIa.] Allowed $2s-2p$ transitions in the Li-like isoelectronic sequence
and allowed $3s-3p$ transitions in the Na-like isoelectronic sequence.
\item[IIb.] Intercombination transitions in the Be-, B-, C-, and Mg-like
isoelectronic sequences.
\end{itemize}
\item[III.] Transitions between the first and second excited configuration.
\end{itemize}
Within each group and subgroup we have derived the average $T_e$ and $EM$.
Categorizing the transitions in this way helps us to better identify any trends
in the $EM$ with respect to the transition type. 
Group~I transitions have been further divided into non--N-like and N-like transitions.
This separation was originally proposed by \citet{Land02a} due to the poor agreement
they found for the $T_e$ derived within each of these transition types.
This is discussed further in Secs.~\ref{sec:fip} and \ref{sec:discuss}.
The subdivision of transition
Group~II is to allow us to investigate a longstanding discrepancy between $EM$s
derived using Li- and Na-like ions and those derived using other isoelectronic
sequences \citep[e.g.,][]{Dupr72a,Feld98a,Land02a}. We also discuss this further in
Secs.~\ref{sec:fip} and~\ref{sec:discuss}.

\section{Method of Calculating Temperature and Emission Measure}
\label{sec:em}

The  intensity of an observed spectral line due to a transition from level $j$ to level $i$ in element X of
ionization state $+m$ can be written as
\begin{equation}
  I_{ji}
  =  \frac{1}{4\pi d^2}\int_{V}G_{ji}(T_e,n_e)n_e^2\,{\rm d}V 
\end{equation}
where $n_e$ is the electron density, 
$V$ is the emitting volume along the line
of sight, and $d$ is the distance to the source. $G_{ji}(T_e,n_e)$ is
the contribution function, which is defined as
\begin{equation}
  \label{eqn:cont func}
  G_{ji}(T_e,n_e)=\frac{n_j({\rm X}^{+m})}{n({\rm X}^{+m})}
                \frac{n({\rm X}^{+m})}{n({\rm X})}
		\frac{n({\rm X})}{n({\rm H})}
		\frac{n({\rm H})}{n_e}
		\frac{A_{ji}}{n_e} 
\end{equation}
where $n_j({\rm X}^{+m})/n({\rm X}^{+m})$ is the population of upper level $j$ 
relative to all levels in X$^{+m}$,
$n({\rm X}^{+m})/n({\rm X})$ is the fractional abundance of ionization stage 
$+m$
relative to the sum of 
all ionization stages of X, $n({\rm X})/n({\rm H})$ is the 
abundance
of element X relative to hydrogen, and $n({\rm H})/n_e$ is the abundance of hydrogen
relative to the electron density. $A_{ji}$ is the spontaneous emission
coefficient for the  transition.

For the observation analyzed here, the emitting plasma was found to be
isothermal by \citet{Feld99a} and \citet{Land02a}.
For the moment we assume this to be correct but we revisit the validity
of the isothermal assumption in Sec.~\ref{sec:discuss groups}.
One can also make the assumption that the region emitting the observed 
line intensities is at a constant density.
While the line-of-sight
of the observation covers plasma where densities vary by orders of
magnitude, the emission is dominated by a region 
with a small range of densities around the peak density.
Only those emission lines that have a strong density sensitivity in this
range will be affected by the density gradient \citep{Lang90a}.
\citet{Feld99a} inferred a density of $1.8\times10^8$~cm$^{-3}$
for this observation.
A density dependent study of the 74 lines observed here is
beyond the scope of our paper.
Here we use the inferred density of \citet{Feld99a} in our analysis.

If we now assume that all the emission comes from the same parcel of gas of
nearly constant temperature, $T_c$, and density, we can approximate
\begin{equation}
  I_{ji}=\frac{G_{ji}(T_c,n_e)}{4\pi d^2} EM
\end{equation}
where the emission measure $EM$ is defined as
\begin{equation}  
   EM=\int n_e^2\,{\rm d}V
\end{equation}
and can be evaluated
from the observed line intensity as
\begin{equation}
  \label{eqn:em}
   EM=4\pi d^2\frac{I_{ji}}{G_{ji}(T_c,n_e)} .
\end{equation}
This has the same value for all transitions if the constant temperature and
density assumption is correct, which we label $EM_c$. 
Thus, from the observed line intensities,
$I_{ji}$, and using accurate data for $G_{ji}(T_e,n_e)$, one can calculate 
the emission measure and electron temperature of the emitting region.  
This is done by plotting the
$EM$ against $T_e$. The resulting curves for
each observed line should intersect at a common point yielding
$[T_c,EM_c]$. But this depends on the assumption of 
constant temperature and density being correct and on the accuracy of the
underlying atomic data. Here, one of the issues
we are investigating is the effect on solar coronal observations
of the newly calculated fractional abundances
\begin{equation}
f^m=\frac{n({\rm X}^{+m})}{n({\rm X})}.
\end{equation}

The units used throughout this paper for $EM$ and $T_e$ are cm$^{-3}$ and
K, respectively.  For ease of reading, we typically drop these units below.

\section{Improved collisional ionization equilibrium (CIE) calculations}
\label{sec:cie}

The plasma conditions of the solar upper atmosphere are often described
as being optically-thin, low-density, dust-free,
and in steady-state or quasi-steady-state.  Under these conditions the
effects of any radiation field can be ignored, three-body collisions are
unimportant, and the ionization balance of the gas is time-independent. 
This is commonly called CIE or coronal equilibrium.
These conditions are not always the case in the solar upper
atmosphere in the event of impulsive heating events but,
given the inactivity and low density of the plasma analyzed here, they
sufficiently describe the observed conditions.
For a thorough discussion of plasma conditions where one must treat the
time scales and density effects more carefully, we direct the reader to 
\citet{Summ06a}.

In CIE, recombination is due primarily to dielectronic recombination (DR) and
radiative recombination (RR).  At the temperature of peak formation in CIE, DR
dominates over RR for most ions. Ionization is primarily a result of electron
impact ionization (EII).  At temperatures low enough for both atoms and ions to
exist, charge transfer (CT) can be both an important recombination and
ionization process \citep{Arna85a, King96a}.
   CT is not expected to be important at solar coronal temperatures
and is not included in the work of \citet{Mazz98a}, \citet{Brya06a},
or the present paper.
Considering all the ions and levels that need to be taken into account, it is
clear that vast quantities of data are needed. Generating them to the accuracy
required pushes atomic theoretical and experimental methods to the edge of what is
currently achievable and often beyond.  For this reason, the CIE data used by
the solar physics and astrophysics communities have gone through numerous updates
over the years as more reliable atomic data have become available.

\subsection{Recombination rate coefficients}
\label{sec:recom}
The DR and RR rate coefficients used to determine the CIE fractional
abundances utilized by \citet{Land02a} were those recommended by
\citet{Mazz98a}. However, there has been a significant improvement in the
recombination rate coefficients since then. \citet{Badn03a} and
\citet{Badn06a,Badn06b,Badn06c} have calculated DR and RR rate coefficients
for all ionization stages from bare through Na-like for all elements from H through Zn and 
\citet{Gu03a,Gu03b,Gu04a} for a subset of these elements.  The methods of
Badnell and Gu are of comparable sophistication and their DR results
for a given ion agree
with one another typically to better than 35\% at the electron temperatures
where the CIE fractional abundance of that ion is $\ge1\%$.  The RR rate
coefficients are in even better agreement, typically within 10\% over this
temperature range. 
These differences for the DR and RR rate coefficients do not appear to be 
systematic in any way \citep{Brya06a}.
For both DR and RR outside this temperature range,
agreement between these two state-of-the-art theories can become significantly worse.
The DR calculations have also been compared to experimental measurements,
where they exist, and found to be in agreement to within 35\% in the
temperature range where the ion forms in CIE.
For a fuller discussion of the agreement between recent theories and 
the agreement between theory and experiment, we direct the reader to
\citet{Brya06a}.

\subsection{Electron impact ionization rate coefficients}
\label{sec:eii}
There have also been recent attempts to improve the state of the EII rate
coefficients used in CIE calculations.  The most complete of these studies
is that of \citet{Dere07a}, who produced recommended rate coefficients for
all ionization stages of the elements H through Zn.  These data are based
on  a combination of laboratory experiments and theoretical calculations. 
In addition, there have been works by \citet{Suno06a} and \citet{Matt07a}
that also address the issue of updating the EII database.  These works are
less complete than that of \citet{Dere07a}.  \citet{Suno06a} provides EII
cross sections for all ionization stages of C.  \citet{Matt07a} provides
EII cross sections for all ionization stages of H through O plus Ne and a
selection of other ions up to Ge.  

Between these recent compilations there remain sizable differences in the
EII rate coefficients for certain elements, often in the temperature range
where an ion forms in CIE.    For the ions important to the present work,
differences between recent recommended rate coefficients
 of up to 50\% are seen.  Larger
differences, of up to a factor of $\sim 4$, are found for other ions
not observed in this SUMER observation.   In short, we do not
see the uniform agreement between recommended sets of EII data
 as we do for the state-of-the-art
DR and RR calculations.

Despite these outstanding issues regarding the accuracy of the various EII
databases, we have used the compilation of \citet{Dere07a} to calculate
fractional CIE abundances.  Of the recent EII compilations,
the Dere database offers the most complete
selection of rate coefficients.
However, given the large differences between the \citet{Dere07a} and
\citet{Matt07a} results, we believe that further analysis of the
EII database is required to resolve these differences.

\subsection{Updated CIE calculations}
\label{sec:cie update}
The new recombination data of \citet{Badn03a} and
\citet{Badn06a,Badn06b,Badn06c} motivated \citet{Brya06a} to calculate
new CIE fractional abundances. Their results show large differences from the
\citet{Mazz98a} data for certain elements.
Here we revise the work of \citet{Brya06a} 
to include these newly recommended EII rate coefficients for all elements from H through Zn
and some further updates to the DR and RR rate coefficients
for selected ions.

We calculate CIE fractional abundances using the EII data of \citet{Dere07a}
for all ions of the elements H through Zn.
We also include
some corrections for Ca-like ions (K.\ P.\ Dere 2007, private communication).
The DR and RR rate coefficients used here
are those of \citet{Brya06a} but updated to 
include recent corrections to the fitting of some of the rate coefficients
\citep{Badn06b}.  We also include  recent DR work
for Mg-like ions of H through Zn 
and for Al- through Ar-like ions of Fe \citep{Altu07a,Badn06b,Badn06d,Badn06e}.
The DR and RR data for all other ions are those
of \citet{Mazz98a}.

\begin{figure}
  \centering
  \includegraphics[angle=90]{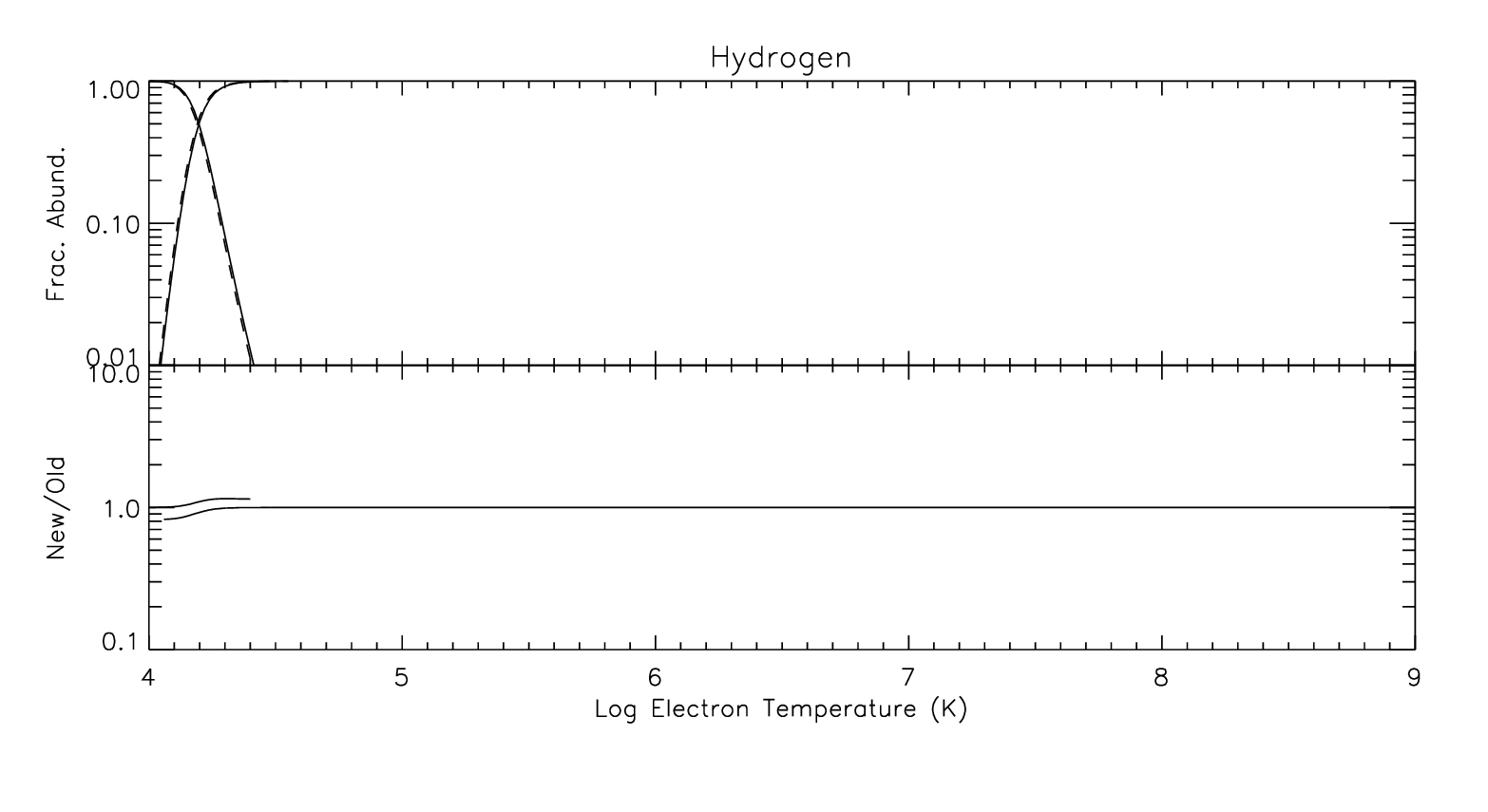}
  \caption[]{Ionization fractional abundance versus electron temperature for 
             all ionization stages of H. 
	     The upper graph shows our
	     results ({\it solid curves}) and the abundances 
	     calculated by Mazzotta et al.\ (1998; {\it dashed curves}).
	     The lower graph shows 
	     the ratio of the calculated abundances.
	     Comparison is made only for fractional abundances greater than
	     $10^{-2}$.
	     We label our results 
	     as ``New'' and those of \protect\citet{Mazz98a} as ``Old''.}
  \label{fig:H Mazz}
\end{figure}

\begin{figure}
  \centering
  \includegraphics[angle=90]{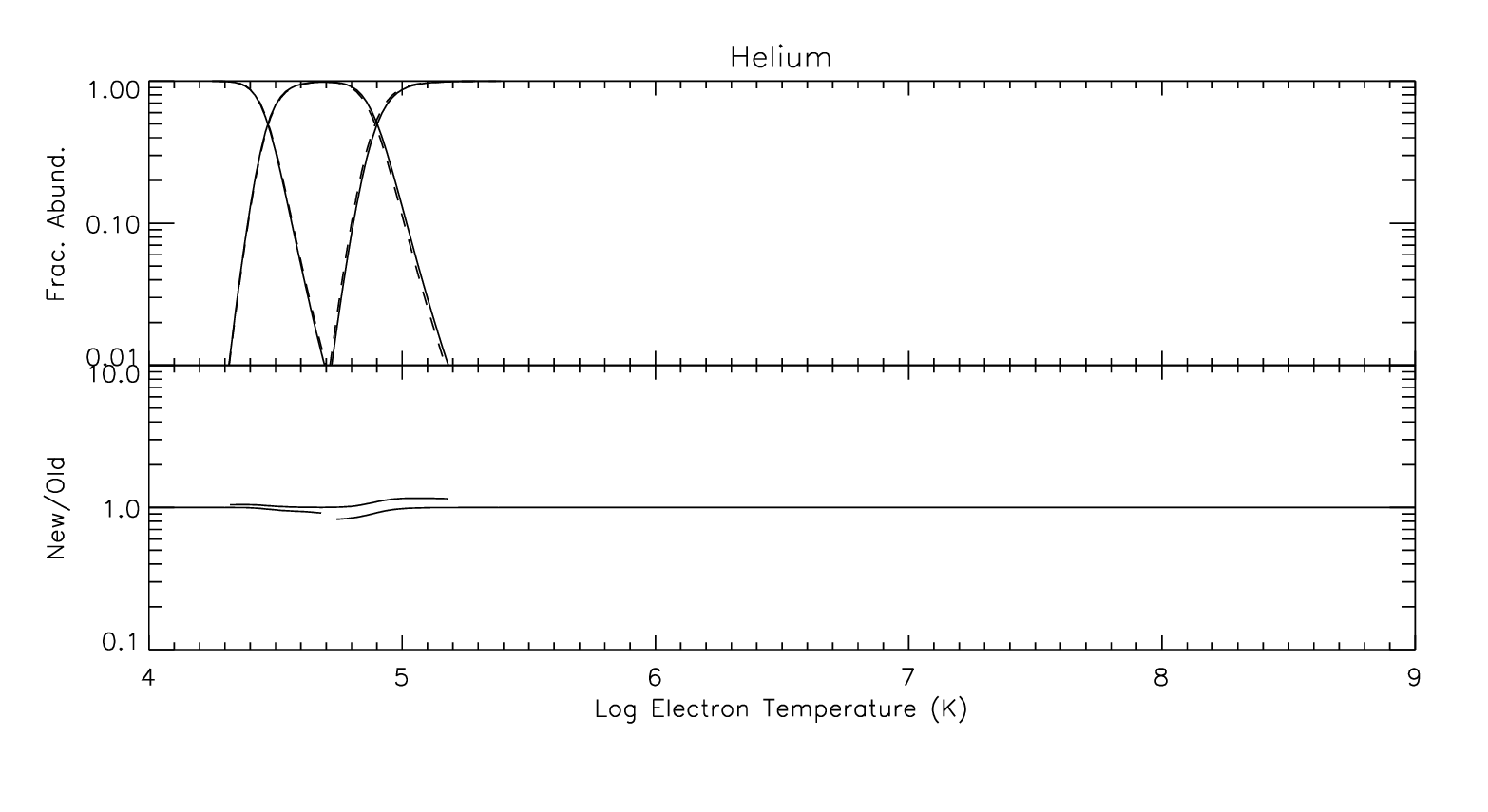}
  \caption[]{Same as Fig.~\protect\ref{fig:H Mazz} but for He.}
  \label{fig:He Mazz}
\end{figure}

\begin{figure}
  \centering
  \includegraphics[angle=90]{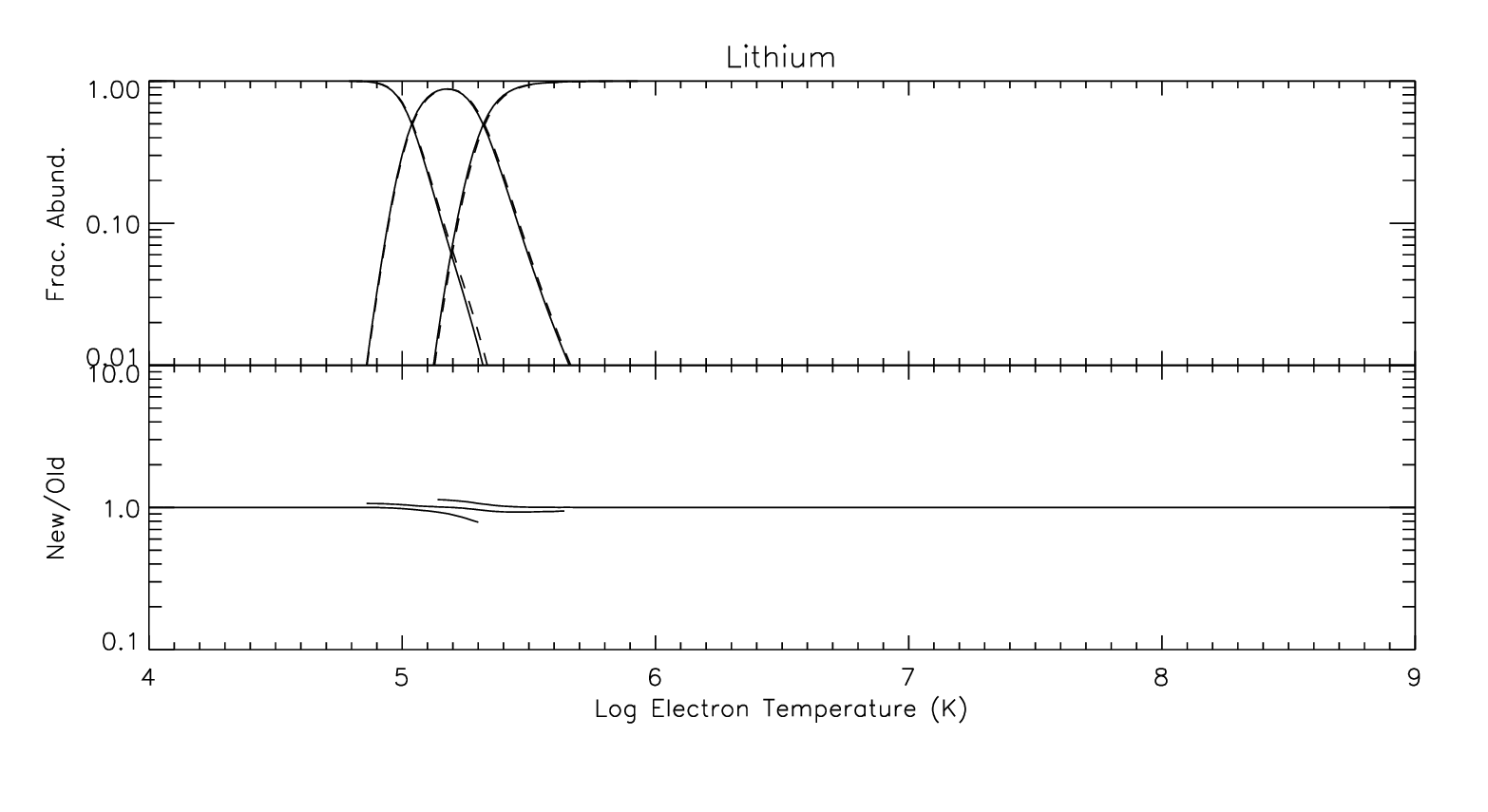}
  \caption[]{Same as Fig.~\protect\ref{fig:H Mazz} but for Li.}
  \label{fig:Li Mazz}
\end{figure}

\begin{figure}
  \centering
  \includegraphics[angle=90]{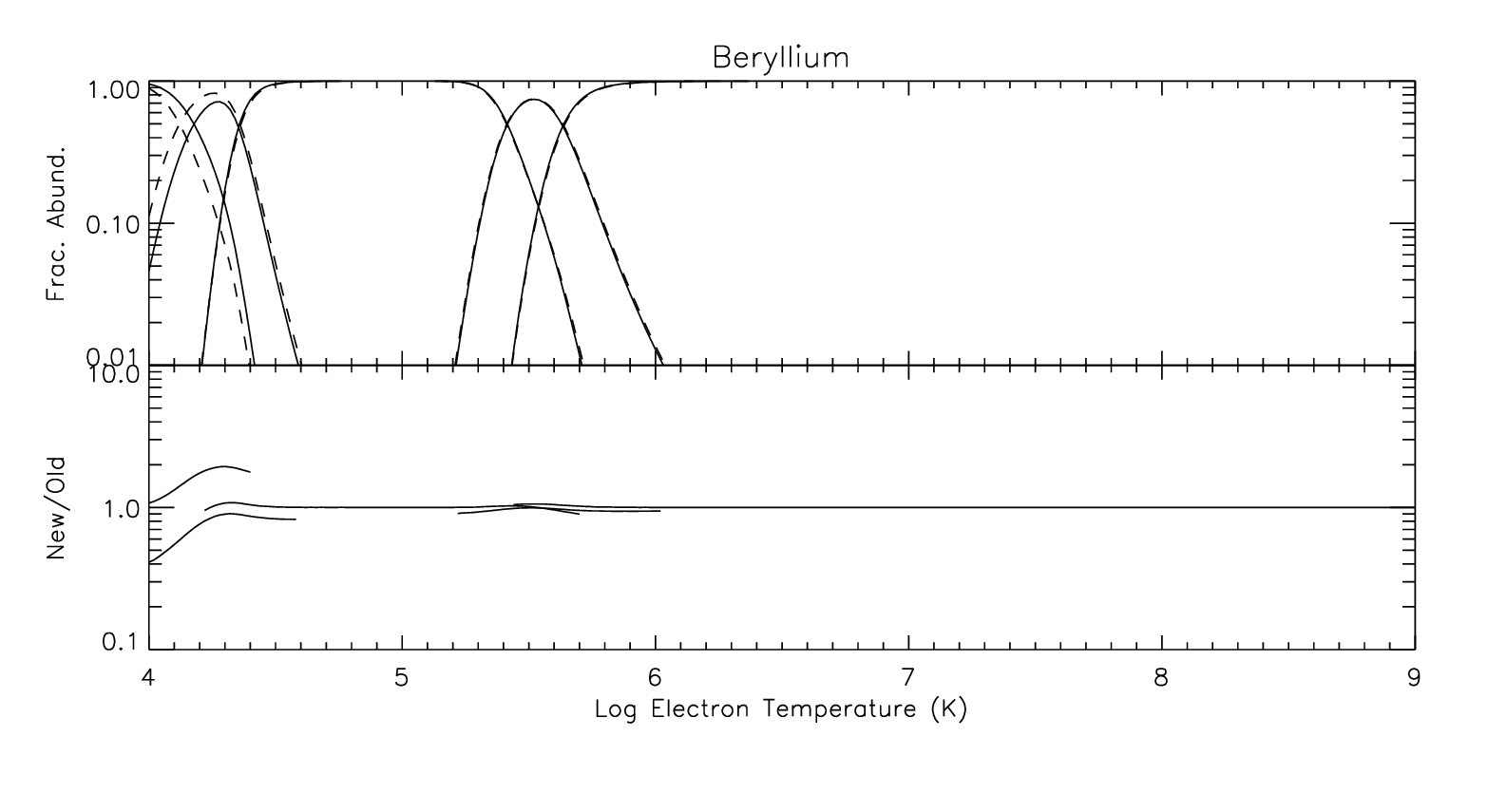}
  \caption[]{Same as Fig.~\protect\ref{fig:H Mazz} but for Be.}
  \label{fig:Be Mazz}
\end{figure}

\begin{figure}
  \centering
  \includegraphics[angle=90]{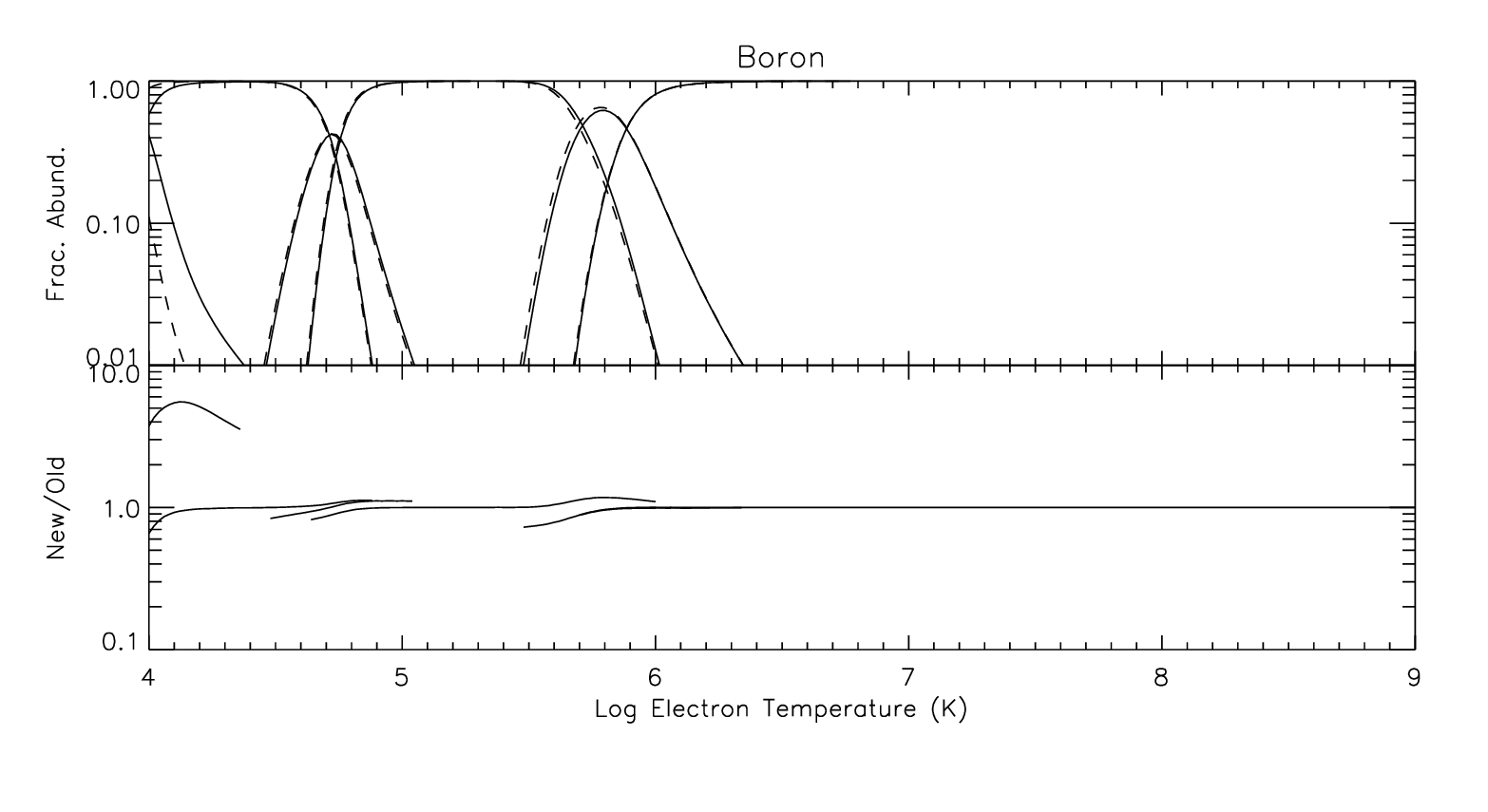}
  \caption[]{Same as Fig.~\protect\ref{fig:H Mazz} but for B.}
  \label{fig:B Mazz}
\end{figure}

\begin{figure}
   \centering
 \includegraphics[angle=90]{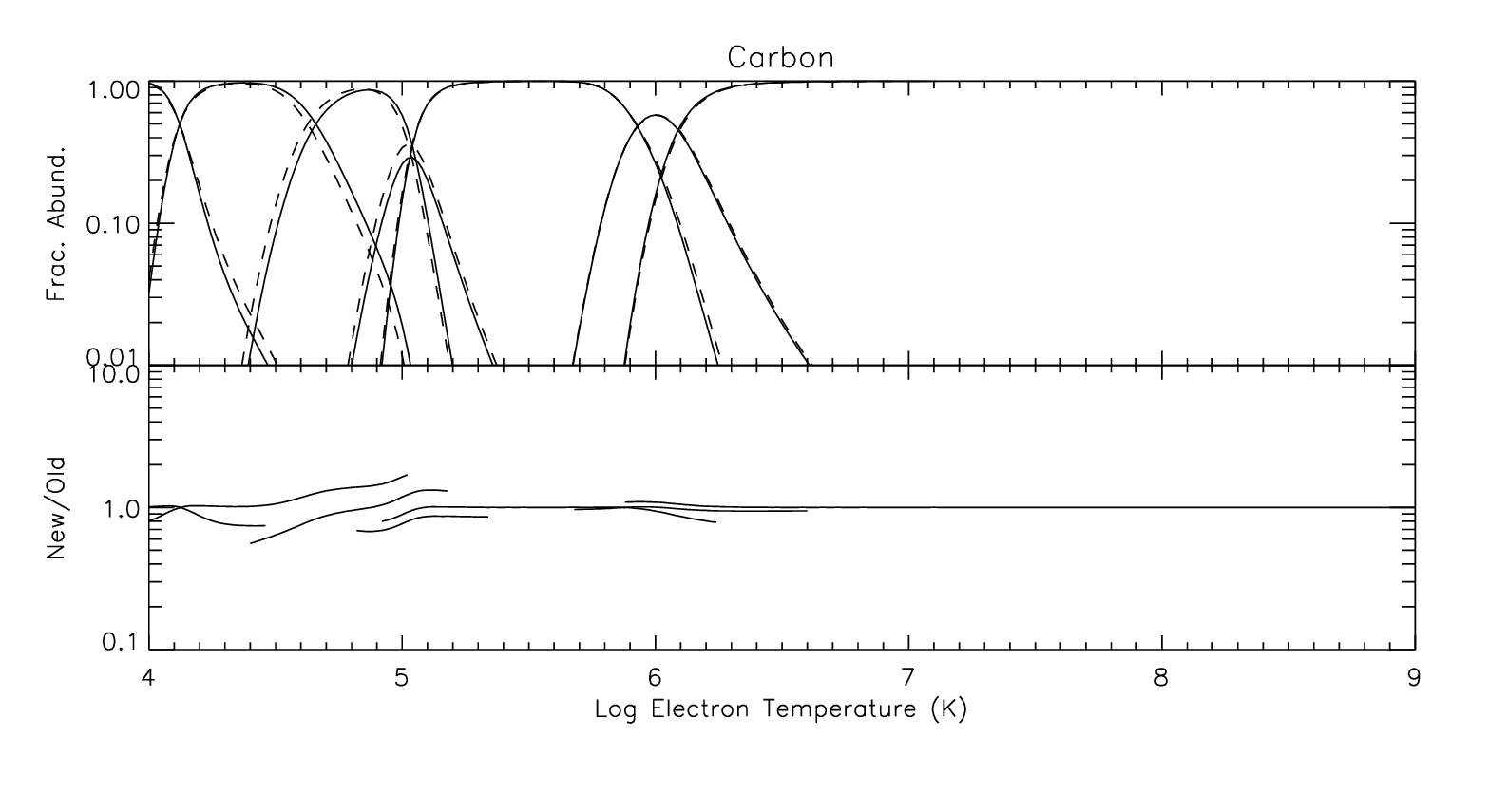}
  \caption[]{Same as Fig.~\protect\ref{fig:H Mazz} but for C.}
  \label{fig:C Mazz}
\end{figure}

\begin{figure}
  \centering
  \includegraphics[angle=90]{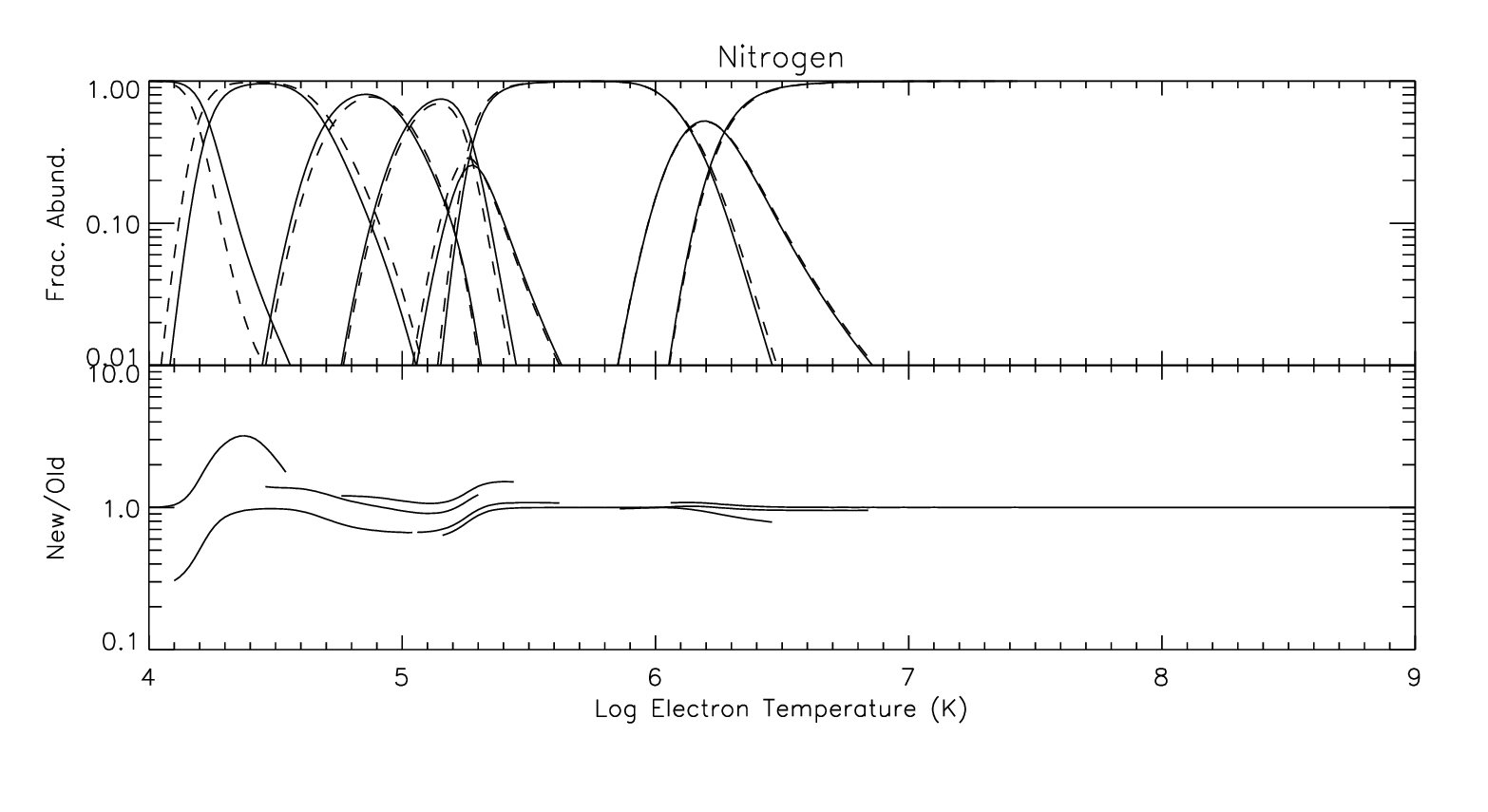}
  \caption[]{Same as Fig.~\protect\ref{fig:H Mazz} but for N.}
  \label{fig:N Mazz}
\end{figure}

\begin{figure}
  \centering
  \includegraphics[angle=90]{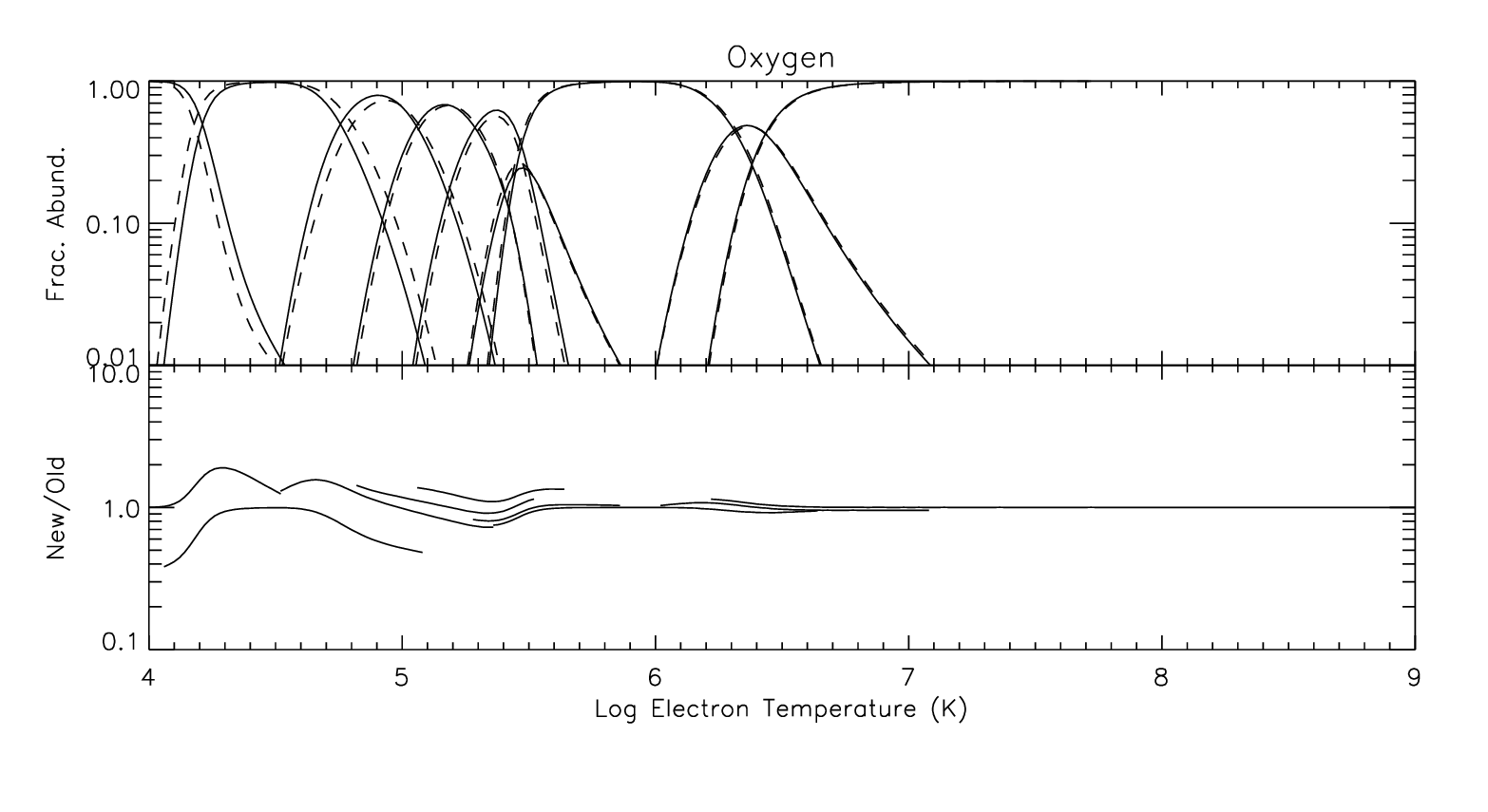}
  \caption[]{Same as Fig.~\protect\ref{fig:H Mazz} but for O.}
  \label{fig:O Mazz}
\end{figure}

\begin{figure}
  \centering
  \includegraphics[angle=90]{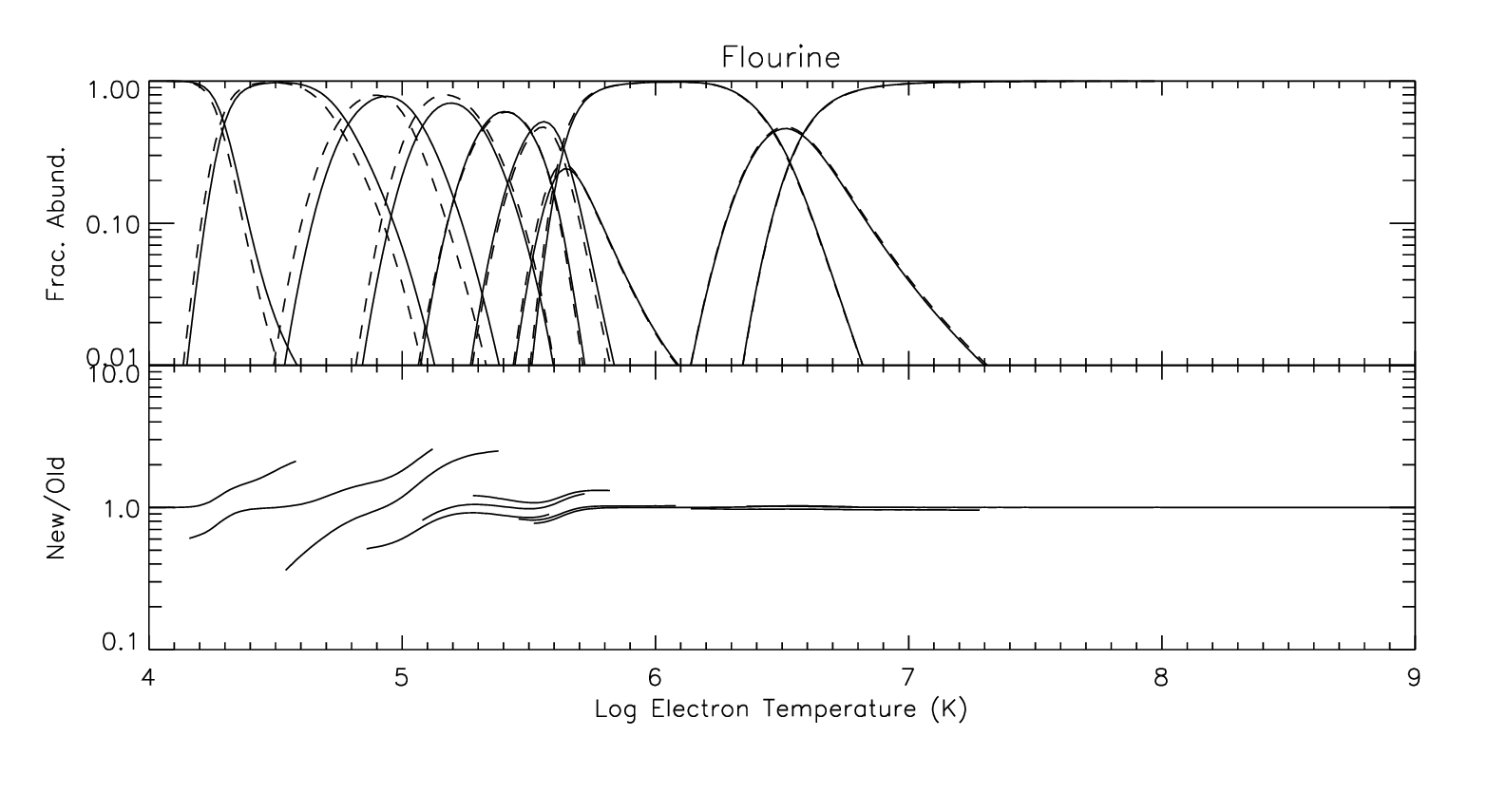}
  \caption[]{Same as Fig.~\protect\ref{fig:H Mazz} but for F.}
  \label{fig:F Mazz}
\end{figure}

\begin{figure}
  \centering
  \includegraphics[angle=90]{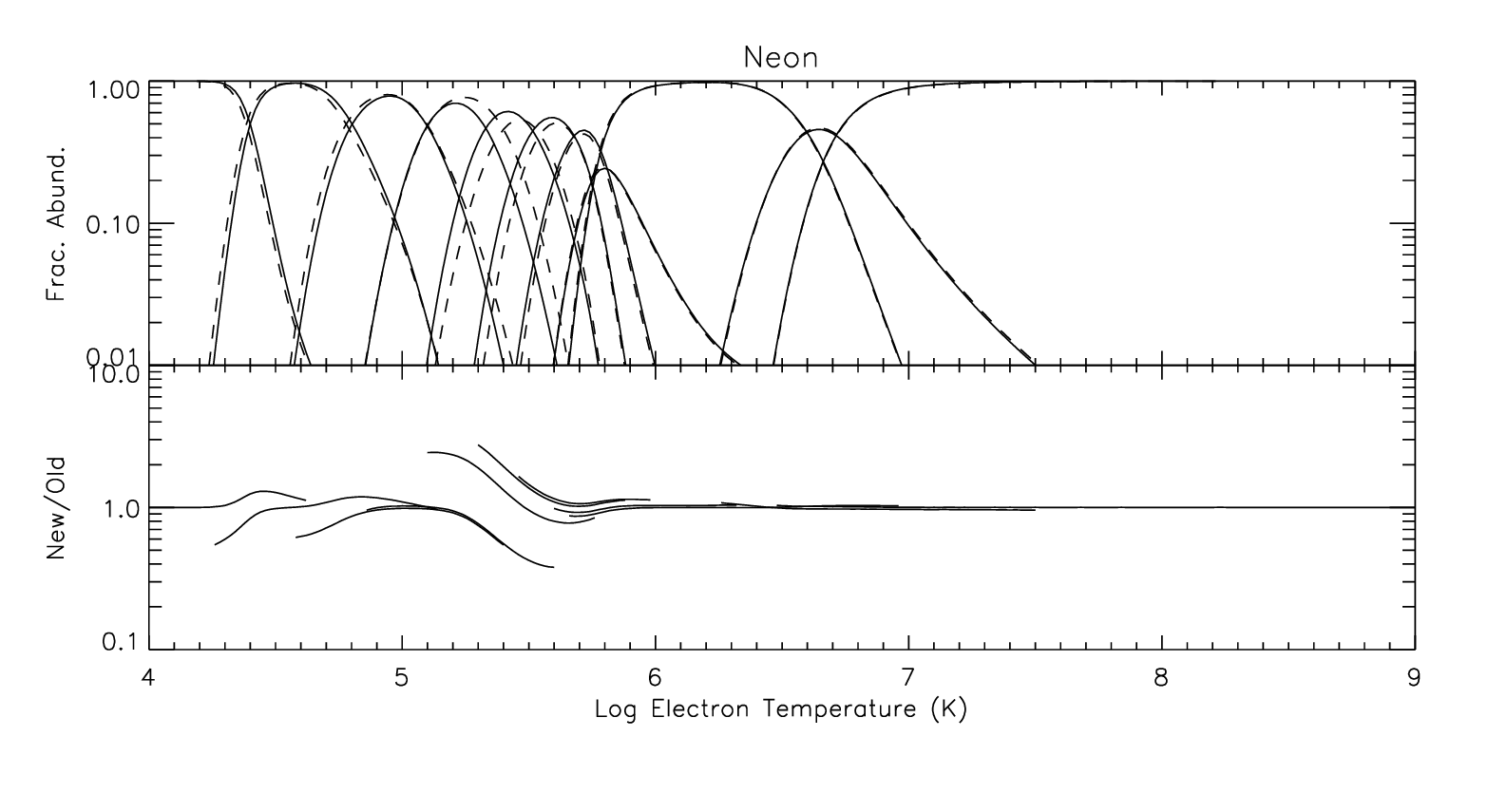}
  \caption[]{Same as Fig.~\protect\ref{fig:H Mazz} but for Ne.}
  \label{fig:Ne Mazz}
\end{figure}

\begin{figure}
  \centering
  \includegraphics[angle=90]{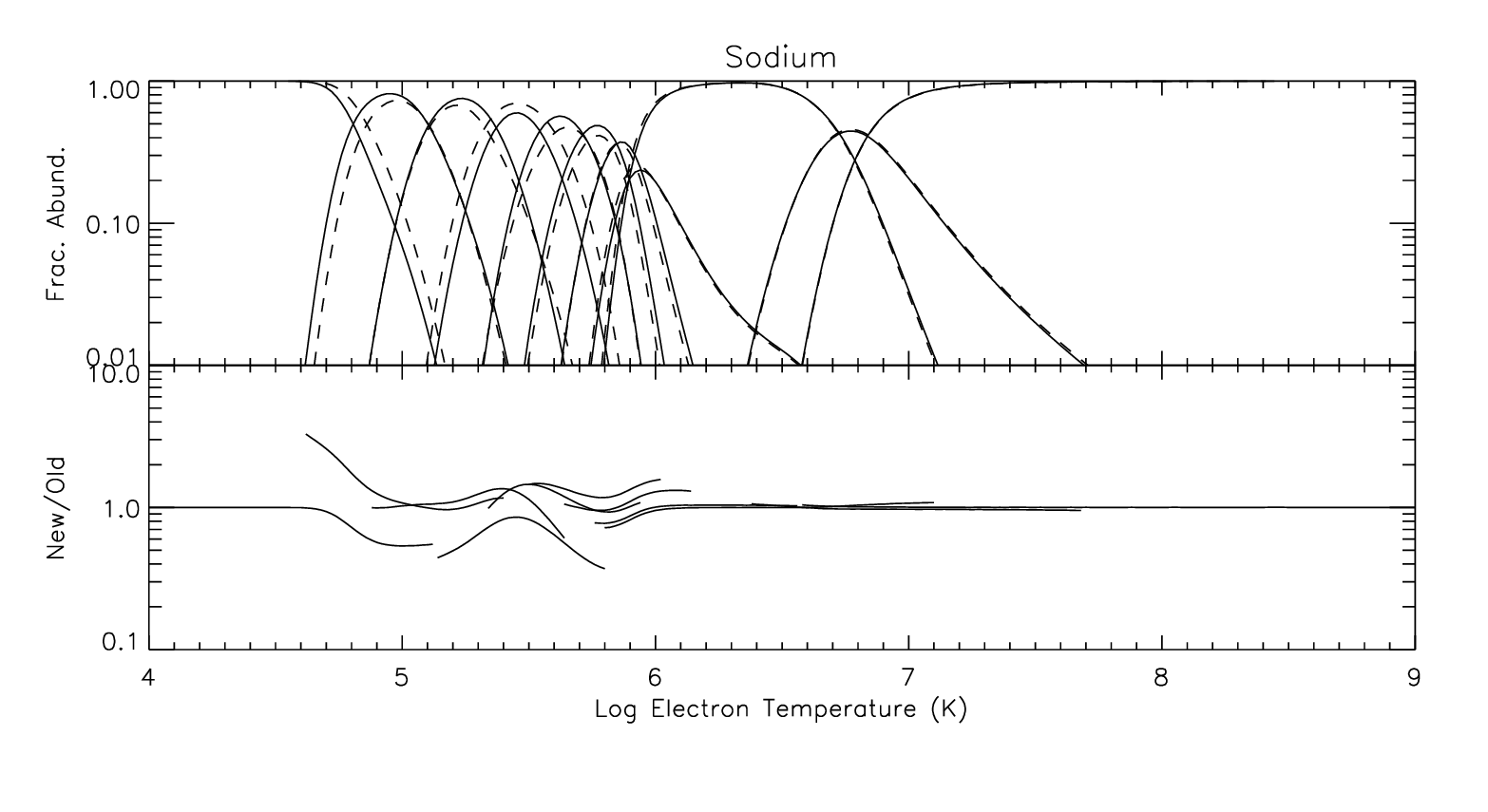}
  \caption[]{Same as Fig.~\protect\ref{fig:H Mazz} but for Na.}
  \label{fig:Na Mazz}
\end{figure}

\begin{figure}
  \centering
  \includegraphics[angle=90]{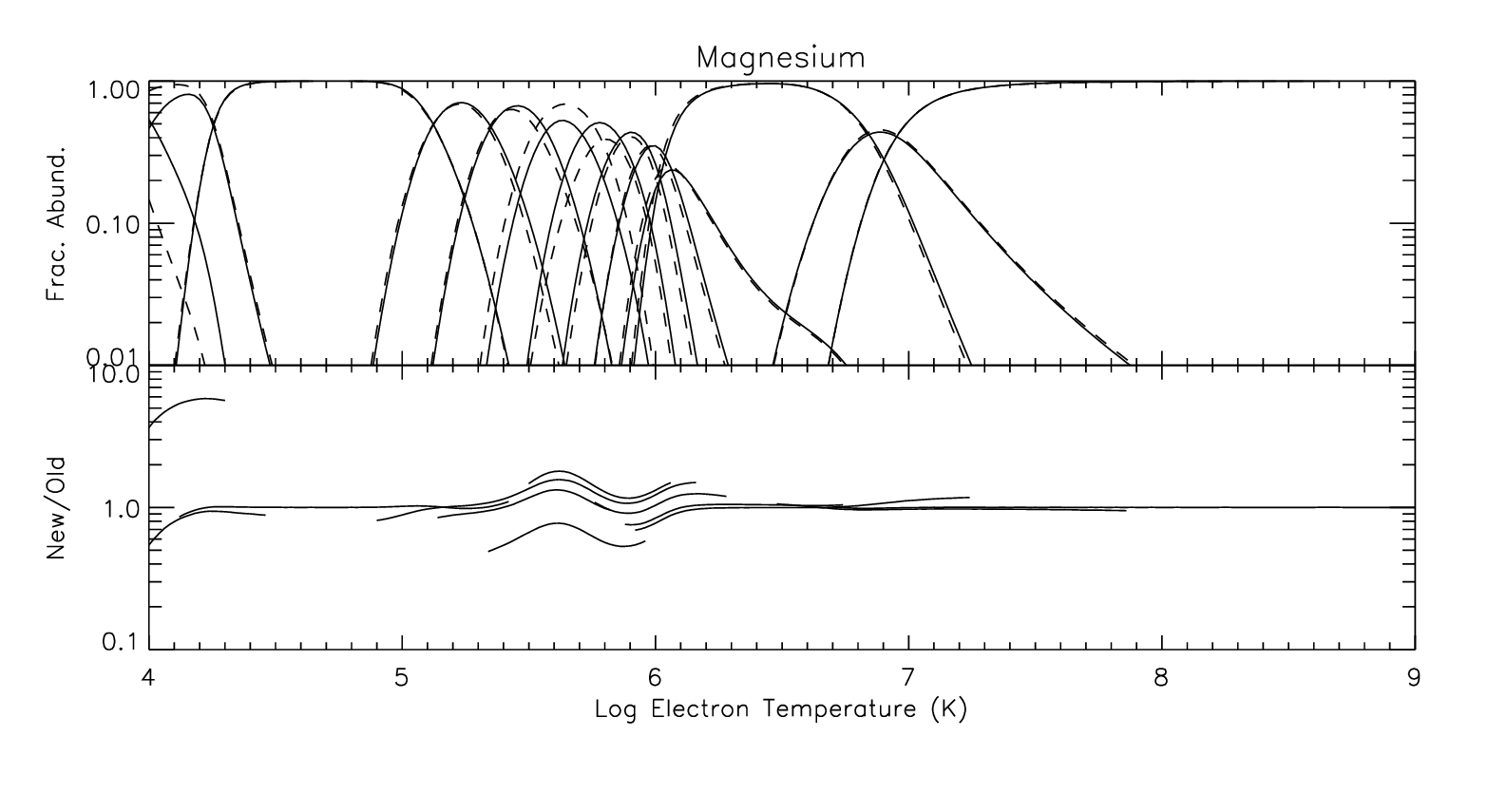}
  \caption[]{Same as Fig.~\protect\ref{fig:H Mazz} but for Mg.}
  \label{fig:Mg Mazz}
\end{figure}

\begin{figure}
  \centering
  \includegraphics[angle=90]{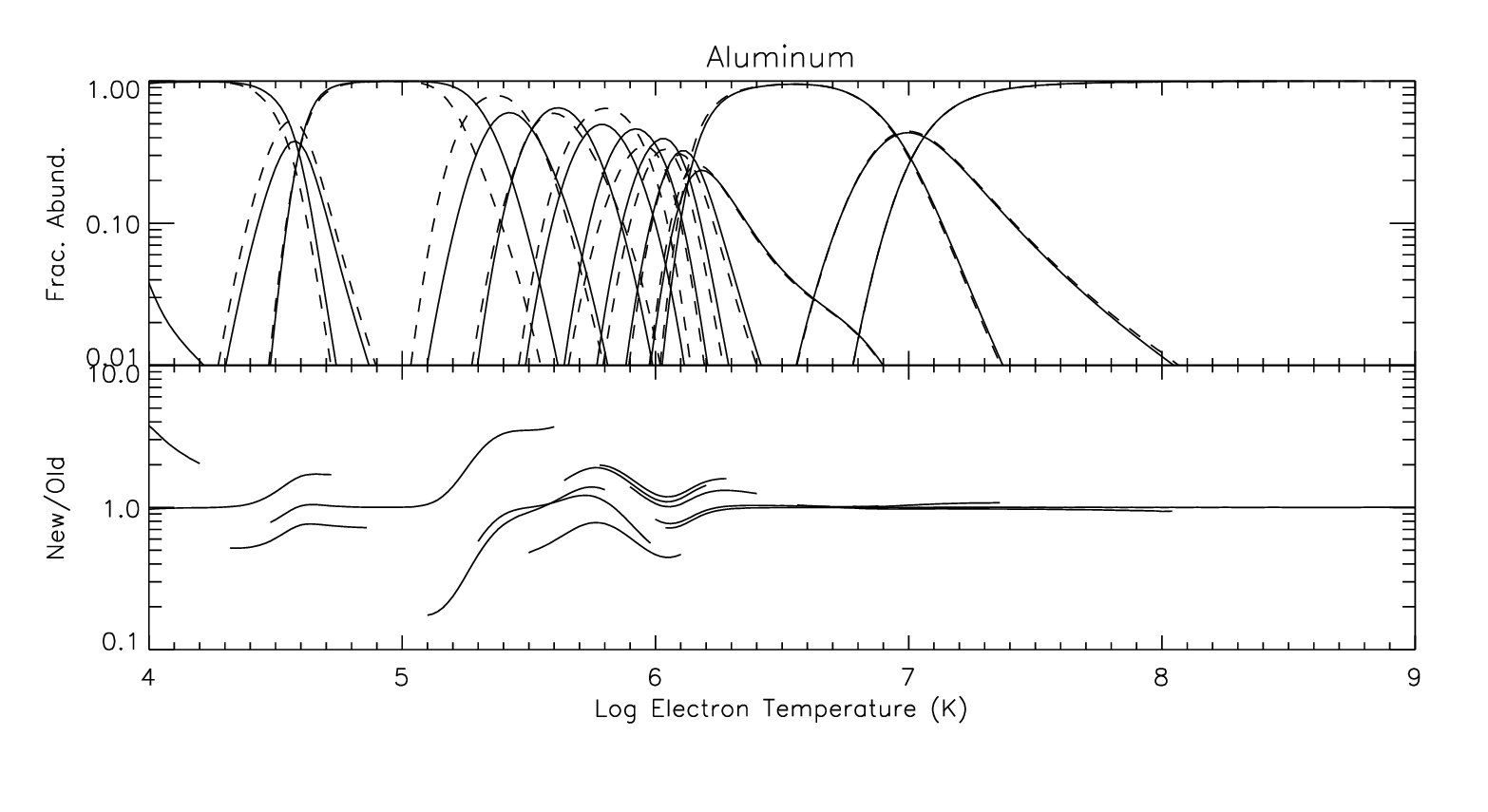}
  \caption[]{Same as Fig.~\protect\ref{fig:H Mazz} but for Al.}
  \label{fig:Al Mazz}
\end{figure}

\begin{figure}
  \centering
  \includegraphics[angle=90]{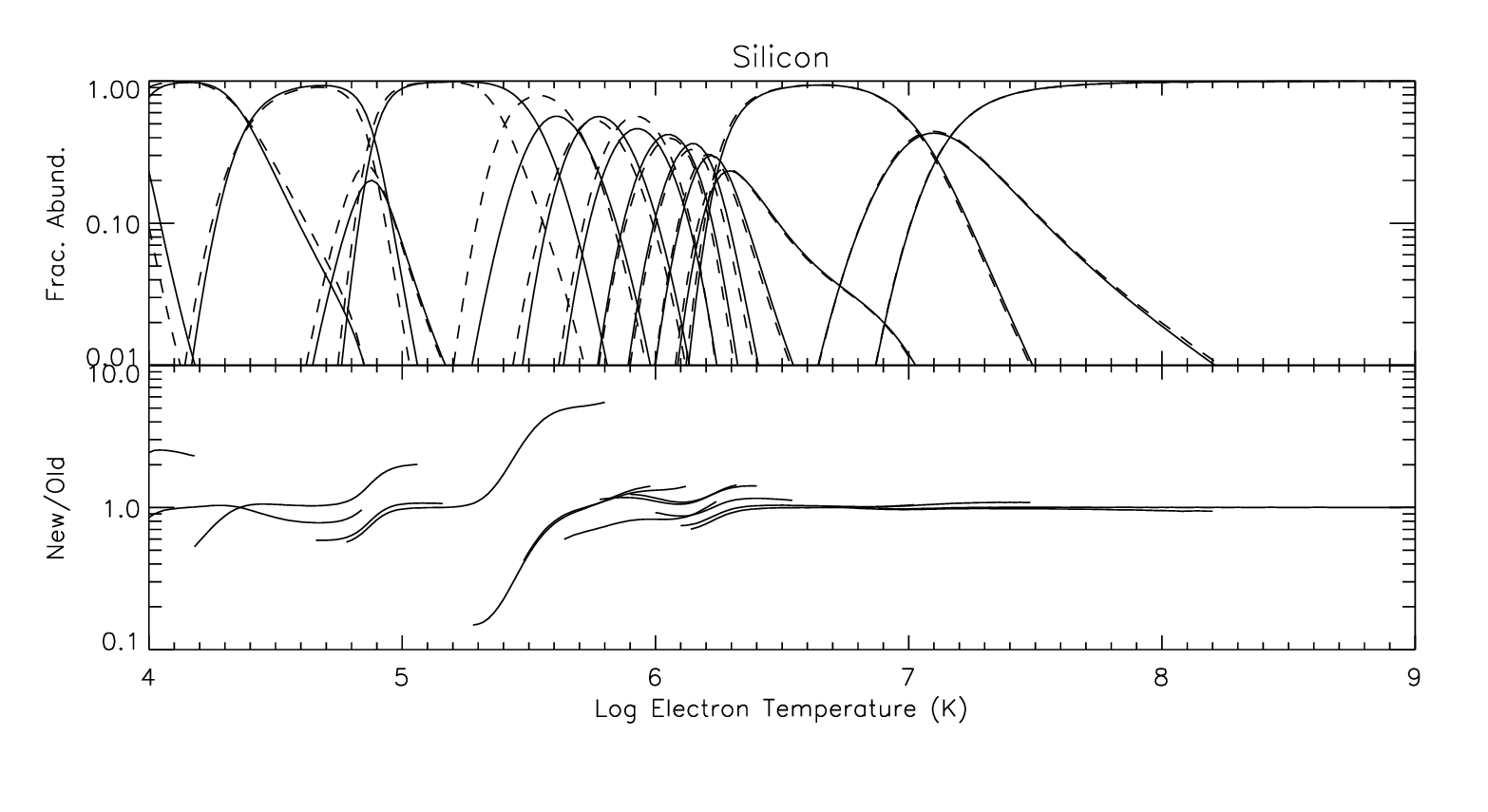}
  \caption[]{Same as Fig.~\protect\ref{fig:H Mazz} but for Si.}
  \label{fig:Si Mazz}
\end{figure}

\begin{figure}
  \centering
  \includegraphics[angle=90]{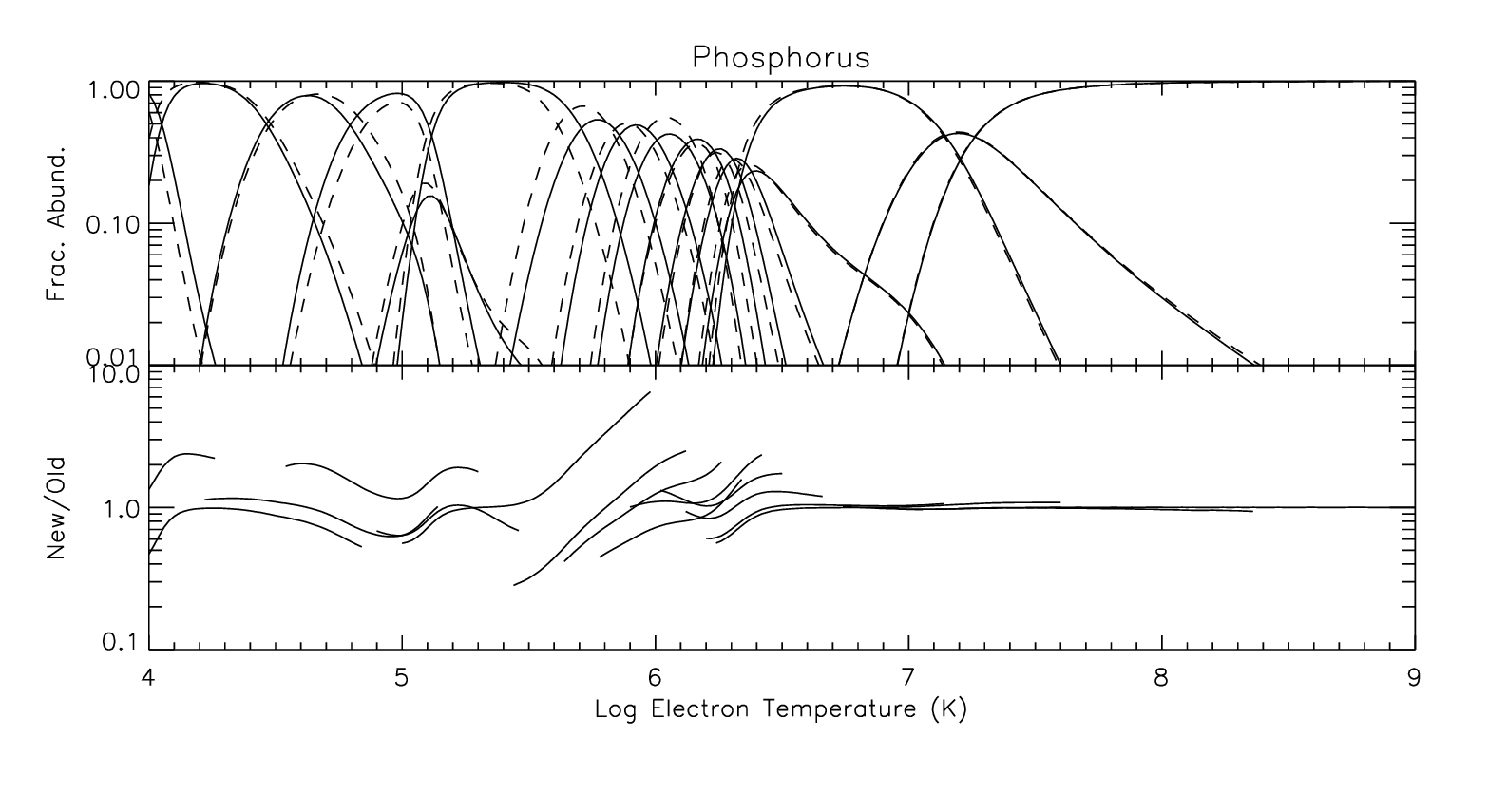}
  \caption[]{Same as Fig.~\protect\ref{fig:H Mazz} but for P.}
  \label{fig:P Mazz}
\end{figure}

\begin{figure}
  \centering
  \includegraphics[angle=90]{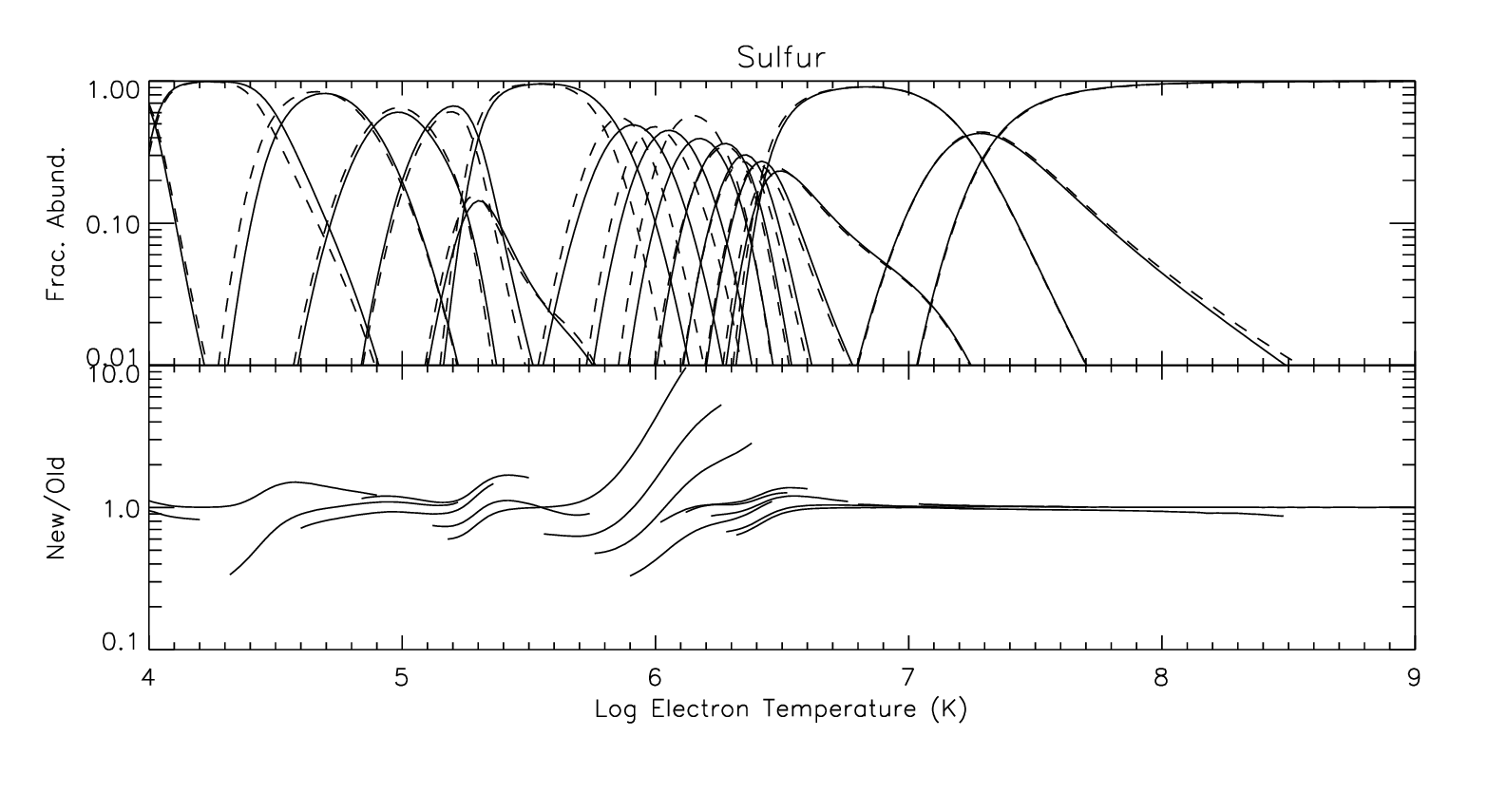}
  \caption[]{Same as Fig.~\protect\ref{fig:H Mazz} but for S.}
  \label{fig:S Mazz}
\end{figure}

\begin{figure}
  \centering
  \includegraphics[angle=90]{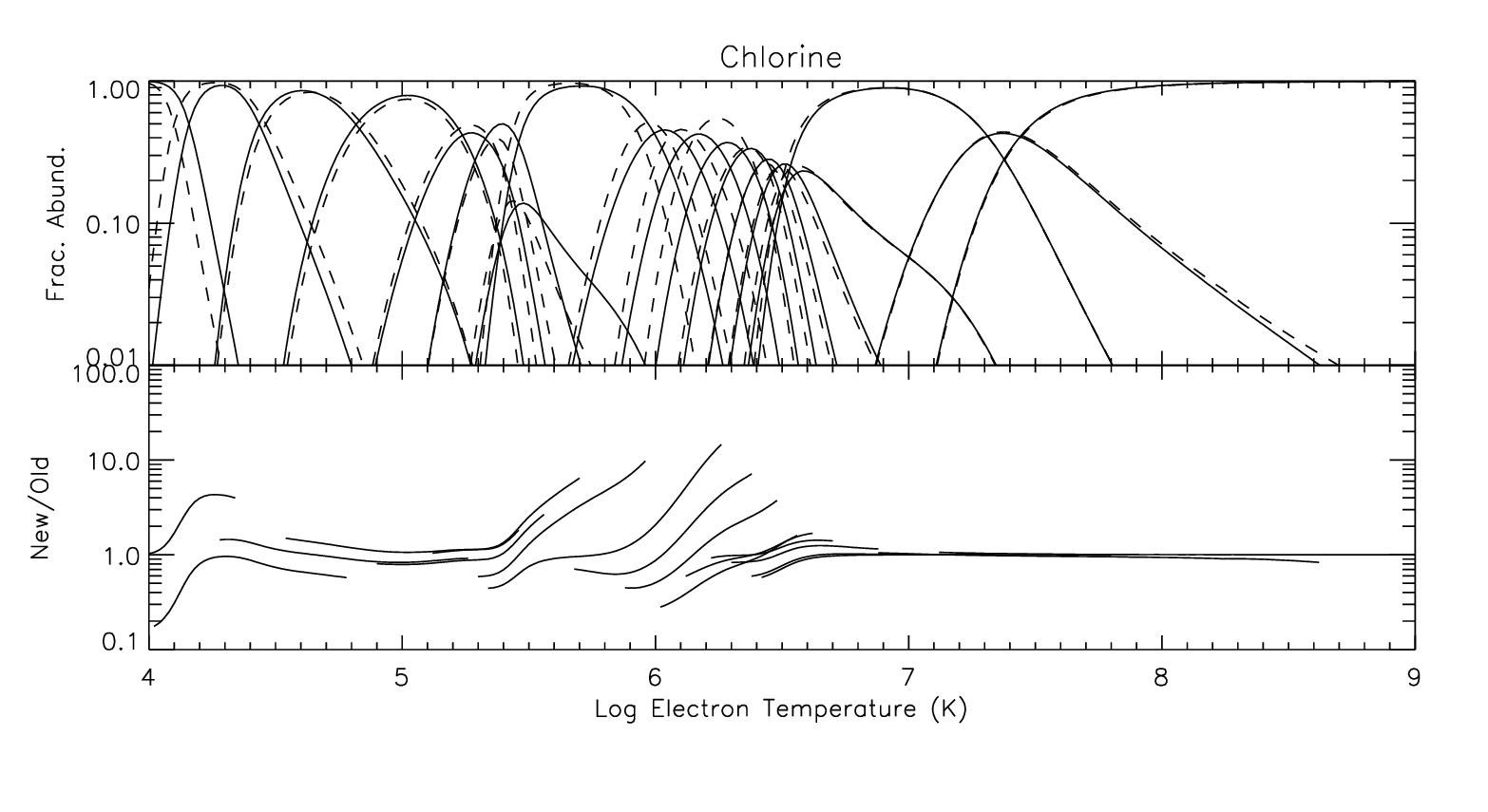}
  \caption[]{Same as Fig.~\protect\ref{fig:H Mazz} but for Cl.}
  \label{fig:Cl Mazz}
\end{figure}

\begin{figure}
  \centering
  \includegraphics[angle=90]{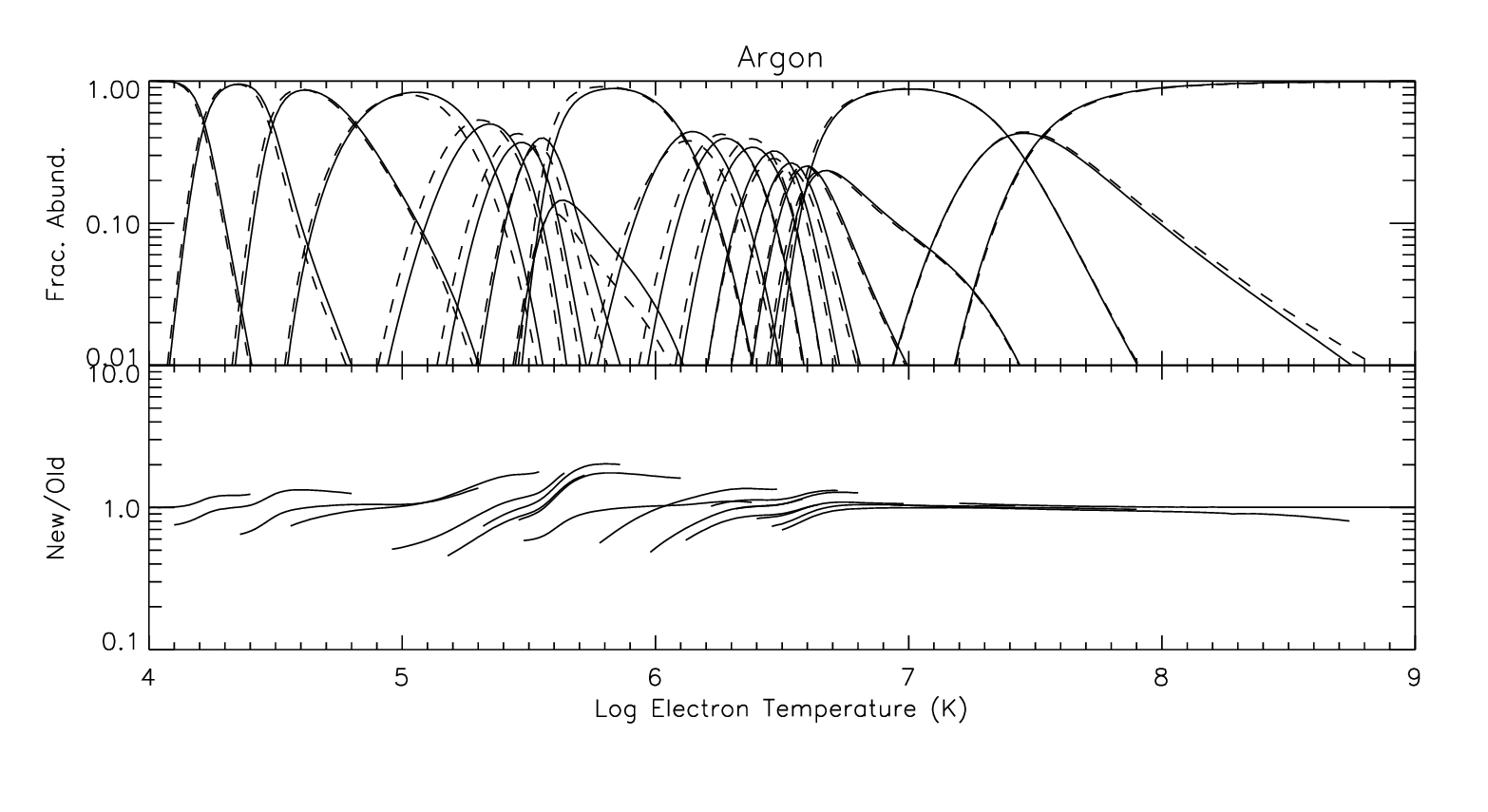}
  \caption[]{Same as Fig.~\protect\ref{fig:H Mazz} but for Ar.}
  \label{fig:Ar Mazz}
\end{figure}

\begin{figure}
  \centering
  \includegraphics[angle=90]{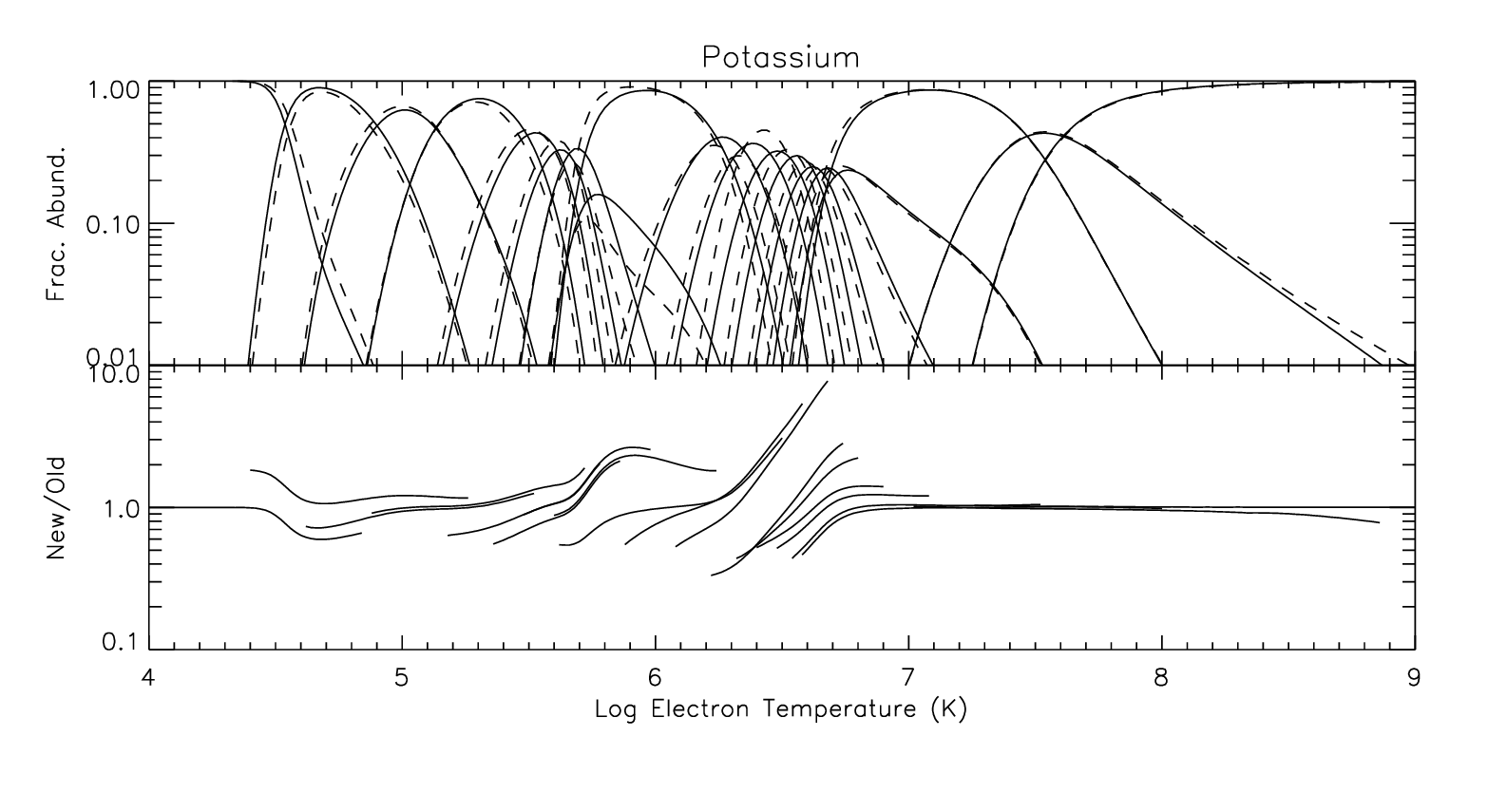}
  \caption[]{Same as Fig.~\protect\ref{fig:H Mazz} but for K.}
  \label{fig:K Mazz}
\end{figure}

\begin{figure}
  \centering
  \includegraphics[angle=90]{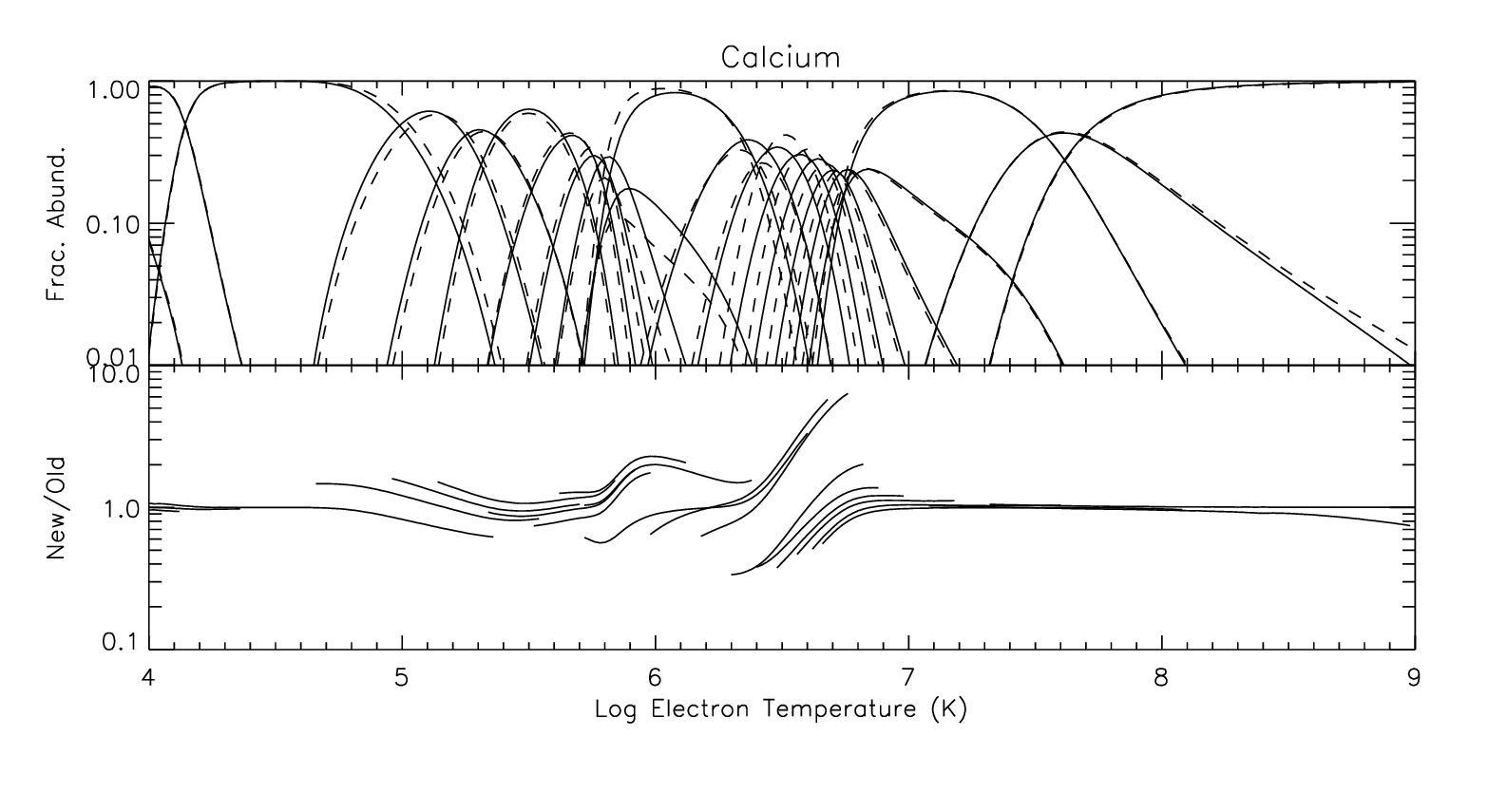}
  \caption[]{Same as Fig.~\protect\ref{fig:H Mazz} but for Ca.}
  \label{fig:Ca Mazz}
\end{figure}

\begin{figure}
  \centering
  \includegraphics[angle=90]{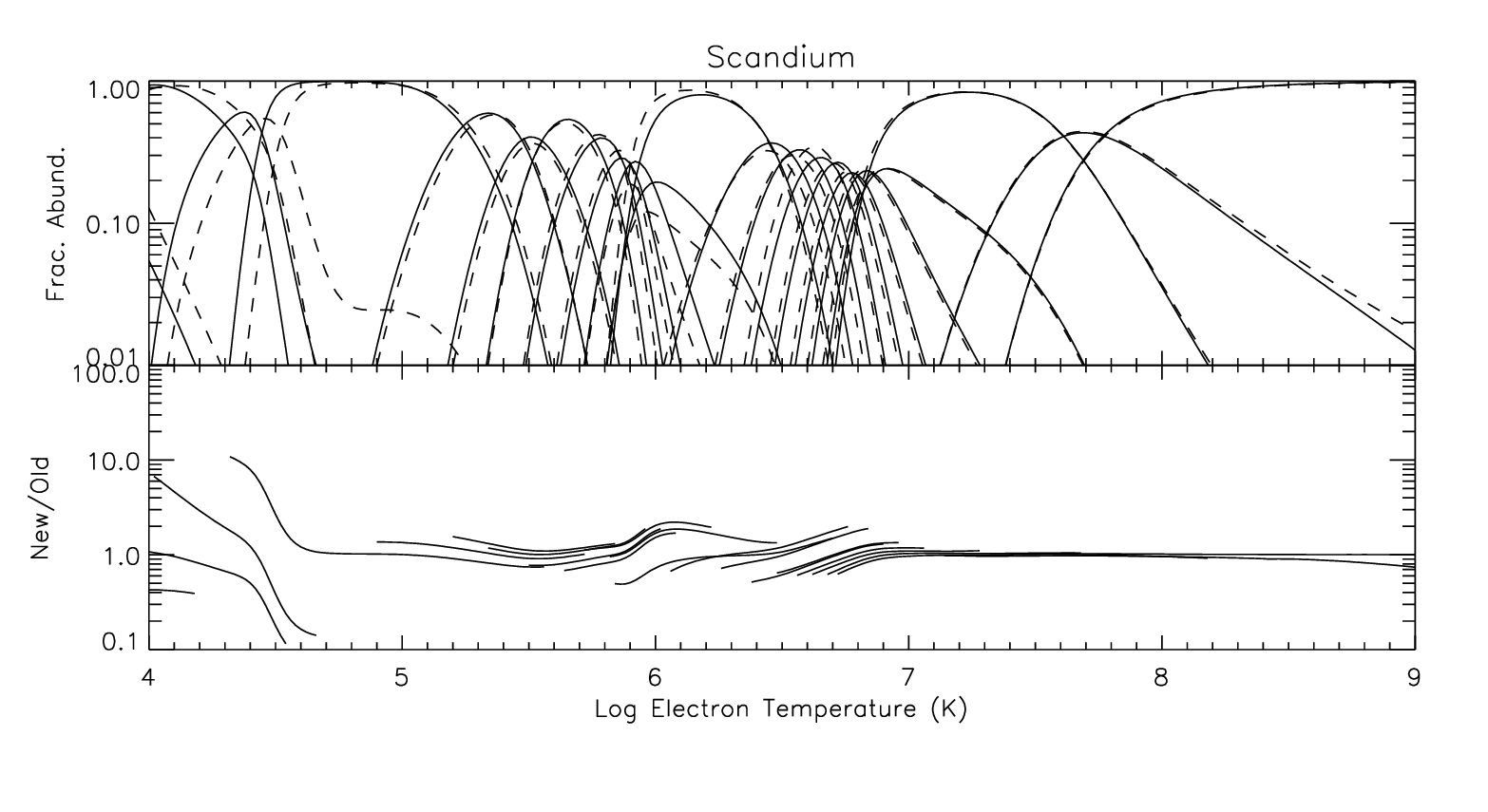}
  \caption[]{Same as Fig.~\protect\ref{fig:H Mazz} but for Sc.}
  \label{fig:Sc Mazz}
\end{figure}

\begin{figure}
  \centering
  \includegraphics[angle=90]{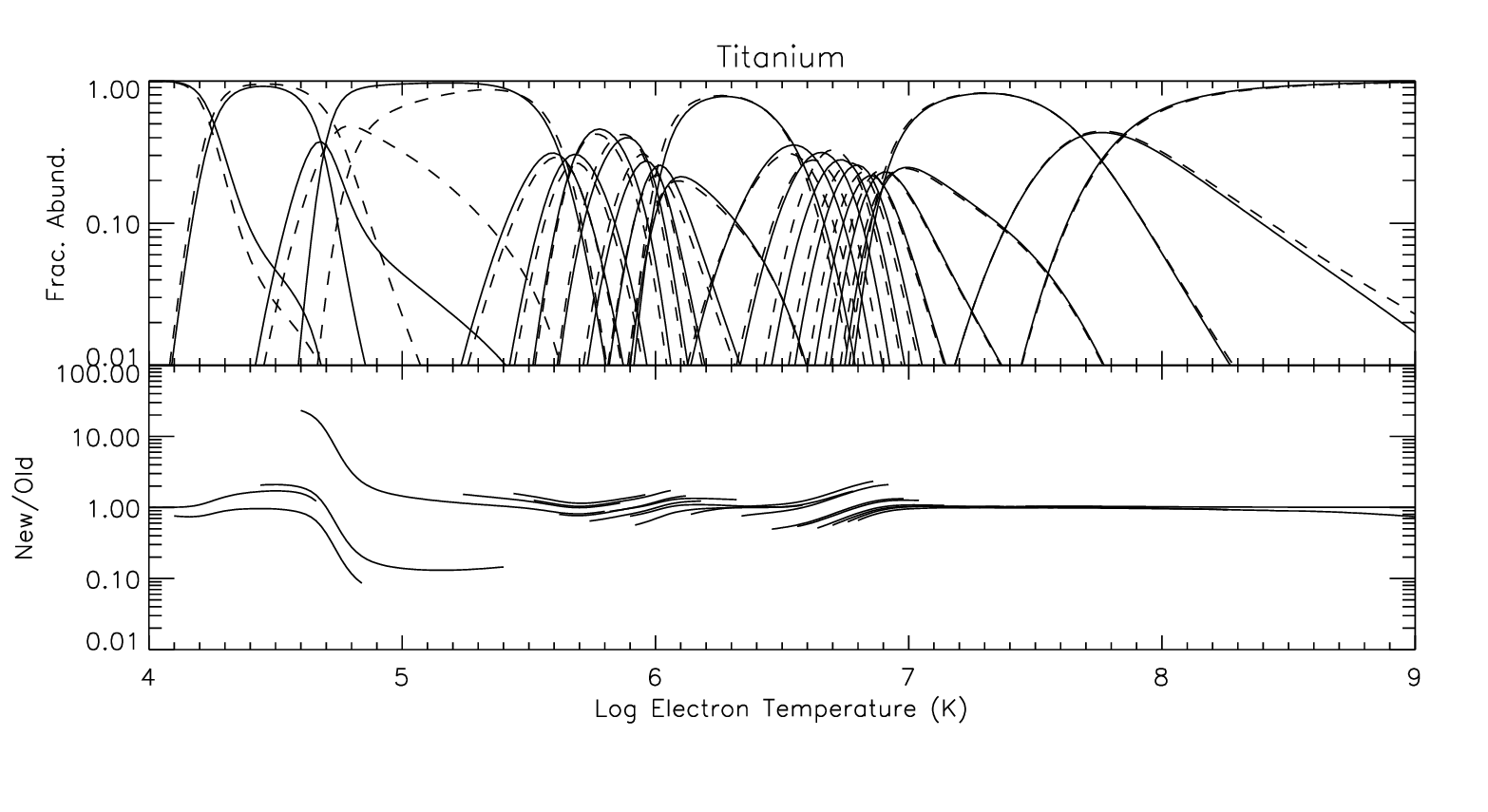}
  \caption[]{Same as Fig.~\protect\ref{fig:H Mazz} but for Ti.}
  \label{fig:Ti Mazz}
\end{figure}

\begin{figure}
  \centering
  \includegraphics[angle=90]{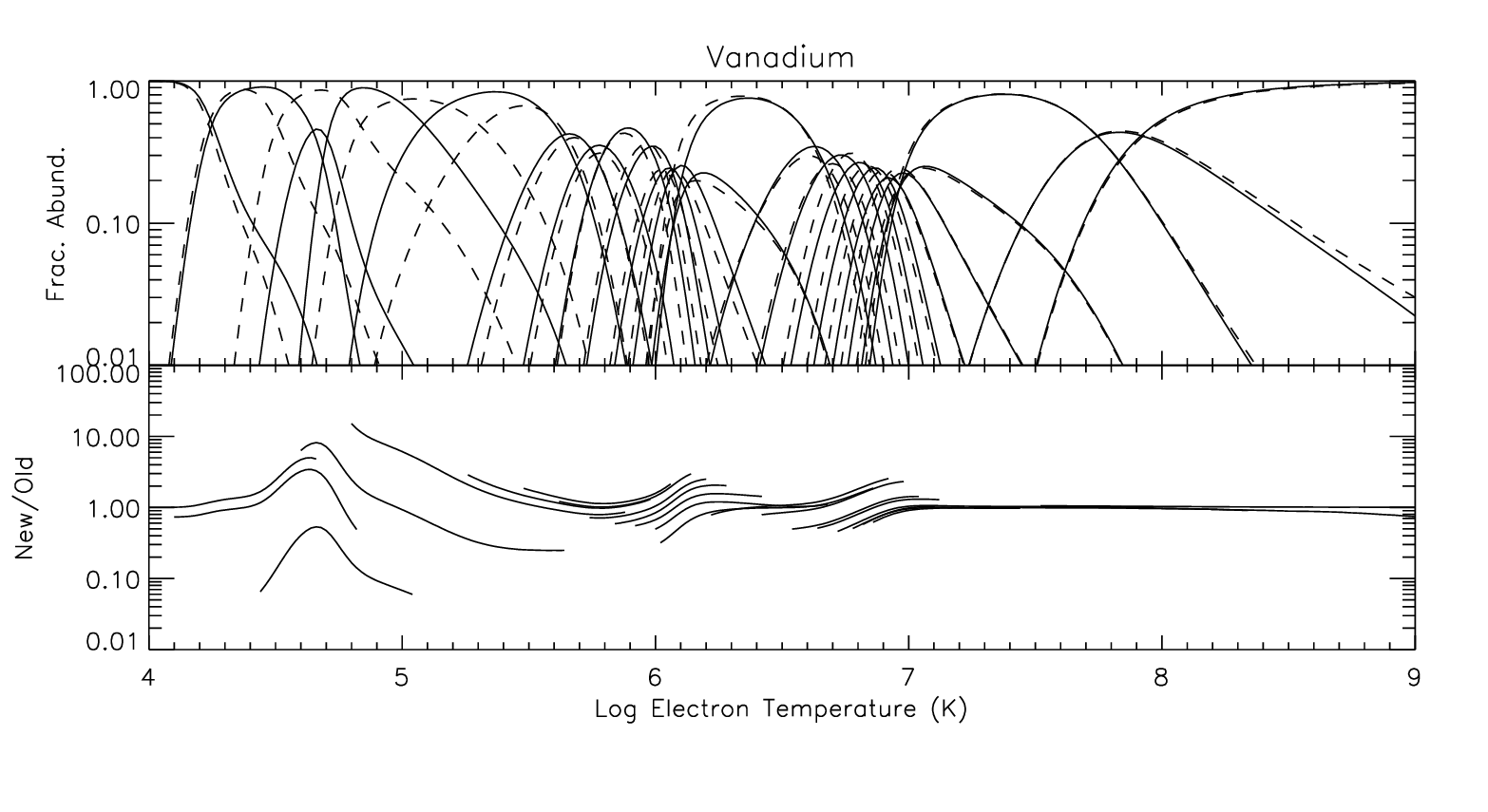}
  \caption[]{Same as Fig.~\protect\ref{fig:H Mazz} but for V.}
  \label{fig:V Mazz}
\end{figure}

\begin{figure}
  \centering
  \includegraphics[angle=90]{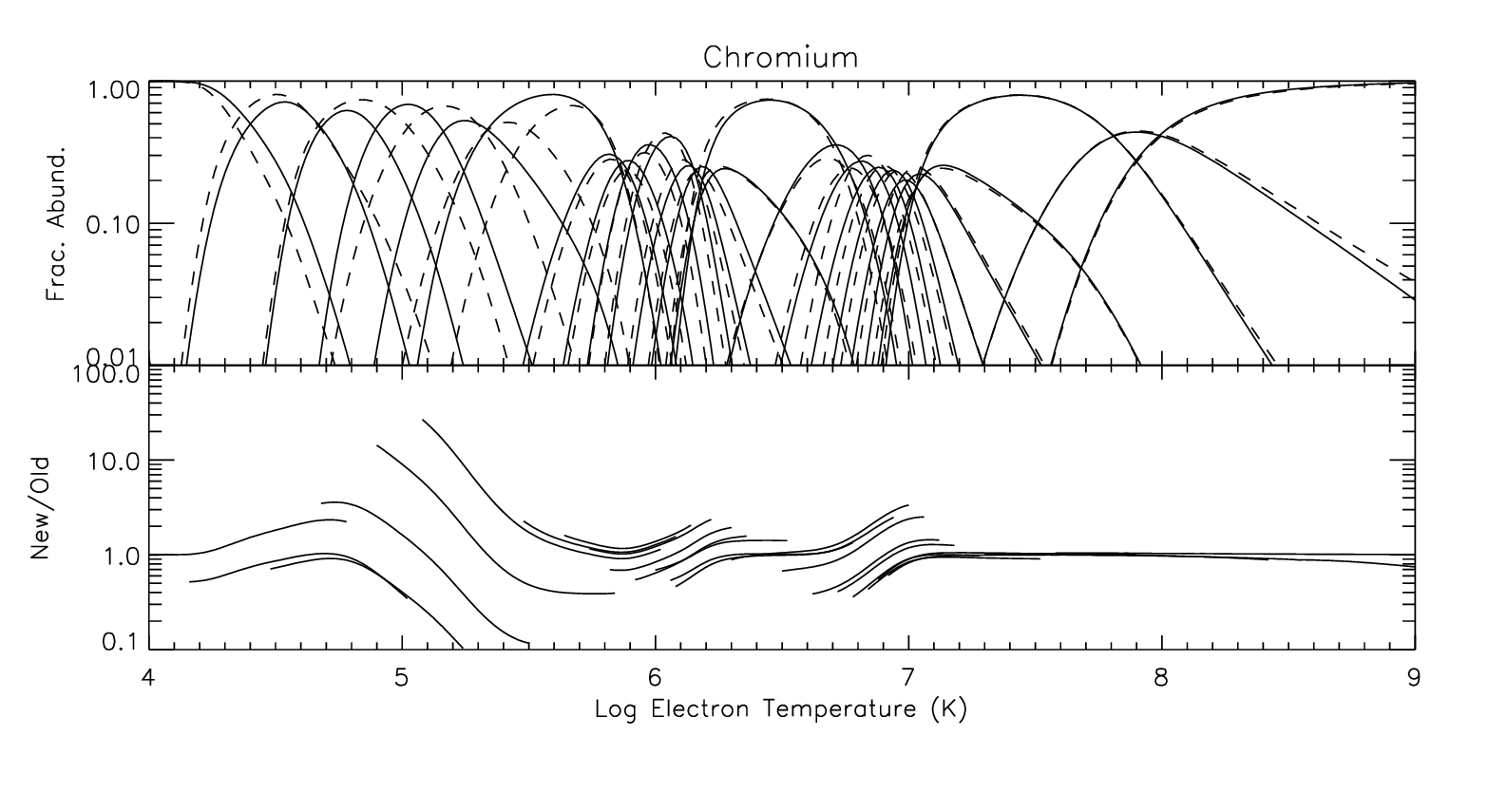}
  \caption[]{Same as Fig.~\protect\ref{fig:H Mazz} but for Cr.}
  \label{fig:Cr Mazz}
\end{figure}

\begin{figure}
  \centering
  \includegraphics[angle=90]{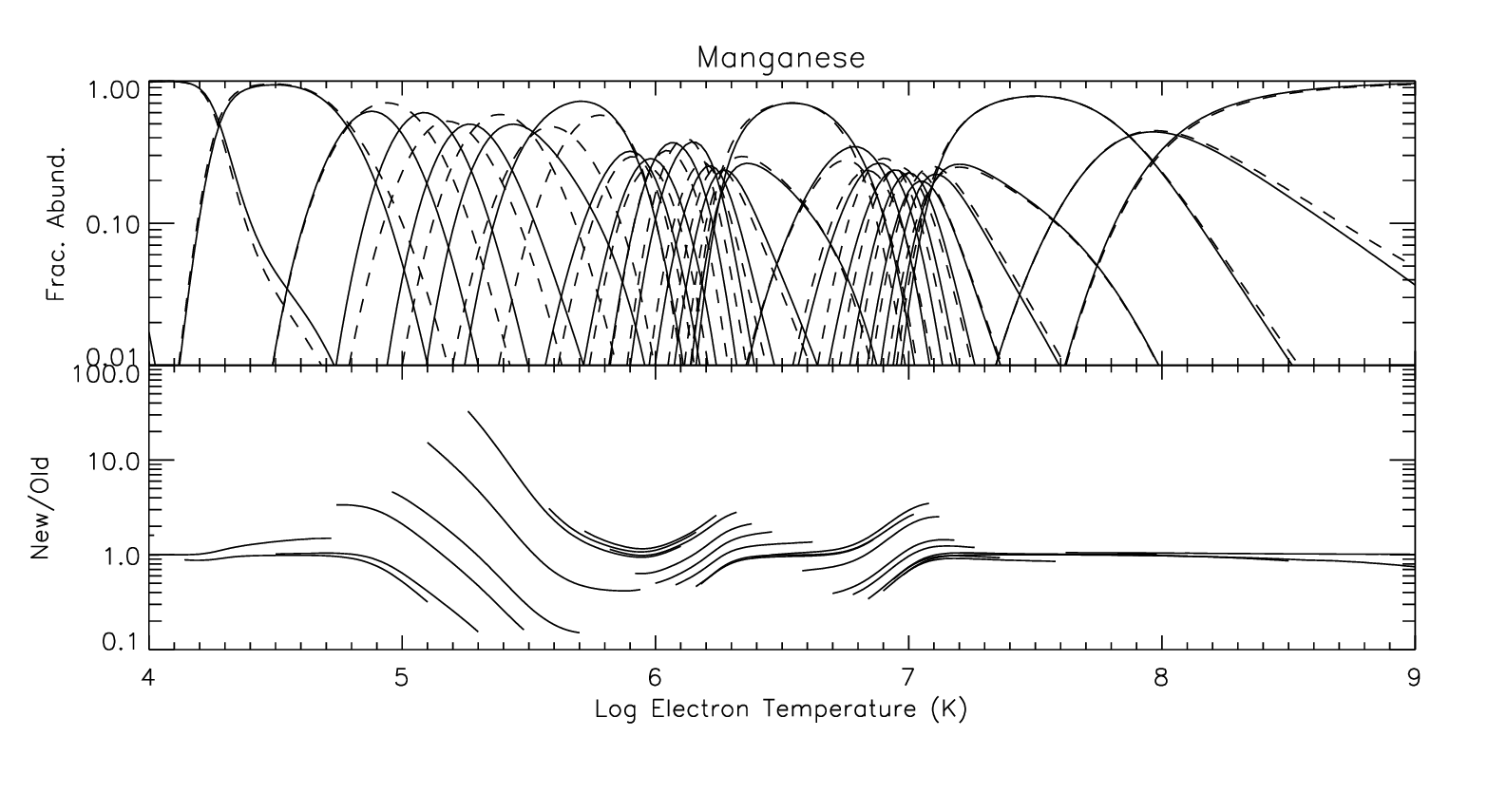}
  \caption[]{Same as Fig.~\protect\ref{fig:H Mazz} but for Mn.}
  \label{fig:Mn Mazz}
\end{figure}

\begin{figure}
  \centering
  \includegraphics[angle=90]{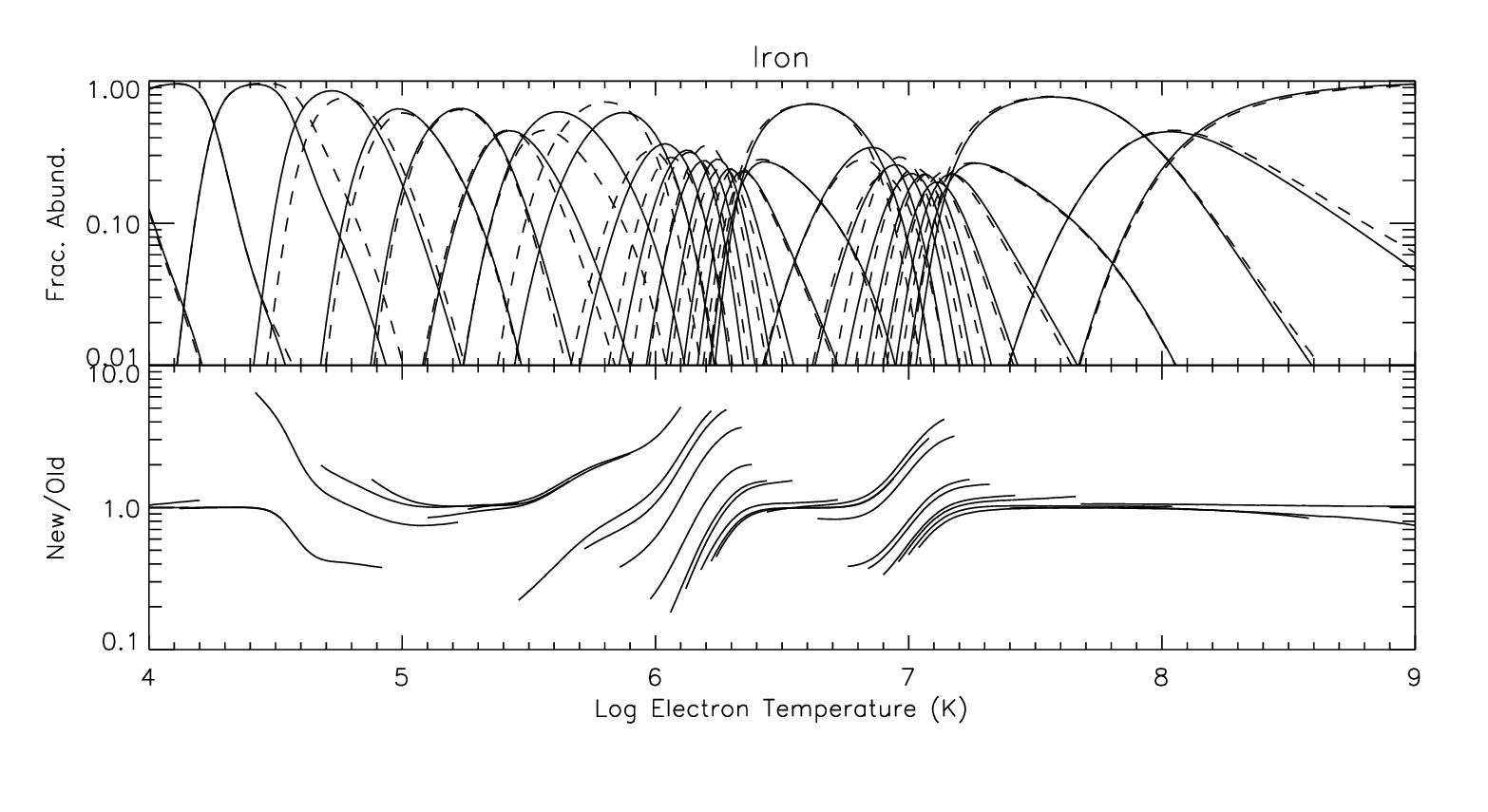}
  \caption[]{Same as Fig.~\protect\ref{fig:H Mazz} but for Fe.}
  \label{fig:Fe Mazz}
\end{figure}

\begin{figure}
  \centering
  \includegraphics[angle=90]{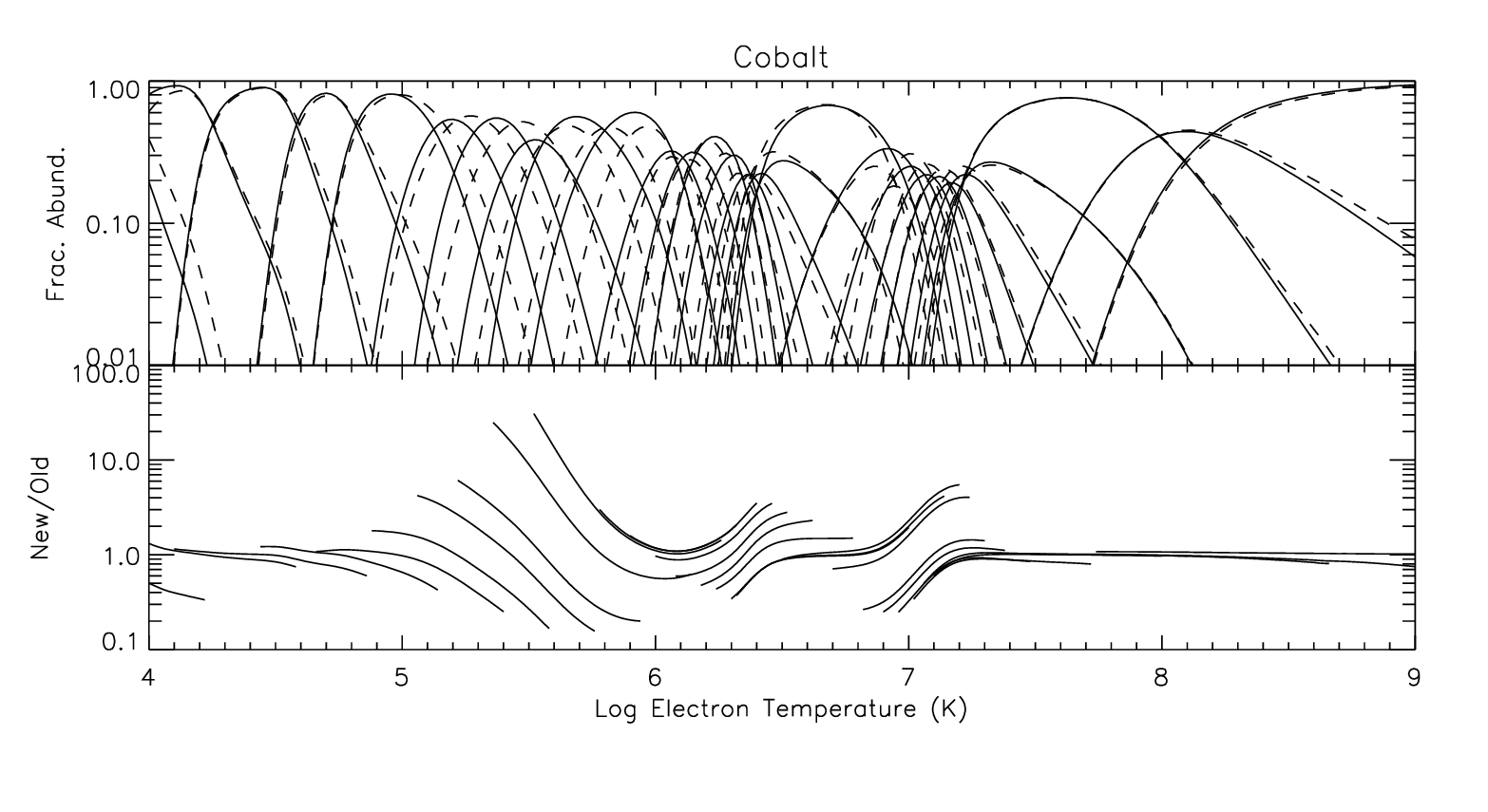}
  \caption[]{Same as Fig.~\protect\ref{fig:H Mazz} but for Co.}
  \label{fig:Co Mazz}
\end{figure}

\begin{figure}
  \centering
  \includegraphics[angle=90]{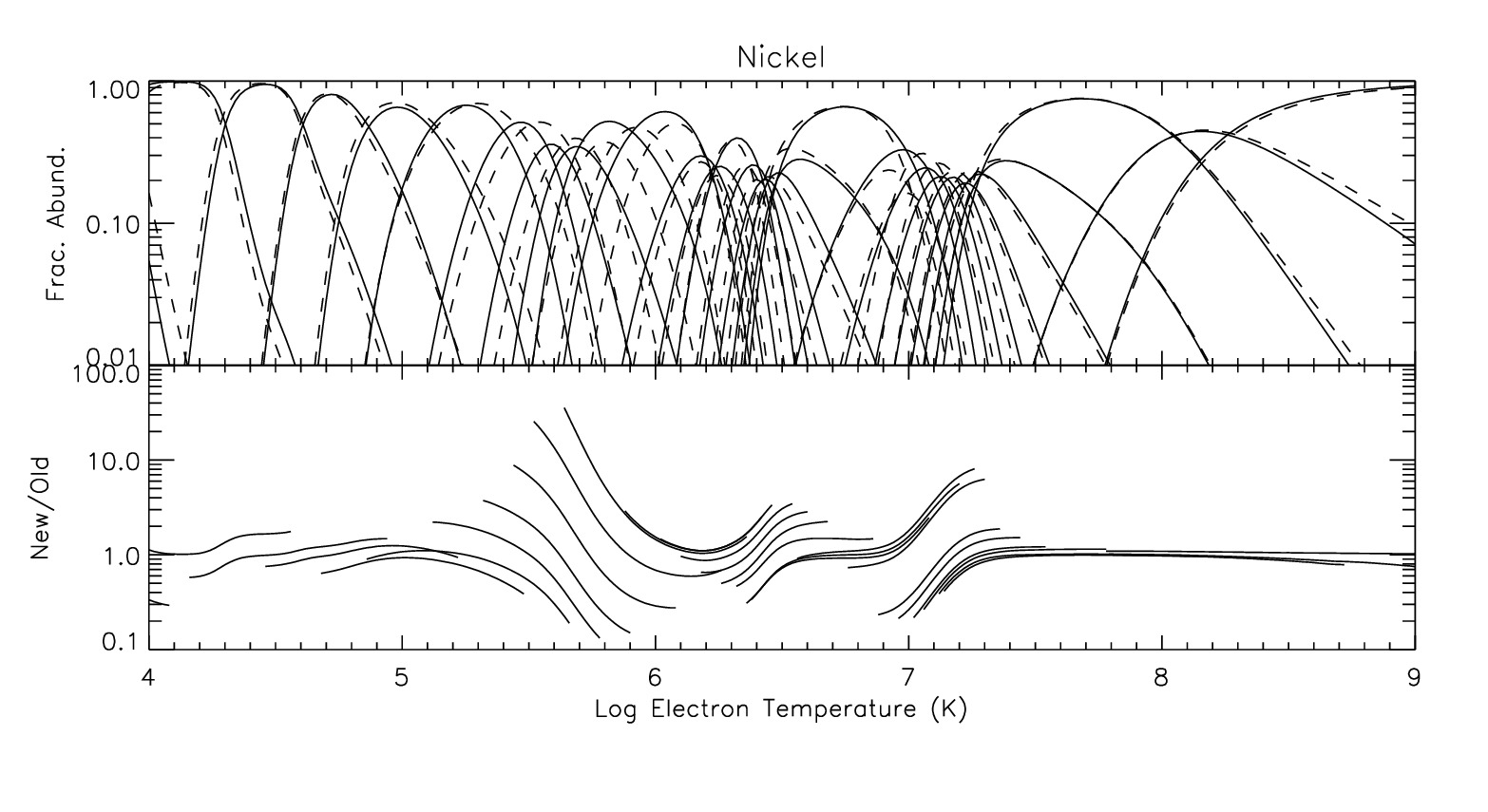}
  \caption[]{Same as Fig.~\protect\ref{fig:H Mazz} but for Ni.}
  \label{fig:Ni Mazz}
\end{figure}

\begin{figure}
  \centering
  \includegraphics[angle=90]{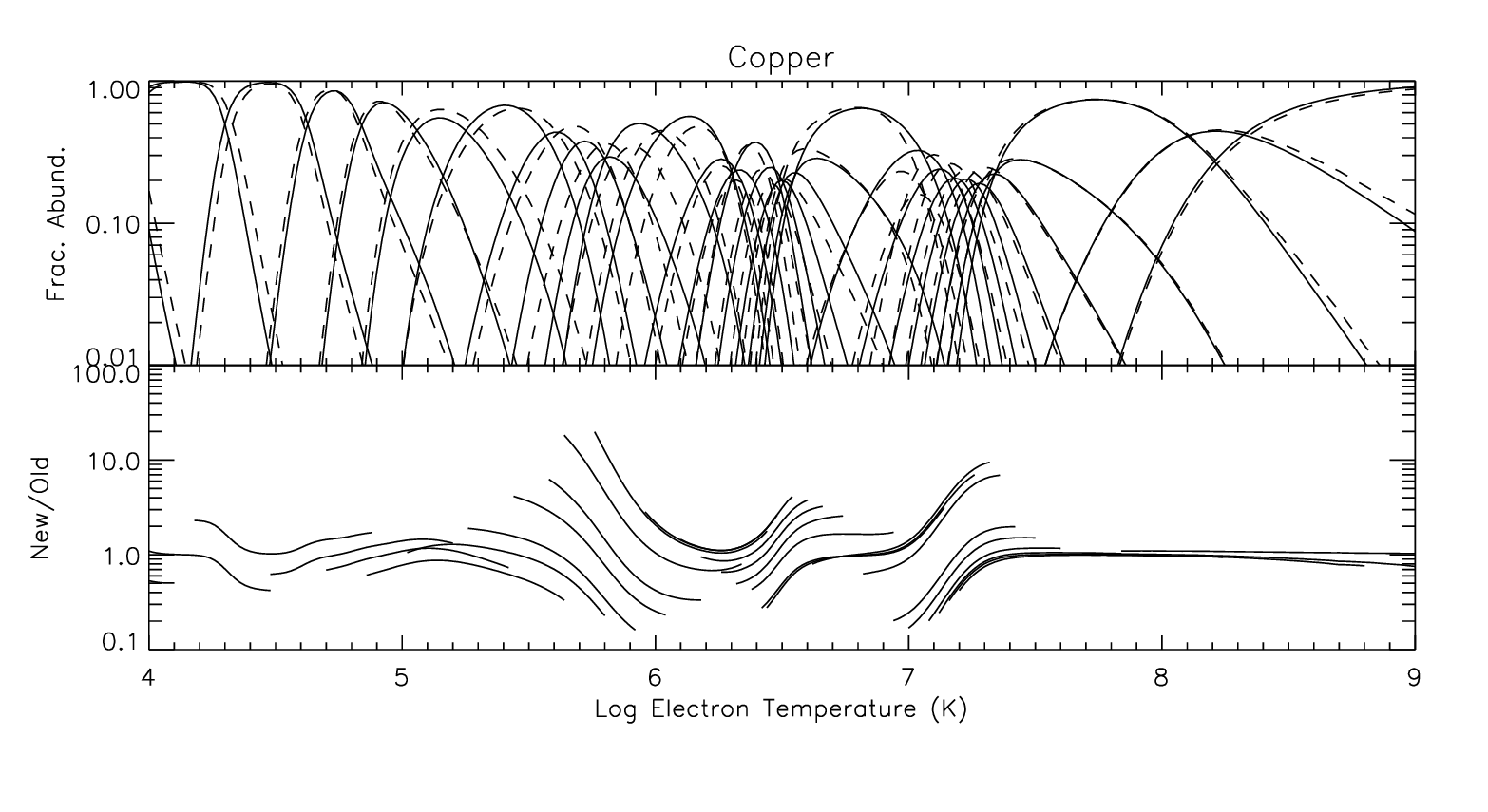}
  \caption[]{Same as Fig.~\protect\ref{fig:H Mazz} but for Cu.}
  \label{fig:Cu Mazz}
\end{figure}

\begin{figure}
  \centering
  \includegraphics[angle=90]{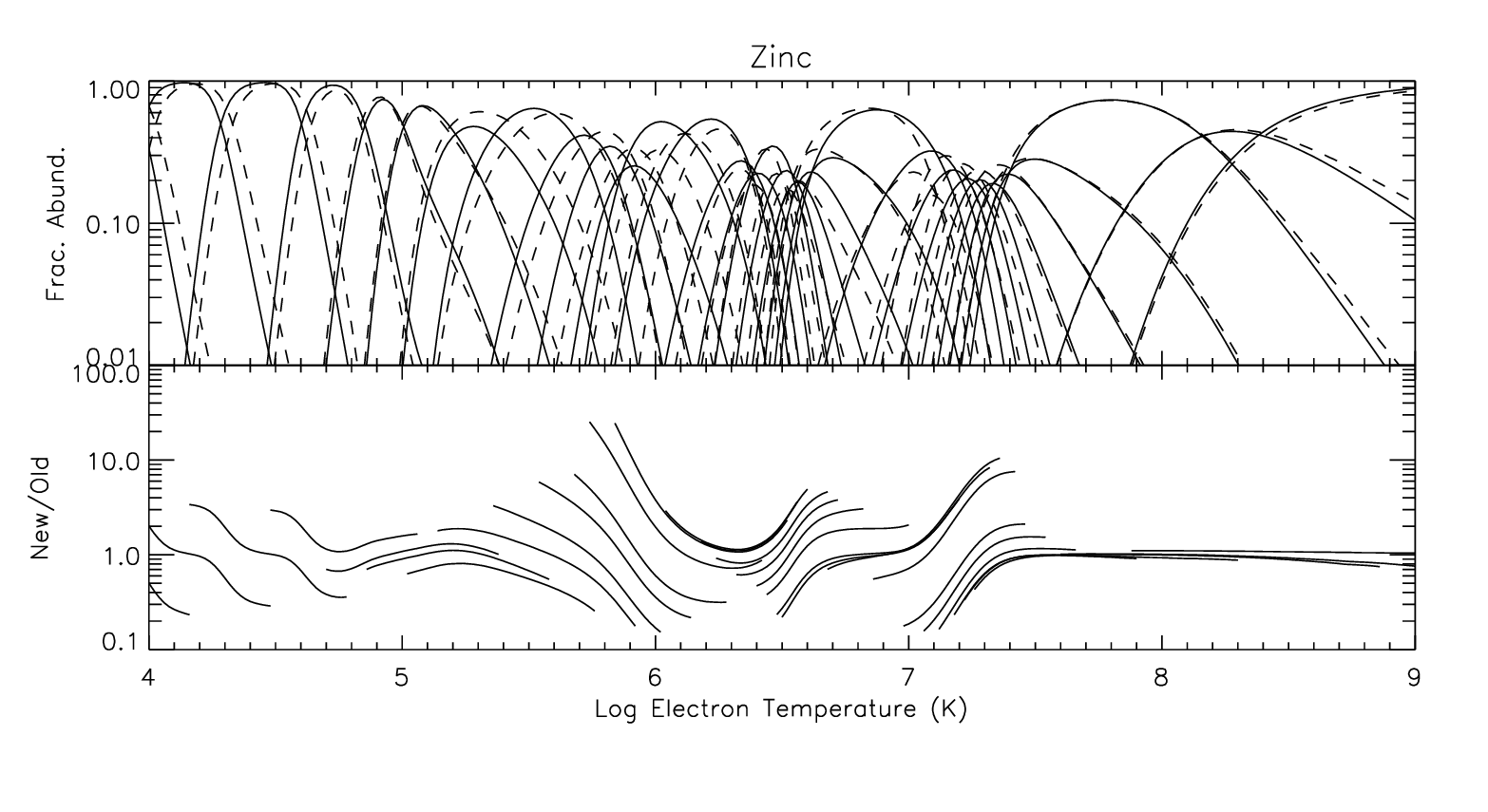}
  \caption[]{Same as Fig.~\protect\ref{fig:H Mazz} but for Zn.}
  \label{fig:Zn Mazz}
\end{figure}

\begin{figure}
  \centering
  \includegraphics[angle=90]{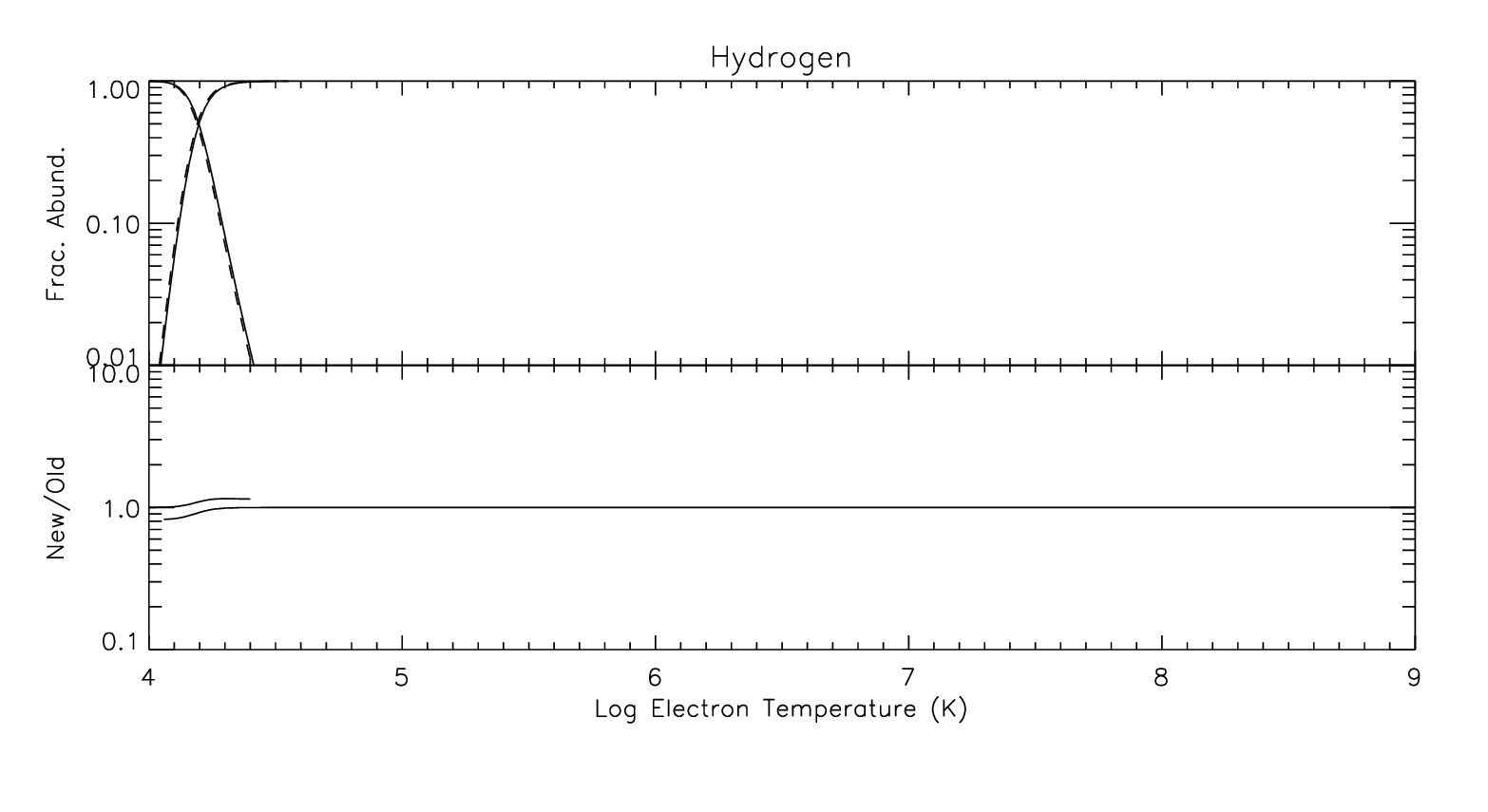}
  \caption[]{Ionization fractional abundance versus electron temperature for 
             all ionization stages of H. 
	     The upper graph shows our
	     results ({\it solid curves}) and the abundances 
	     calculated by Bryans et al.\ (2006; {\it dashed curves}).
	     The lower graph shows 
	     the ratio of the calculated abundances.
	     Comparison is made only for fractional abundances greater than
	     $10^{-2}$.
	     We label our results 
	     as ``New'' and those of \protect\citet{Brya06a} as ``Old''.}
  \label{fig:H Bryans}
\end{figure}

\begin{figure}
  \centering
  \includegraphics[angle=90]{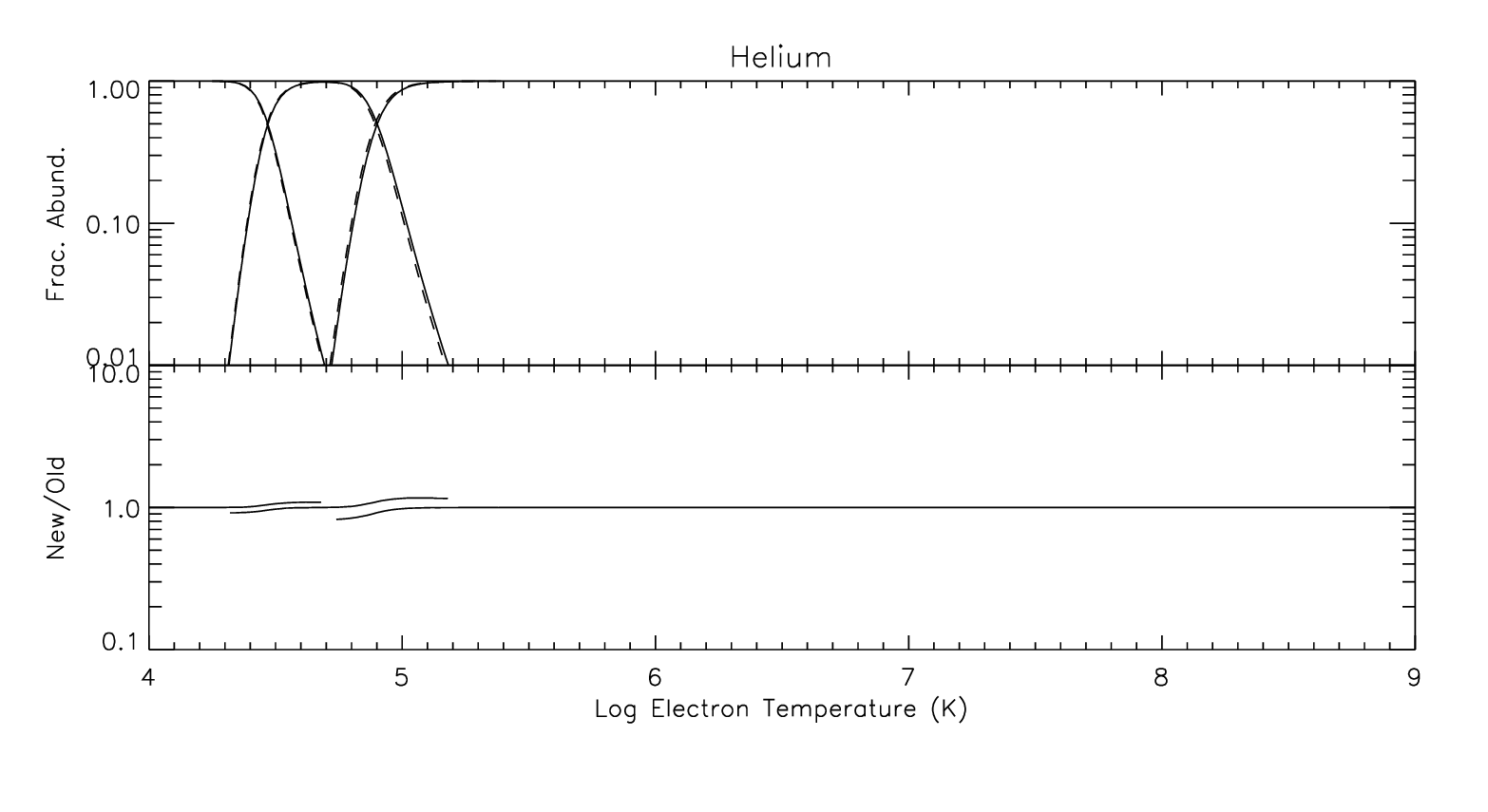}
  \caption[]{Same as Fig.~\protect\ref{fig:H Bryans} but for He.}
  \label{fig:He Bryans}
\end{figure}

\begin{figure}
  \centering
  \includegraphics[angle=90]{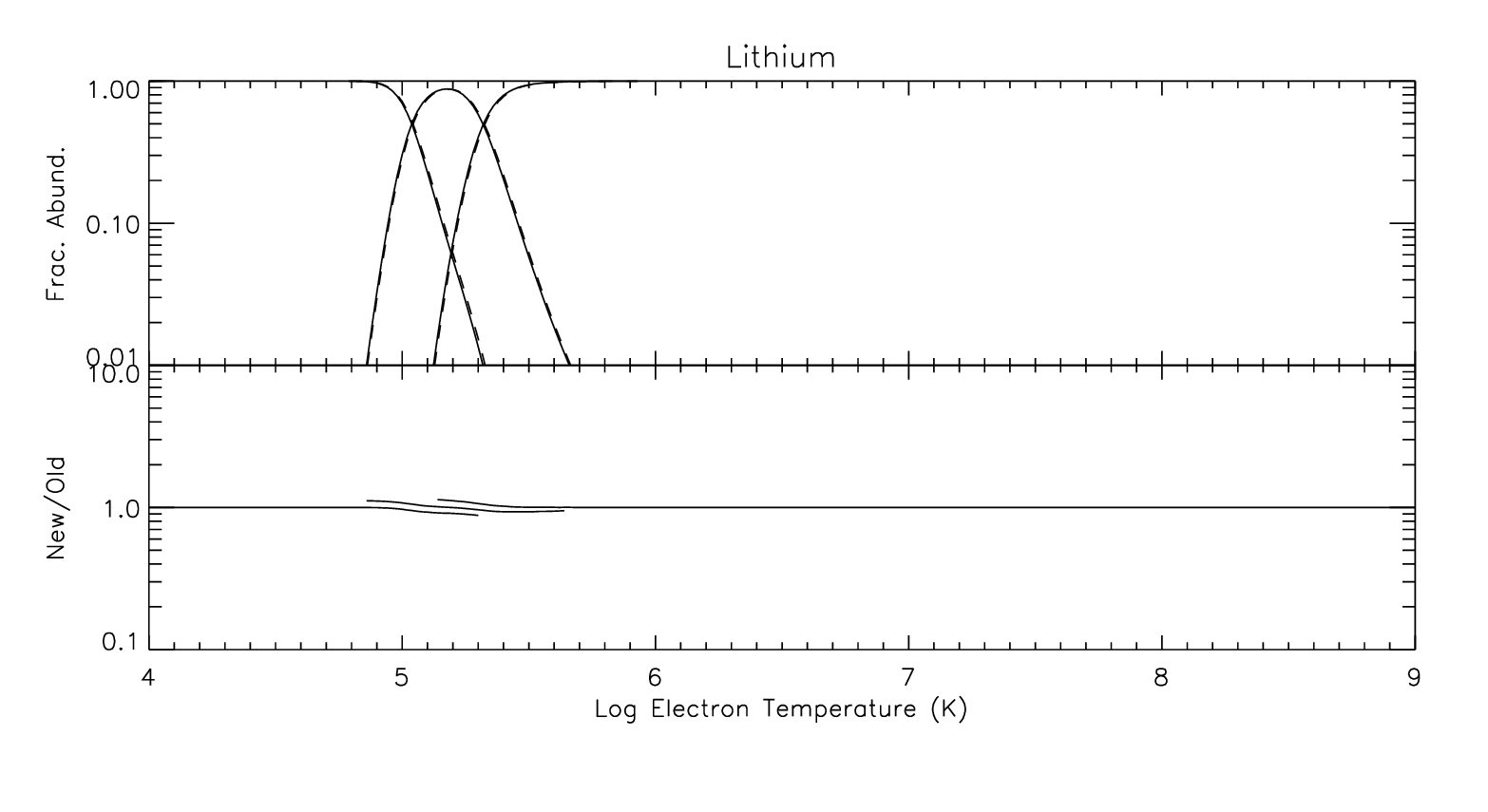}
  \caption[]{Same as Fig.~\protect\ref{fig:H Bryans} but for Li.}
  \label{fig:Li Bryans}
\end{figure}

\begin{figure}
  \centering
  \includegraphics[angle=90]{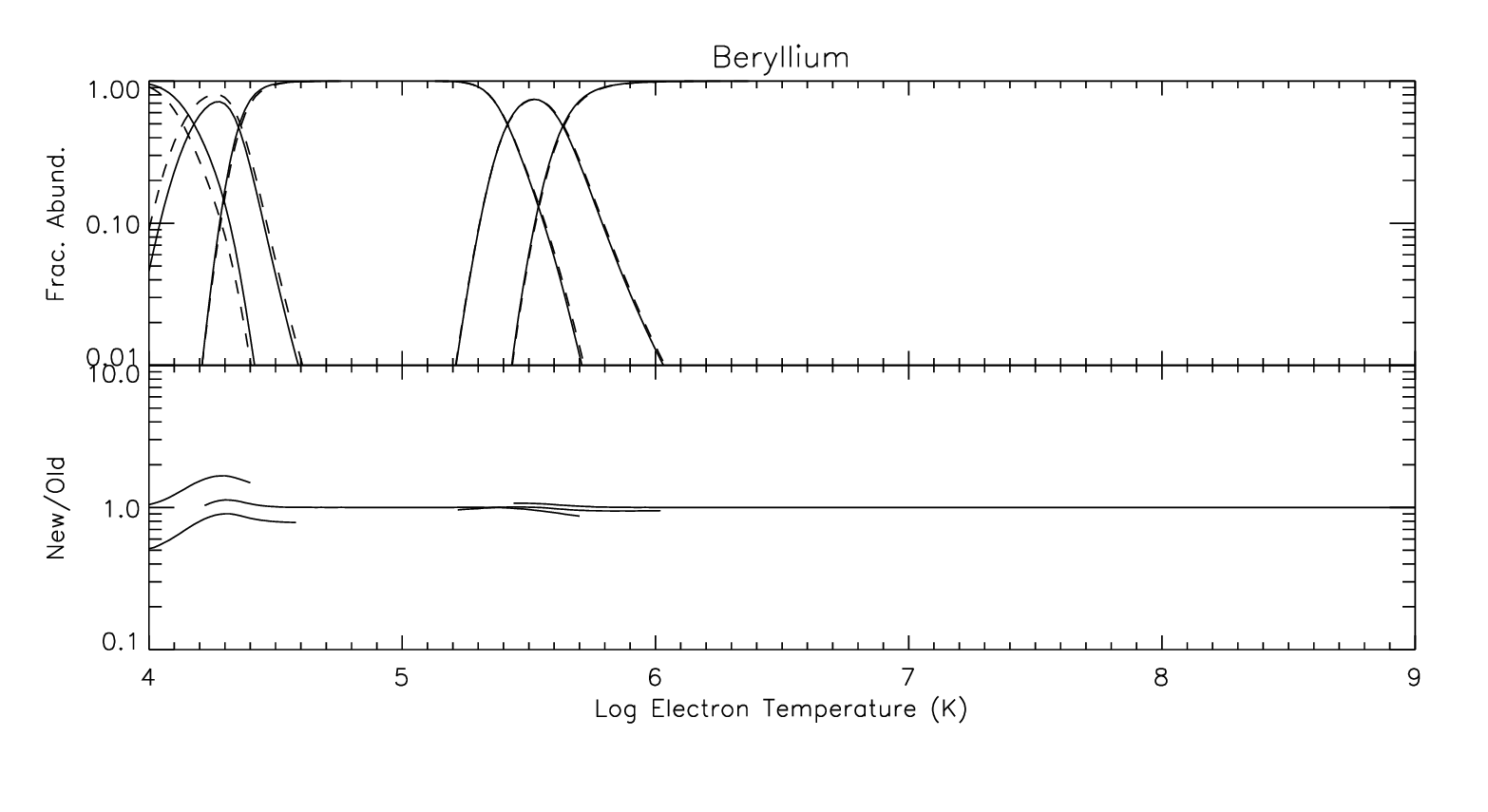}
  \caption[]{Same as Fig.~\protect\ref{fig:H Bryans} but for Be.}
  \label{fig:Be Bryans}
\end{figure}

\begin{figure}
  \centering
  \includegraphics[angle=90]{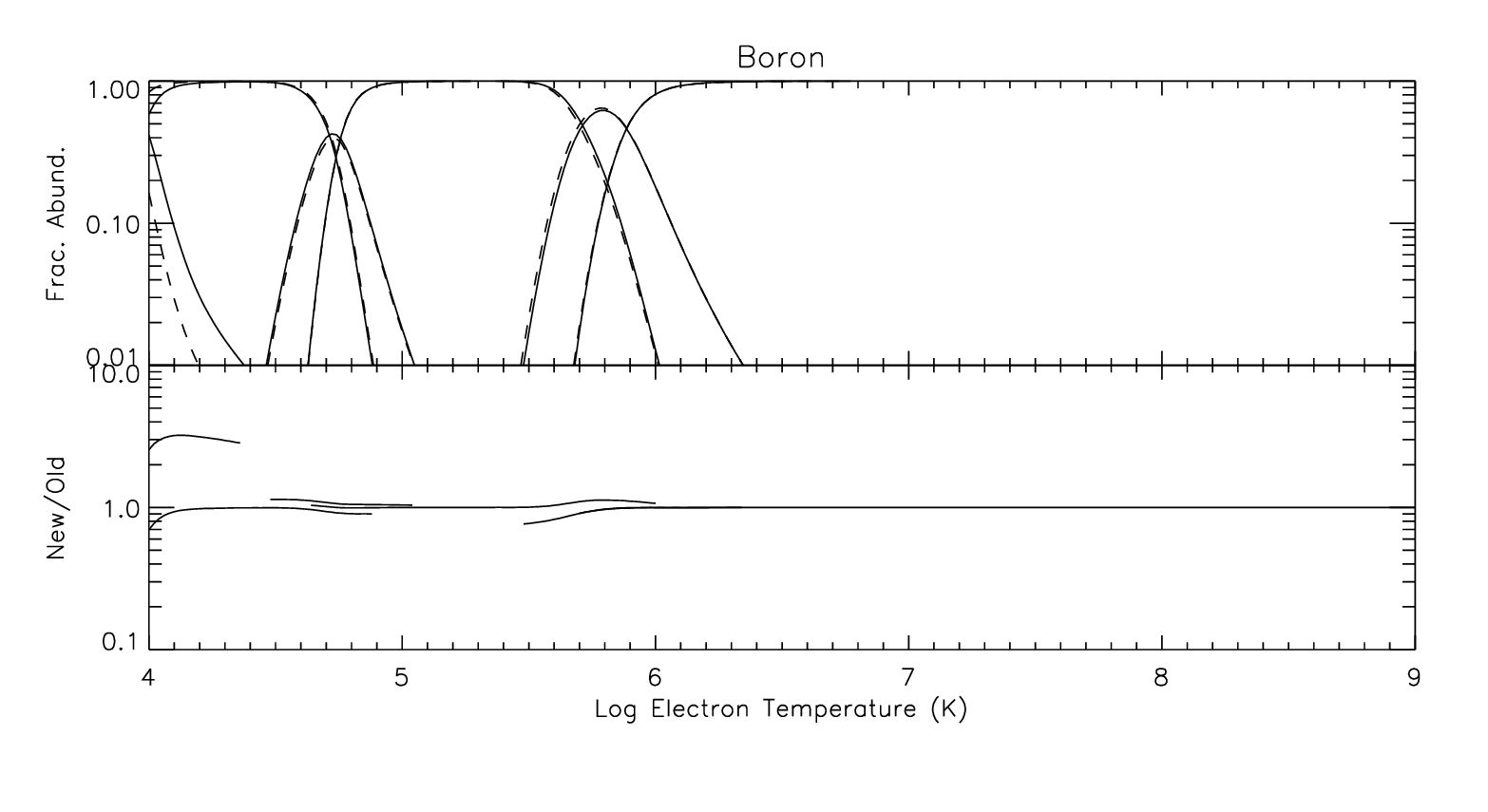}
  \caption[]{Same as Fig.~\protect\ref{fig:H Bryans} but for B.}
  \label{fig:B Bryans}
\end{figure}

\begin{figure}
  \centering
  \includegraphics[angle=90]{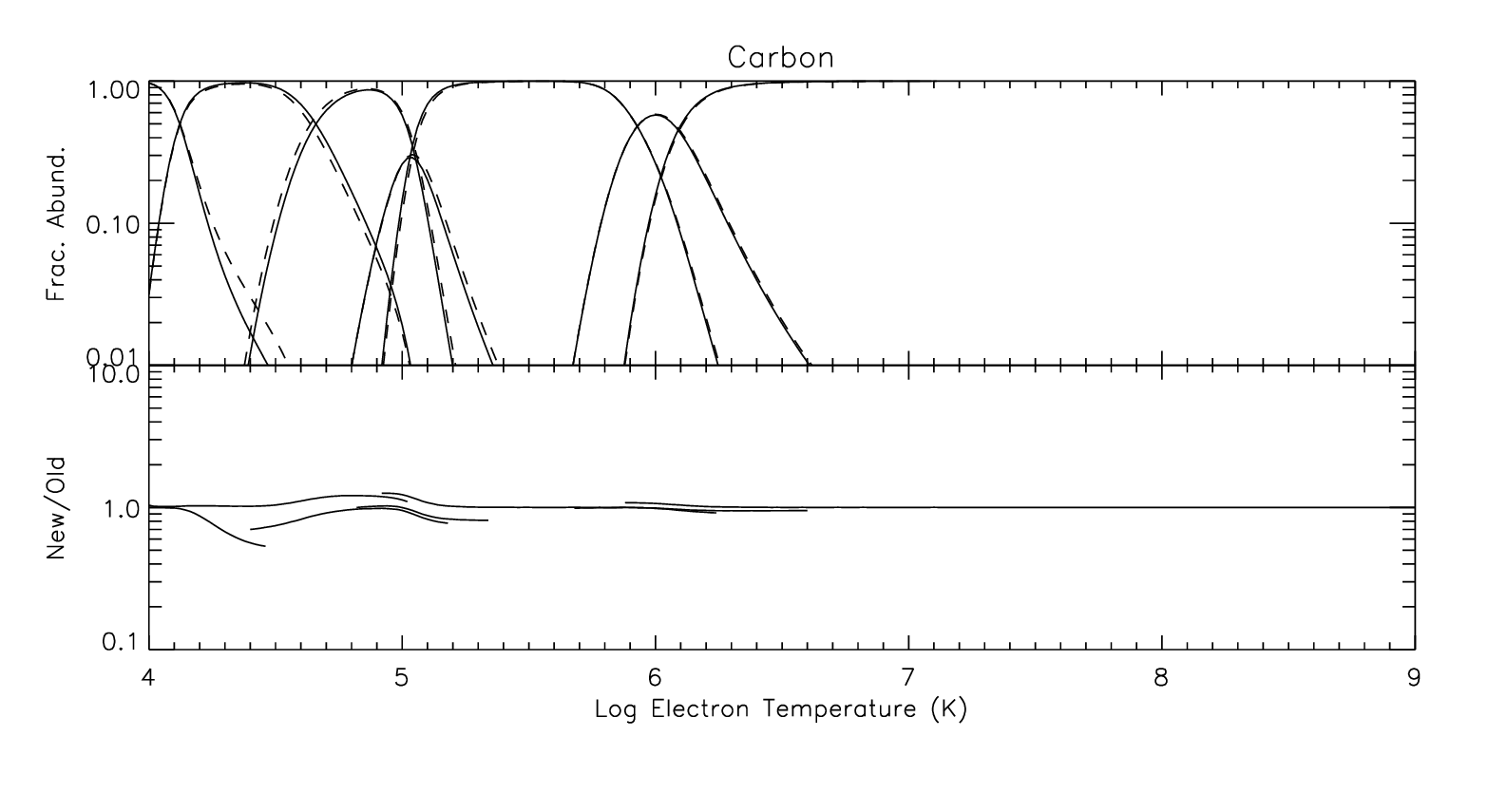}
  \caption[]{Same as Fig.~\protect\ref{fig:H Bryans} but for C.}
  \label{fig:C Bryans}
\end{figure}

\begin{figure}
  \centering
  \includegraphics[angle=90]{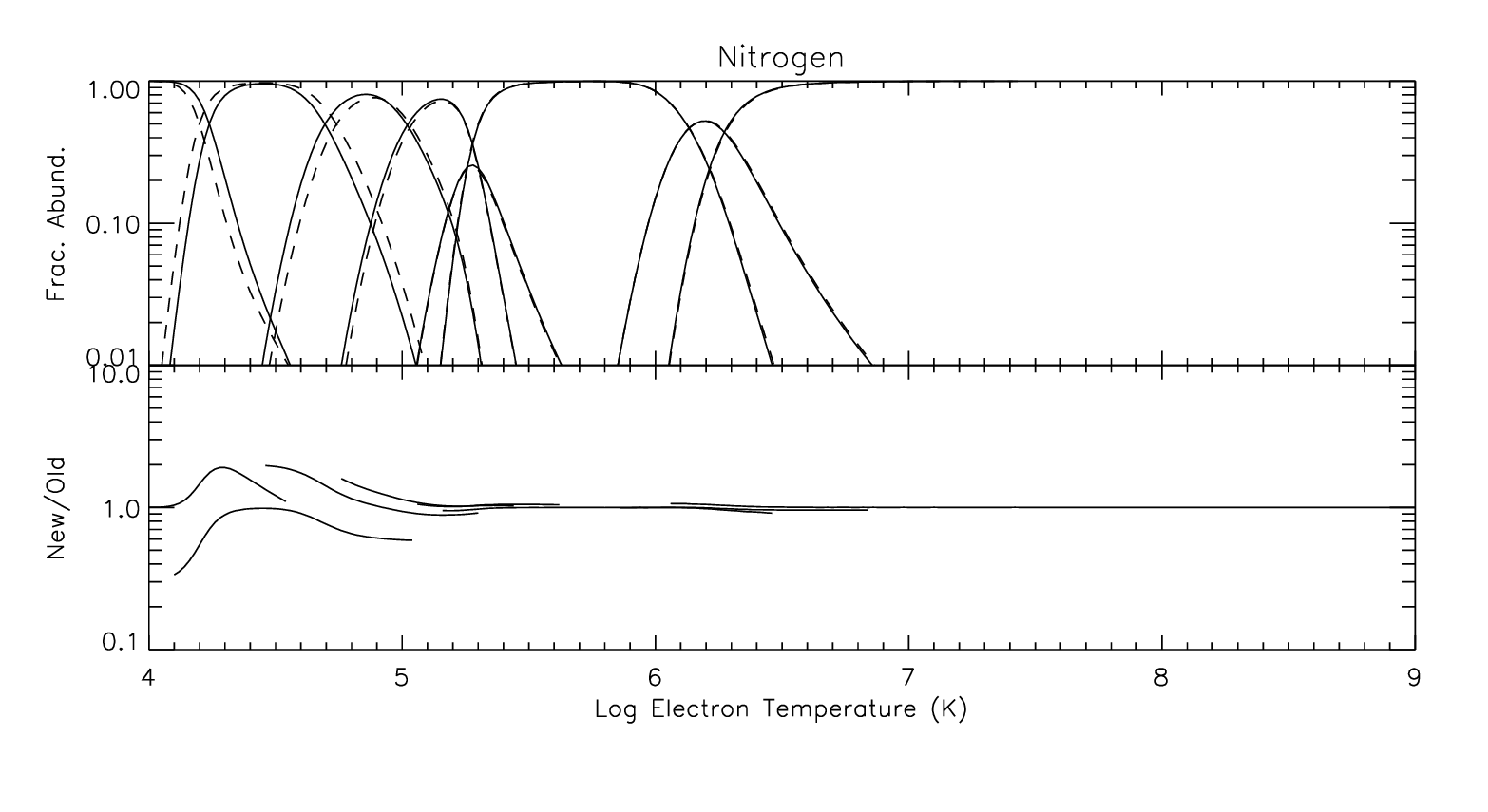}
  \caption[]{Same as Fig.~\protect\ref{fig:H Bryans} but for N.}
  \label{fig:N Bryans}
\end{figure}

\begin{figure}
  \centering
  \includegraphics[angle=90]{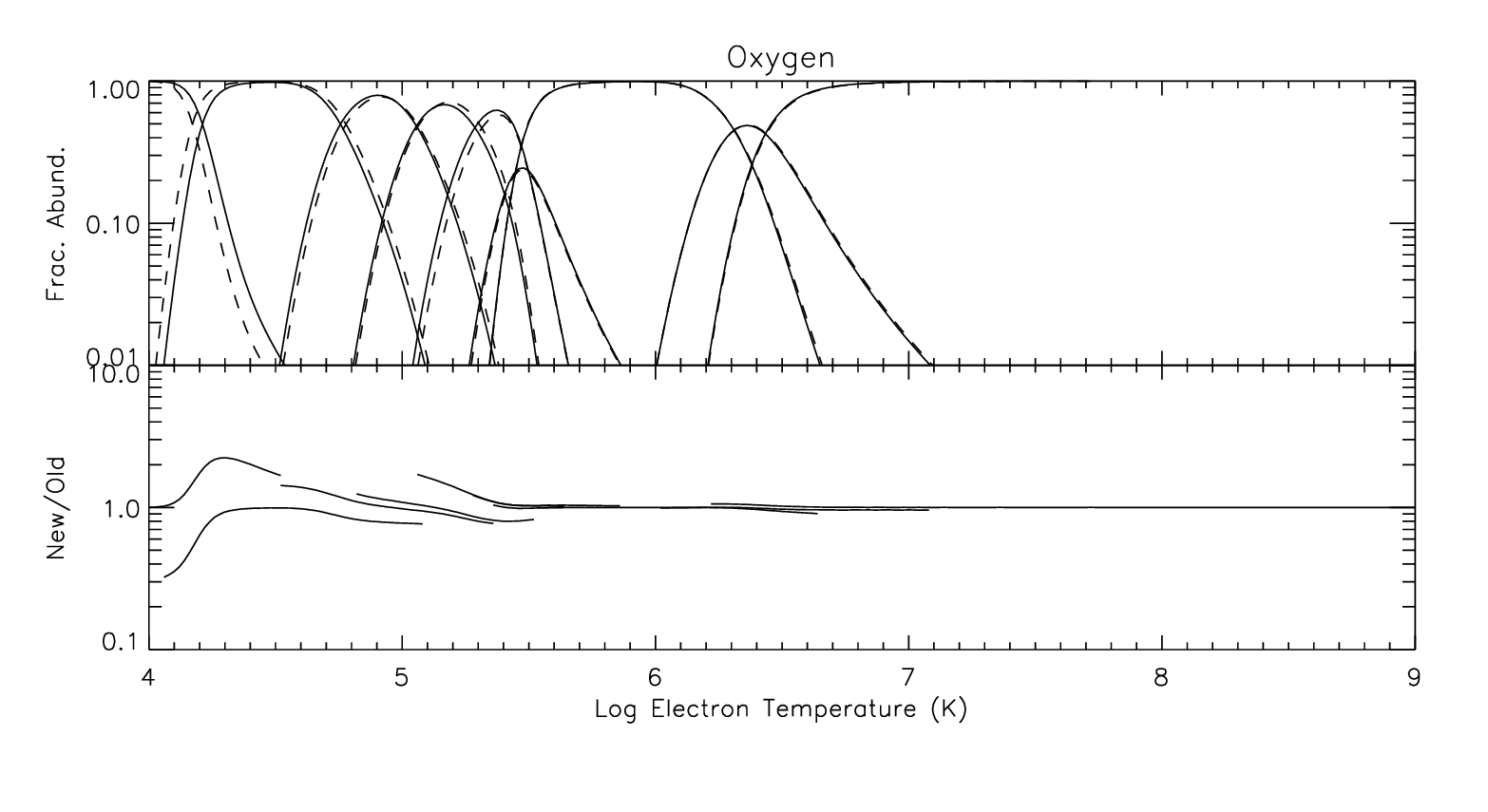}
  \caption[]{Same as Fig.~\protect\ref{fig:H Bryans} but for O.}
  \label{fig:O Bryans}
\end{figure}

\begin{figure}
  \centering
  \includegraphics[angle=90]{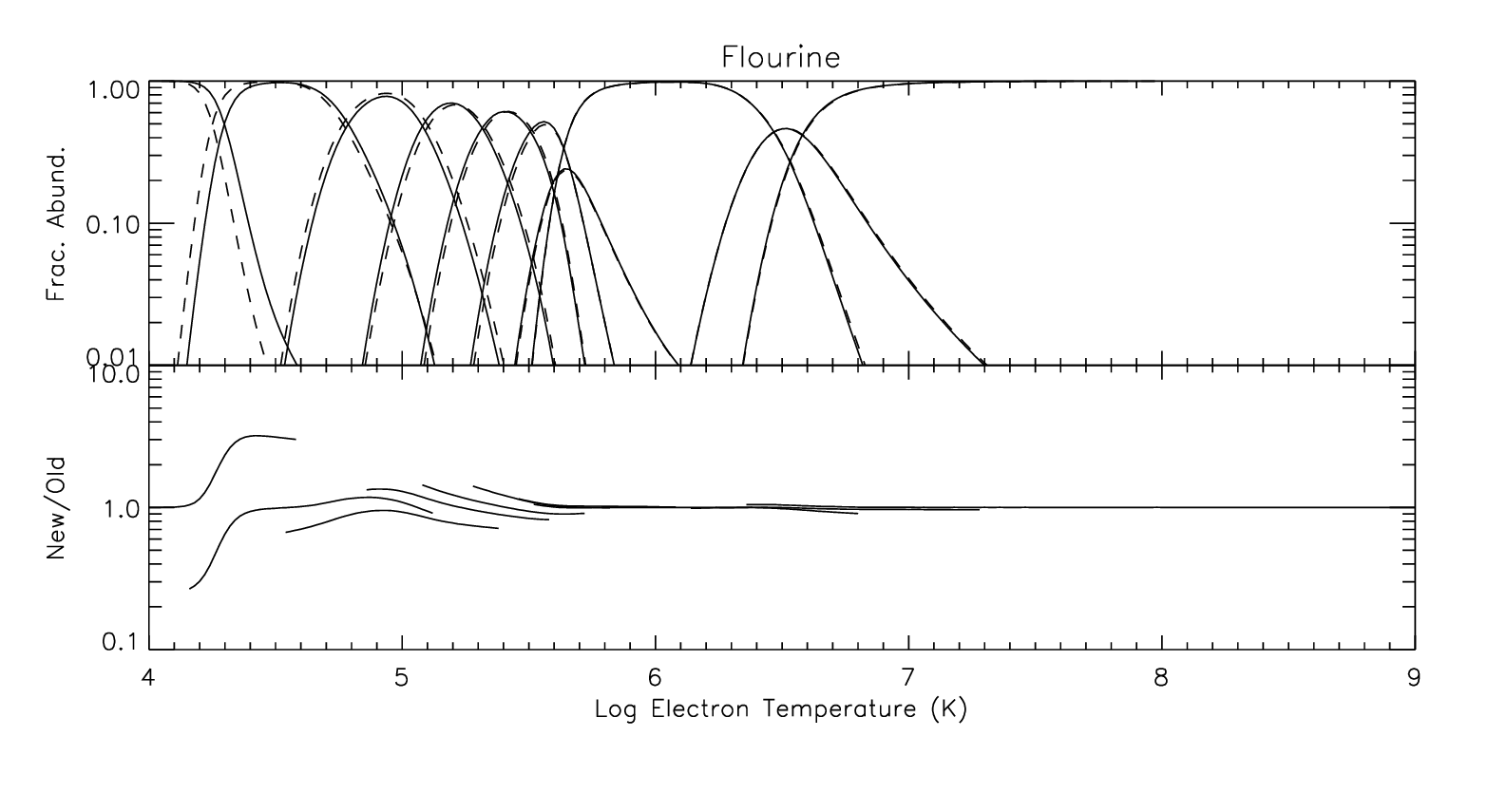}
  \caption[]{Same as Fig.~\protect\ref{fig:H Bryans} but for F.}
  \label{fig:F Bryans}
\end{figure}

\begin{figure}
  \centering
  \includegraphics[angle=90]{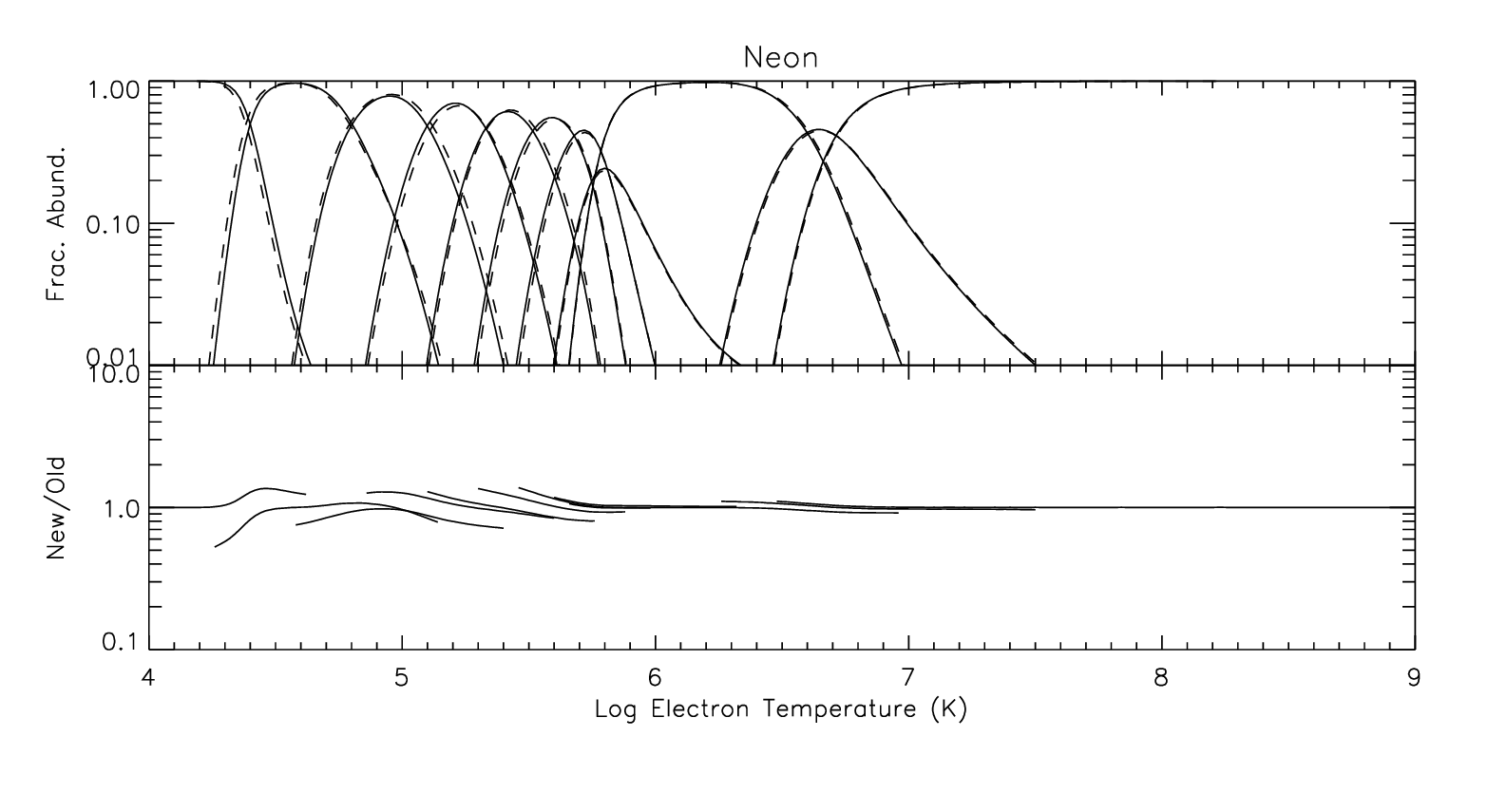}
  \caption[]{Same as Fig.~\protect\ref{fig:H Bryans} but for Ne.}
  \label{fig:Ne Bryans}
\end{figure}

\begin{figure}
  \centering
  \includegraphics[angle=90]{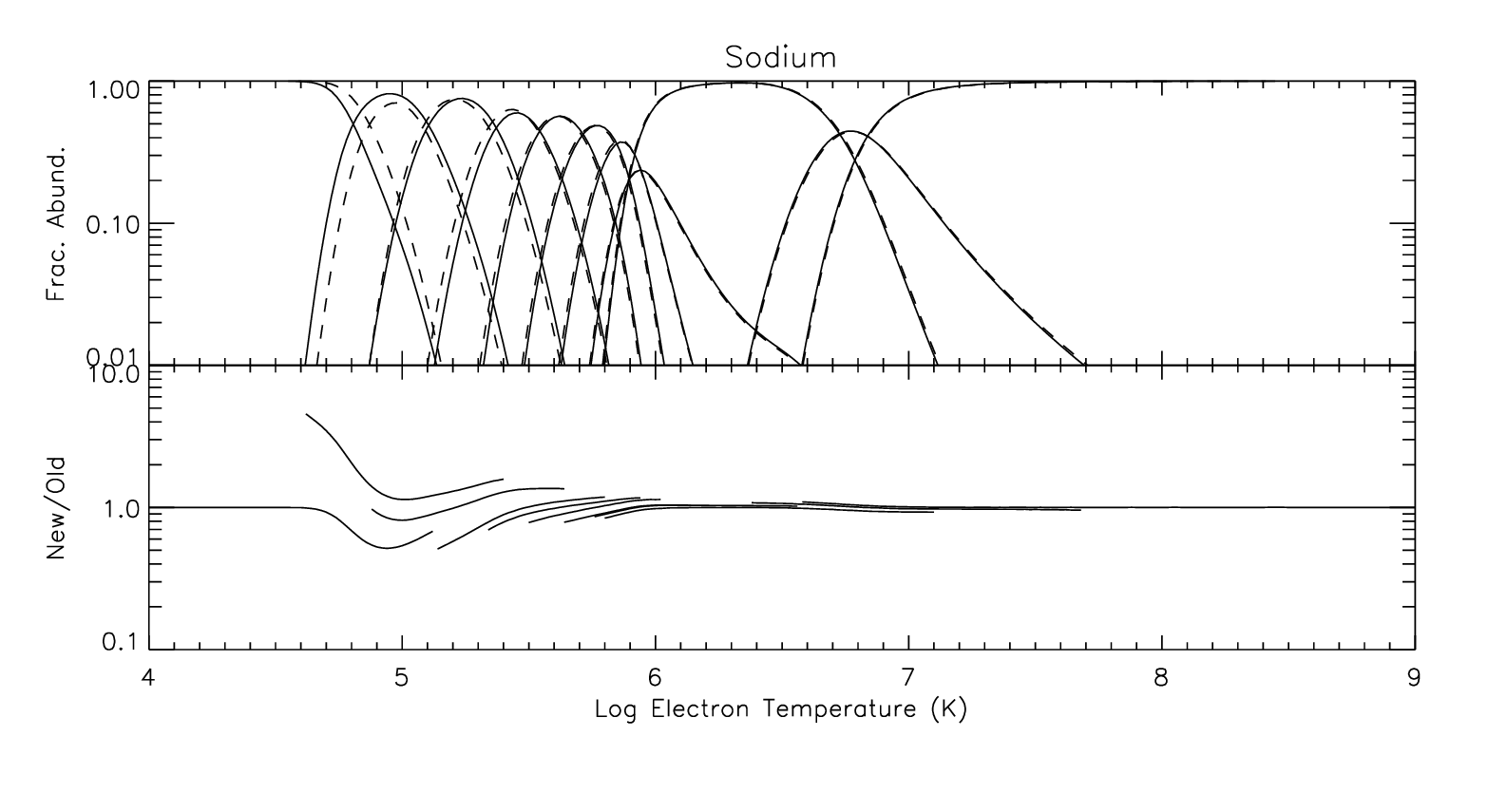}
  \caption[]{Same as Fig.~\protect\ref{fig:H Bryans} but for Na.}
  \label{fig:Na Bryans}
\end{figure}

\begin{figure}
  \centering
  \includegraphics[angle=90]{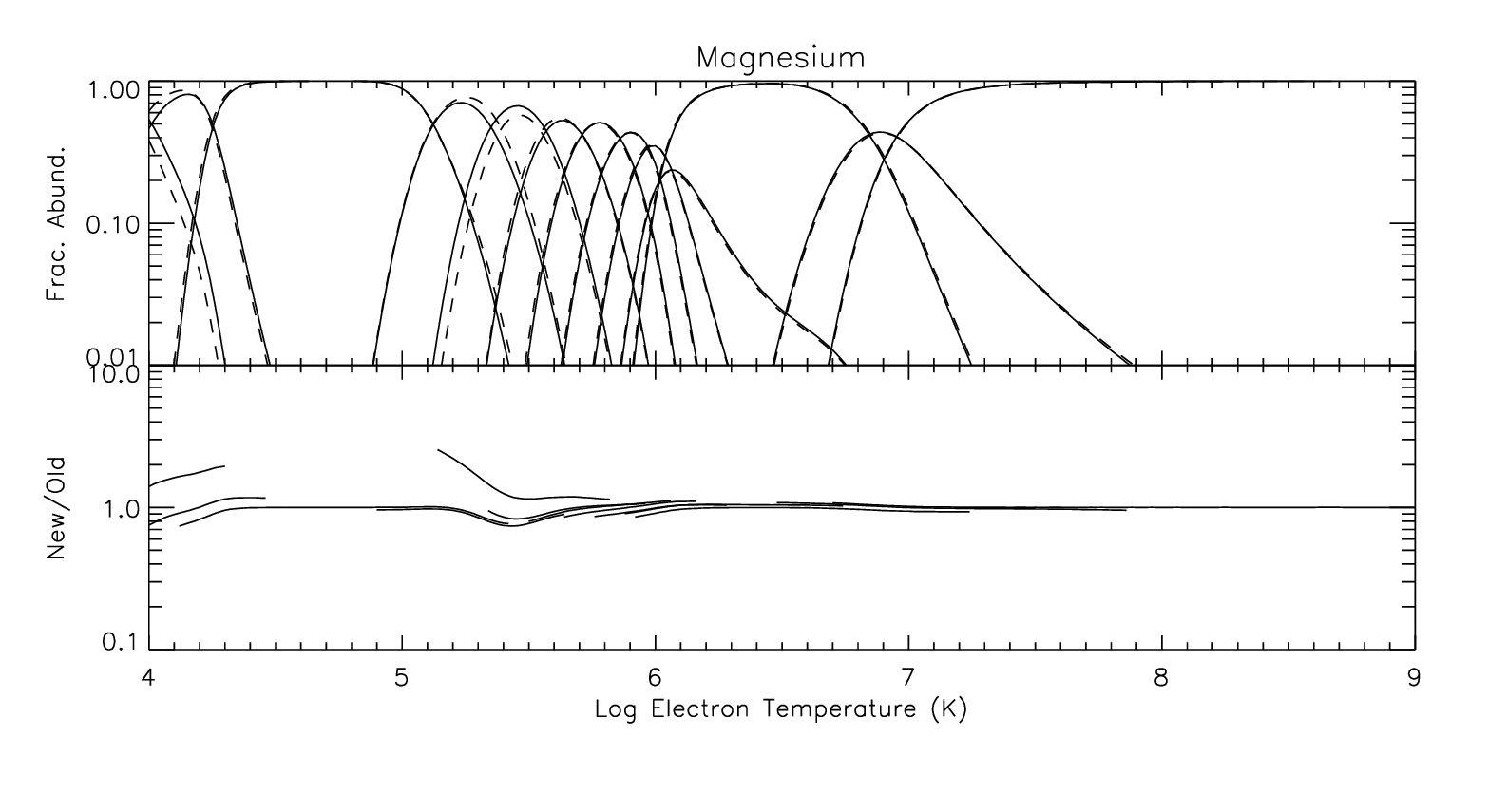}
  \caption[]{Same as Fig.~\protect\ref{fig:H Bryans} but for Mg.}
  \label{fig:Mg Bryans}
\end{figure}

\begin{figure}
  \centering
  \includegraphics[angle=90]{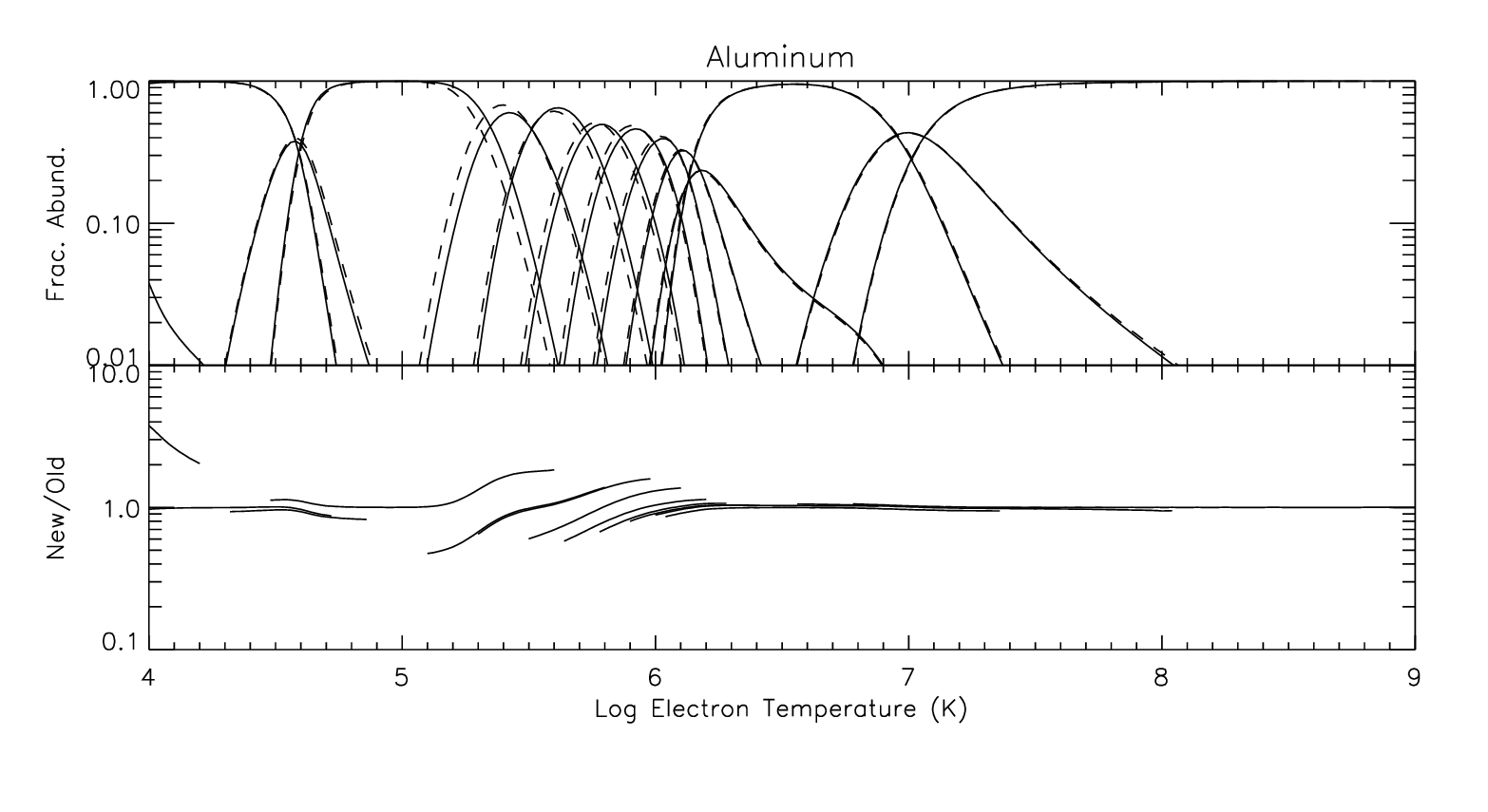}
  \caption[]{Same as Fig.~\protect\ref{fig:H Bryans} but for Al.}
  \label{fig:Al Bryans}
\end{figure}

\begin{figure}
  \centering
  \includegraphics[angle=90]{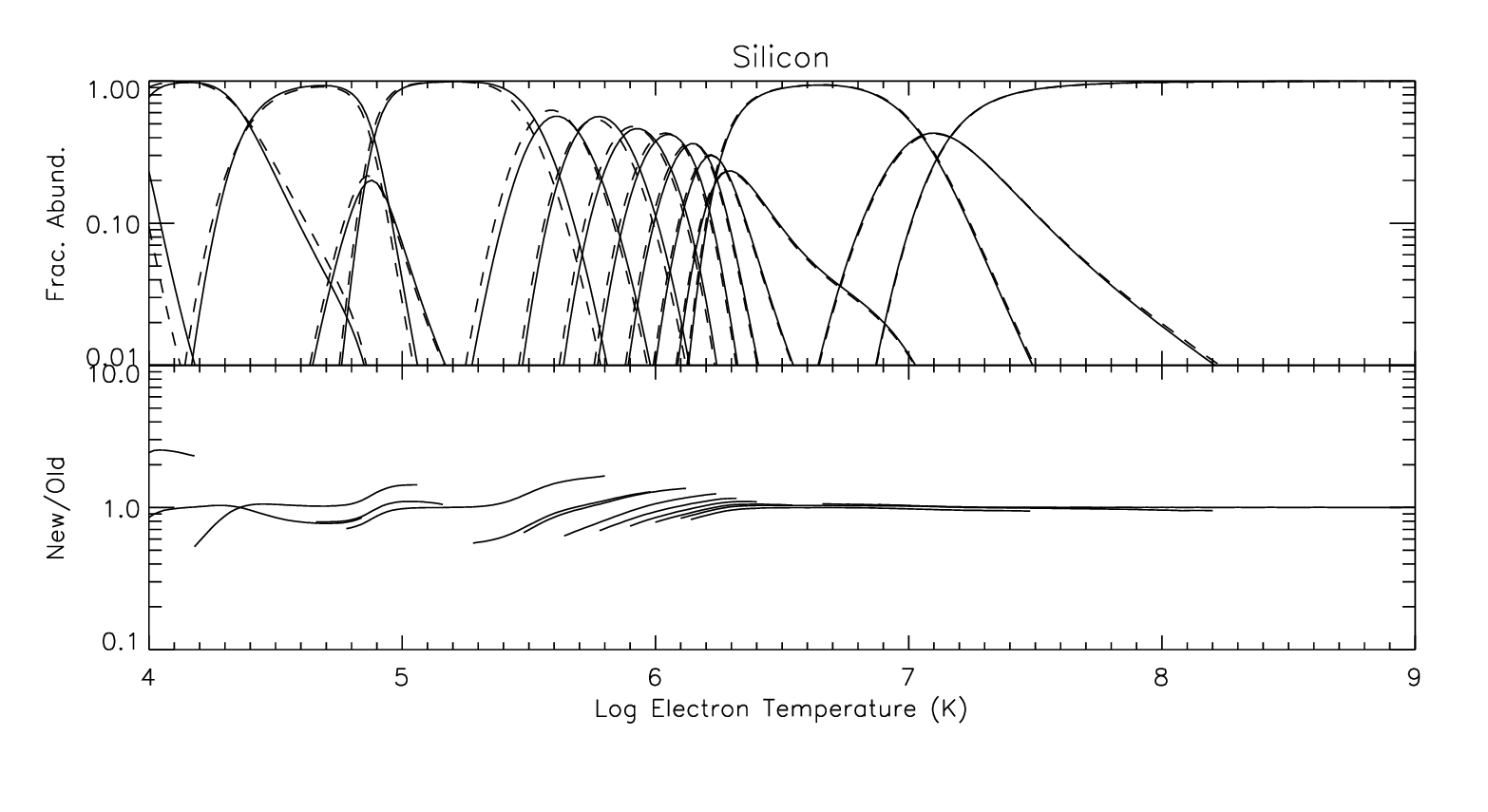}
  \caption[]{Same as Fig.~\protect\ref{fig:H Bryans} but for Si.}
  \label{fig:Si Bryans}
\end{figure}

\begin{figure}
  \centering
  \includegraphics[angle=90]{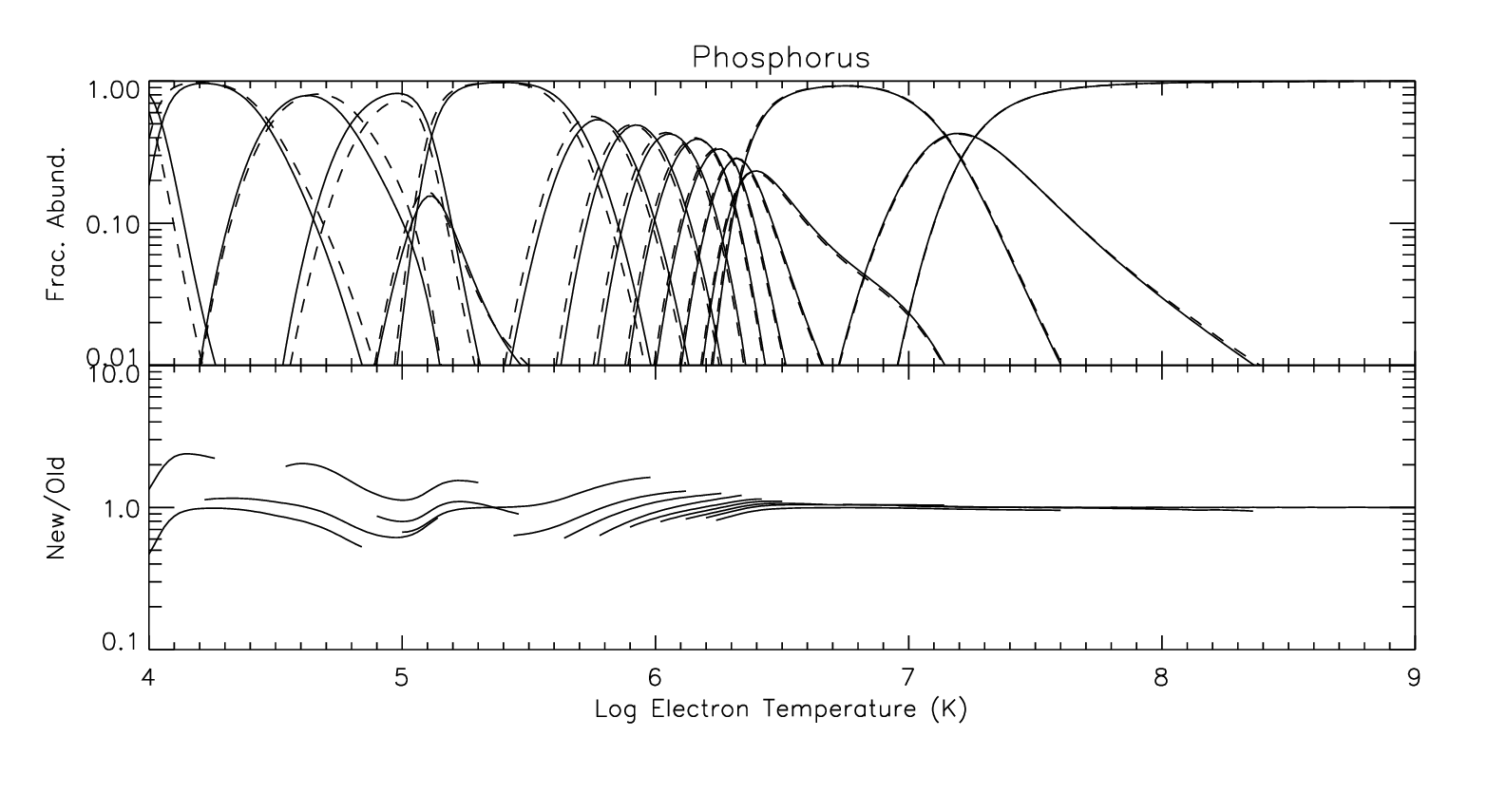}
  \caption[]{Same as Fig.~\protect\ref{fig:H Bryans} but for P.}
  \label{fig:P Bryans}
\end{figure}

\begin{figure}
  \centering
  \includegraphics[angle=90]{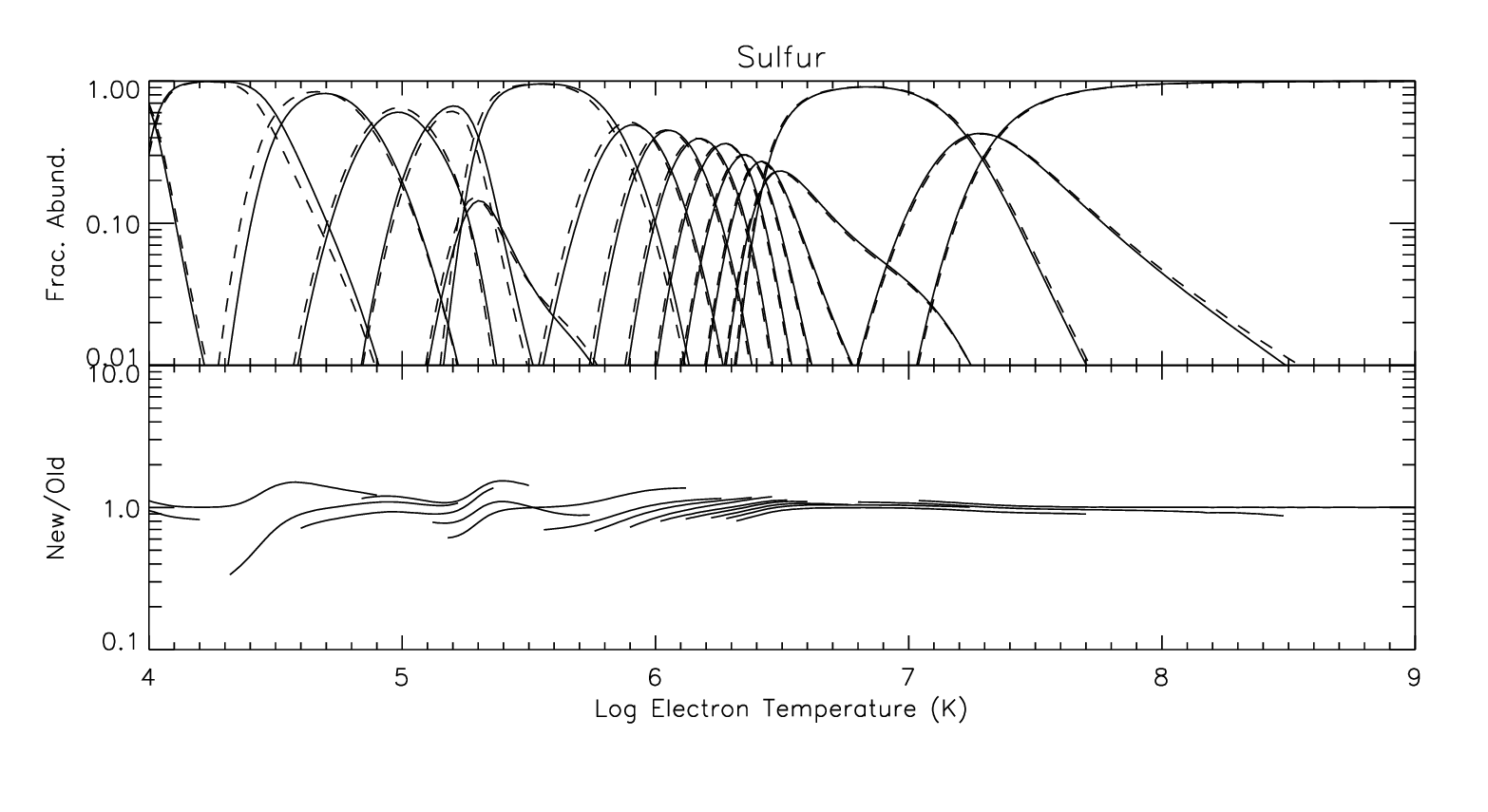}
  \caption[]{Same as Fig.~\protect\ref{fig:H Bryans} but for S.}
  \label{fig:S Bryans}
\end{figure}

\begin{figure}
  \centering
  \includegraphics[angle=90]{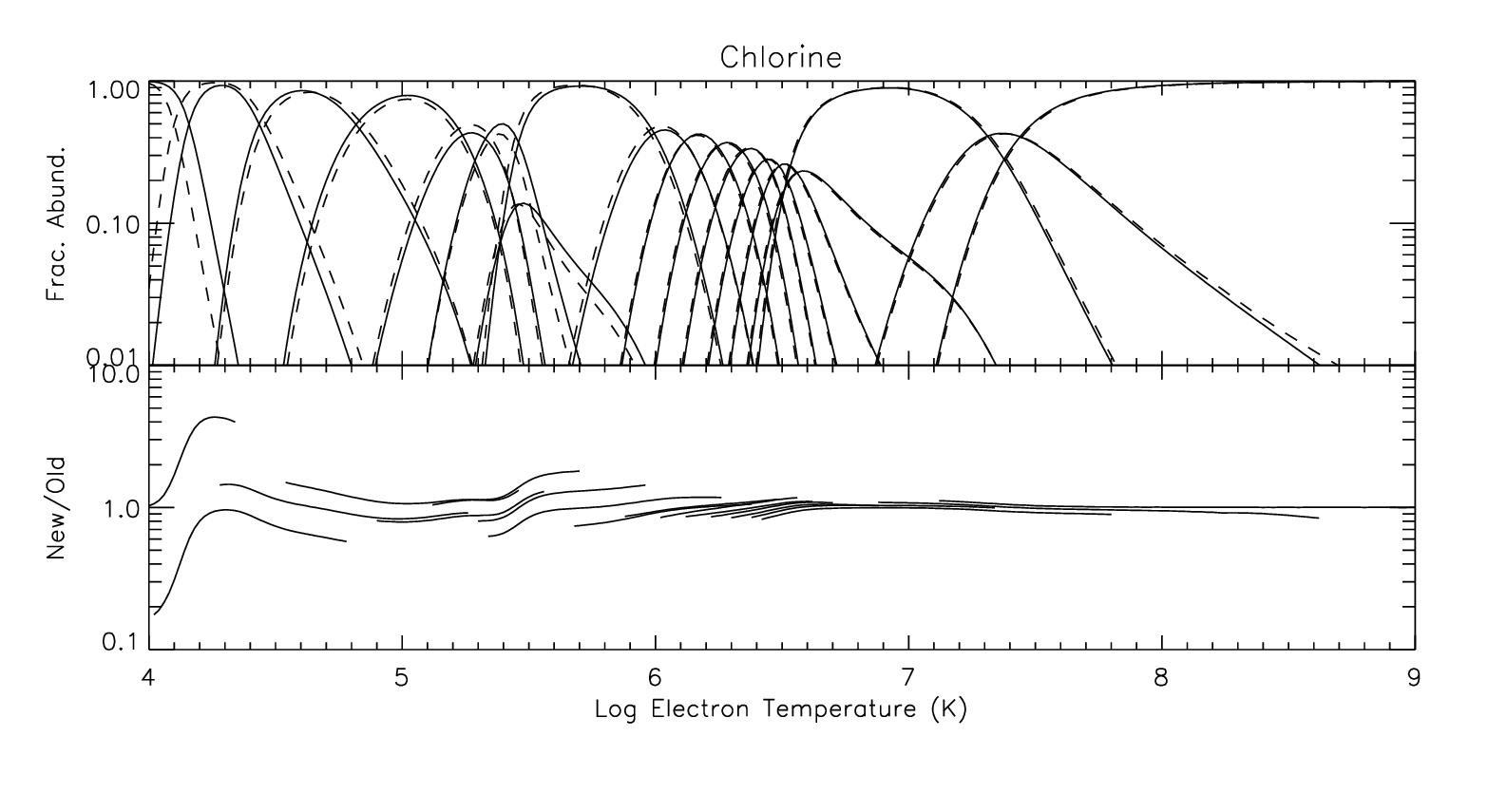}
  \caption[]{Same as Fig.~\protect\ref{fig:H Bryans} but for Cl.}
  \label{fig:Cl Bryans}
\end{figure}

\begin{figure}
  \centering
  \includegraphics[angle=90]{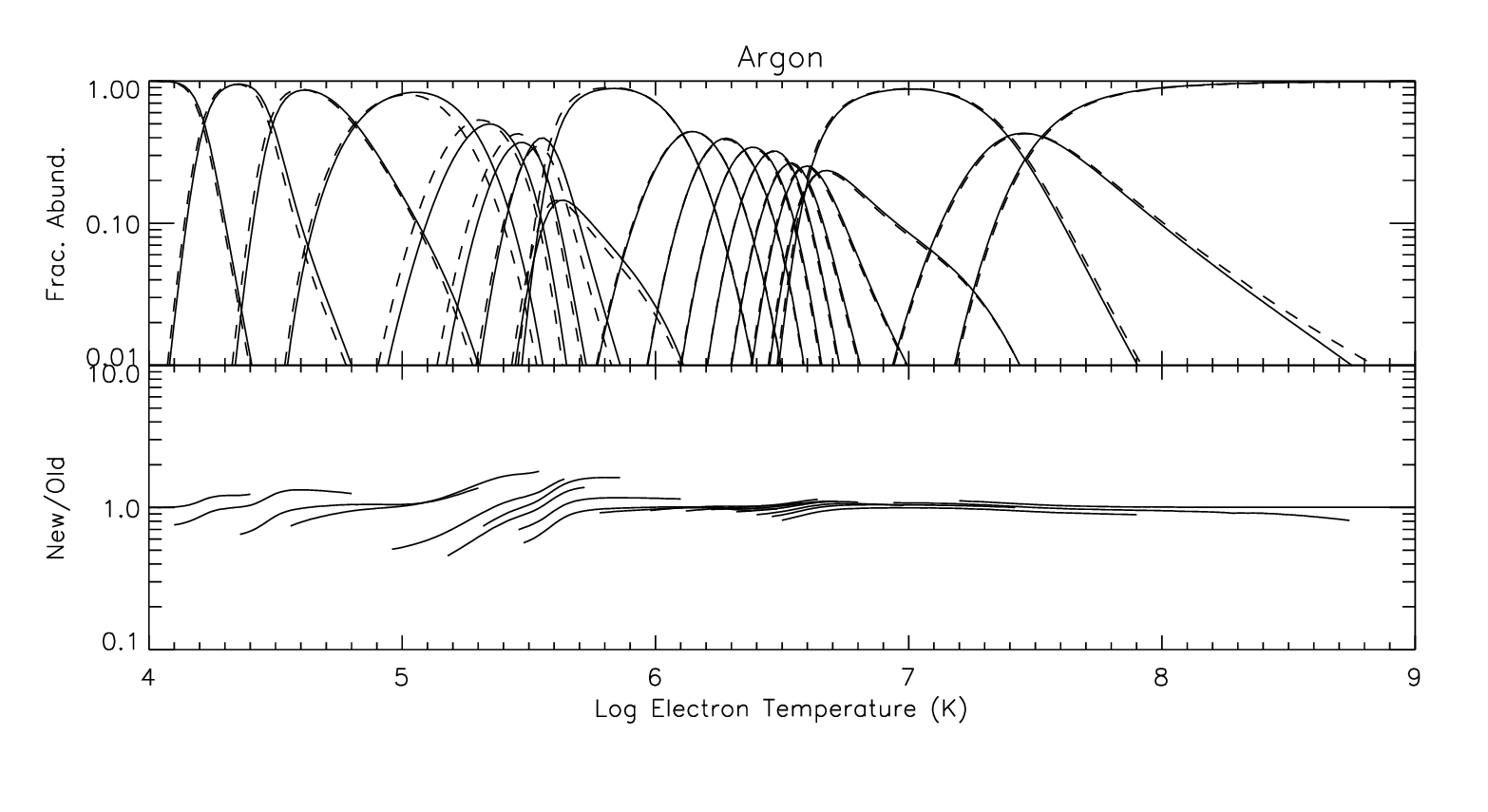}
  \caption[]{Same as Fig.~\protect\ref{fig:H Bryans} but for Ar.}
  \label{fig:Ar Bryans}
\end{figure}

\begin{figure}
  \centering
  \includegraphics[angle=90]{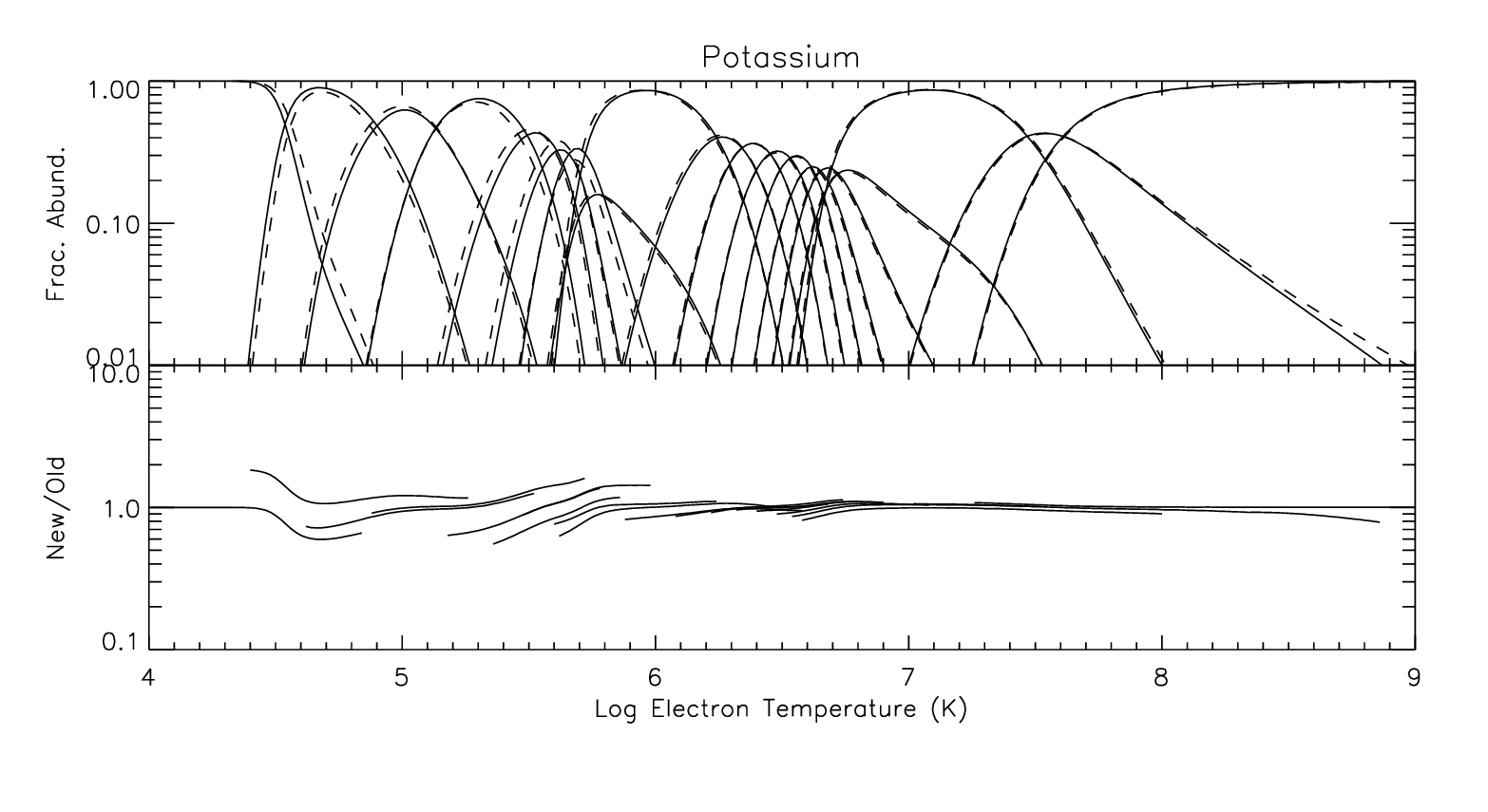}
  \caption[]{Same as Fig.~\protect\ref{fig:H Bryans} but for K.}
  \label{fig:K Bryans}
\end{figure}

\begin{figure}
  \centering
  \includegraphics[angle=90]{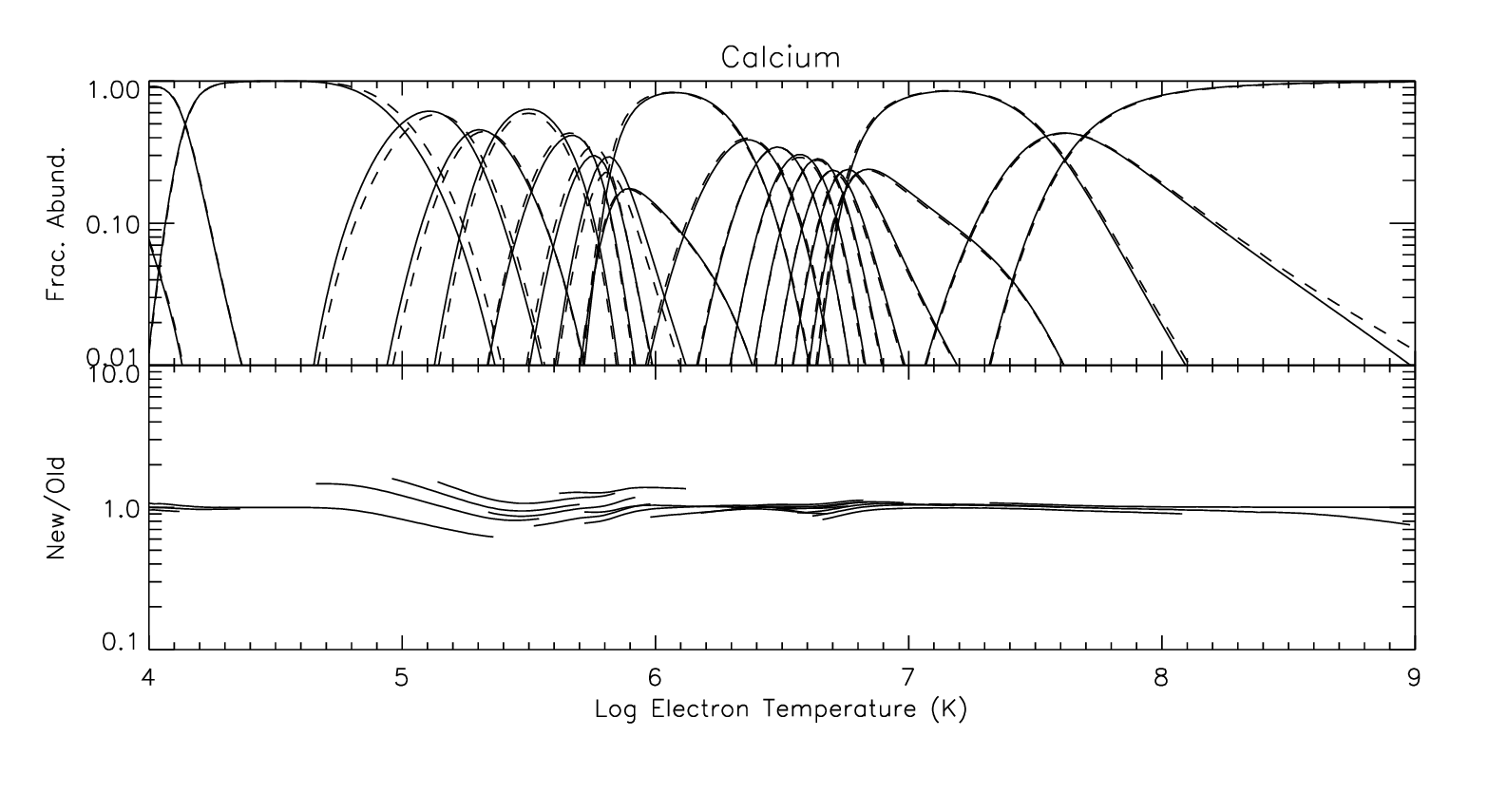}
  \caption[]{Same as Fig.~\protect\ref{fig:H Bryans} but for Ca.}
  \label{fig:Ca Bryans}
\end{figure}

\begin{figure}
  \centering
  \includegraphics[angle=90]{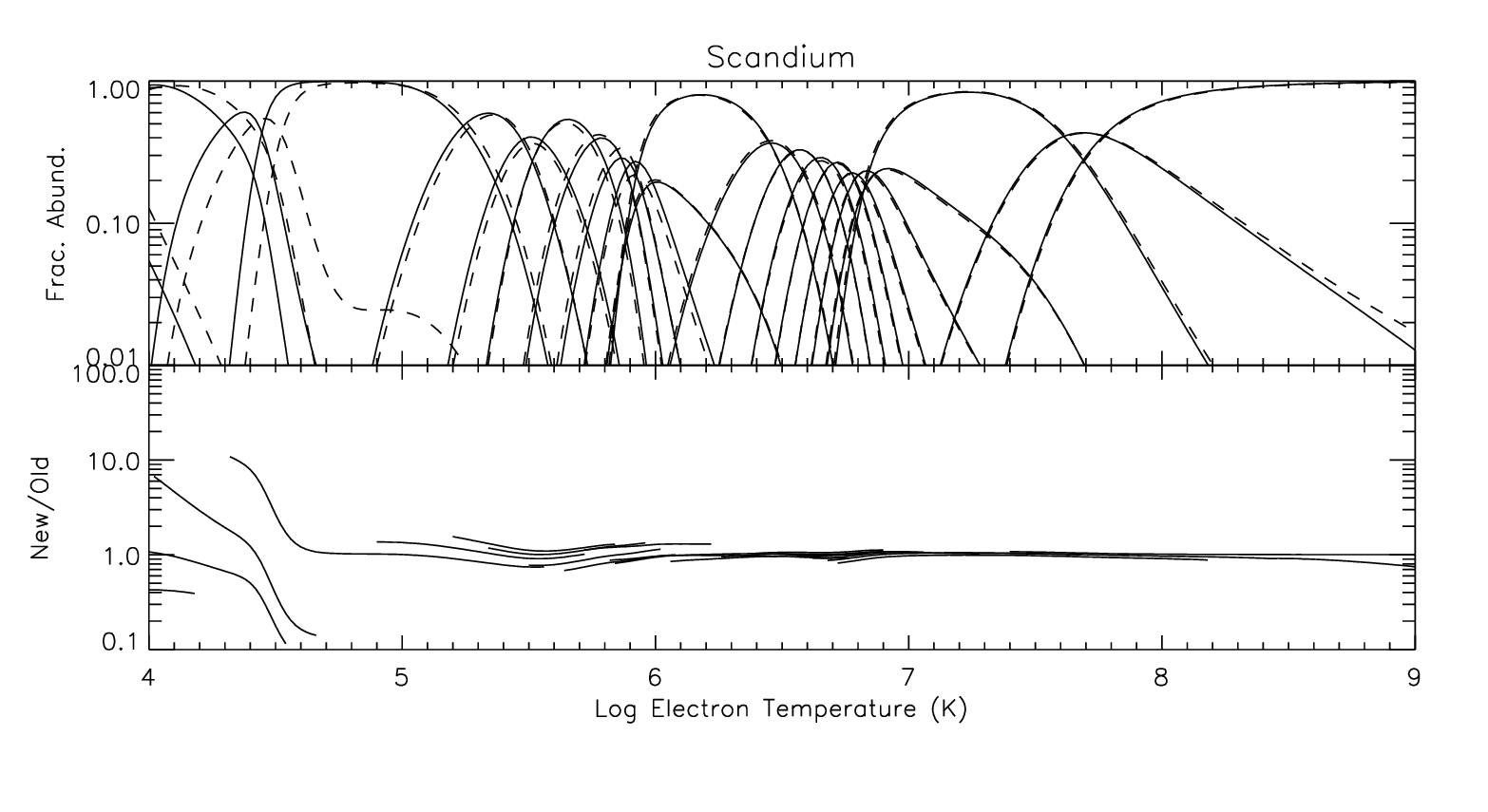}
  \caption[]{Same as Fig.~\protect\ref{fig:H Bryans} but for Sc.}
  \label{fig:Sc Bryans}
\end{figure}

\begin{figure}
  \centering
  \includegraphics[angle=90]{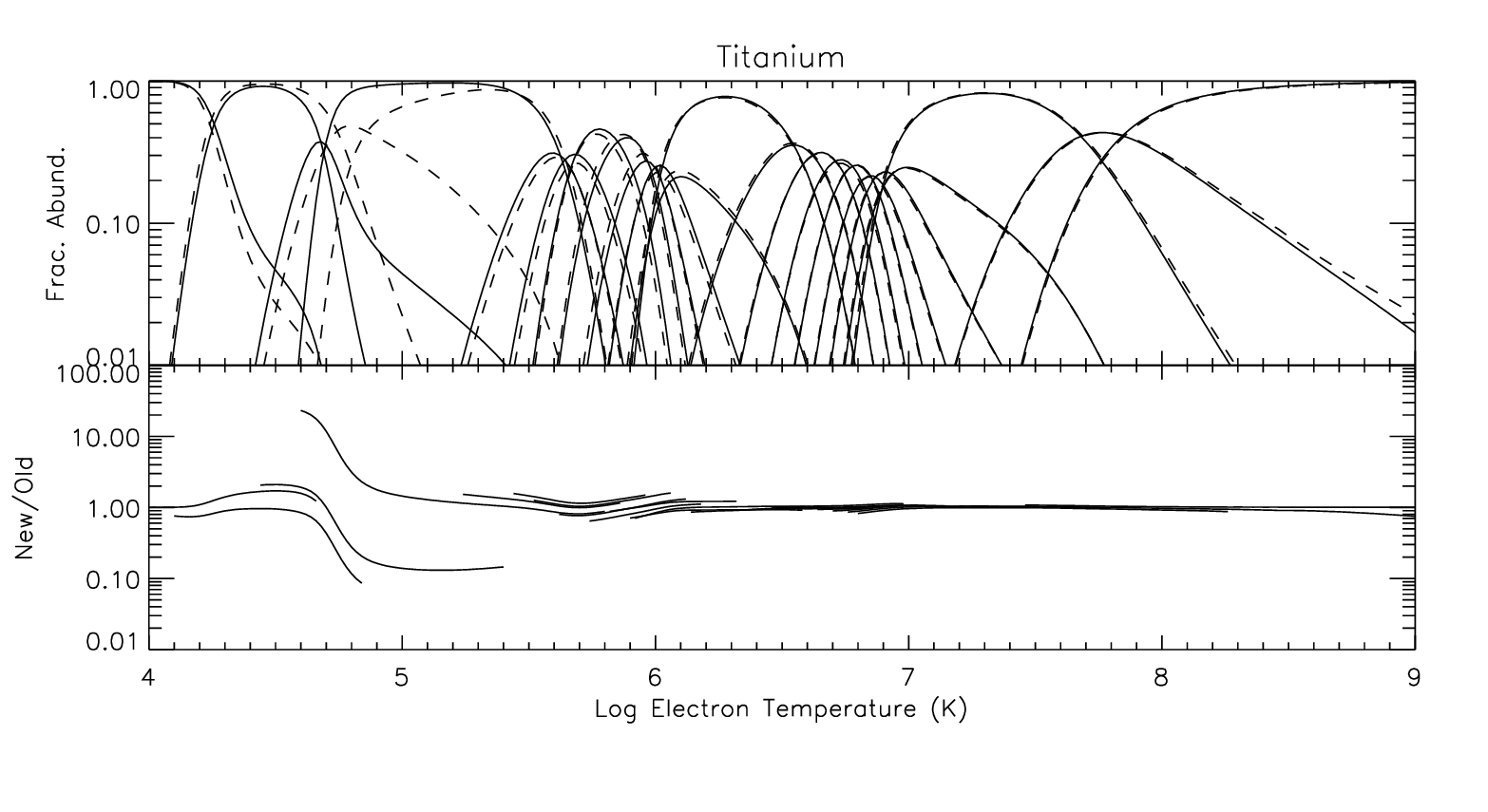}
  \caption[]{Same as Fig.~\protect\ref{fig:H Bryans} but for Ti.}
  \label{fig:Ti Bryans}
\end{figure}

\begin{figure}
  \centering
  \includegraphics[angle=90]{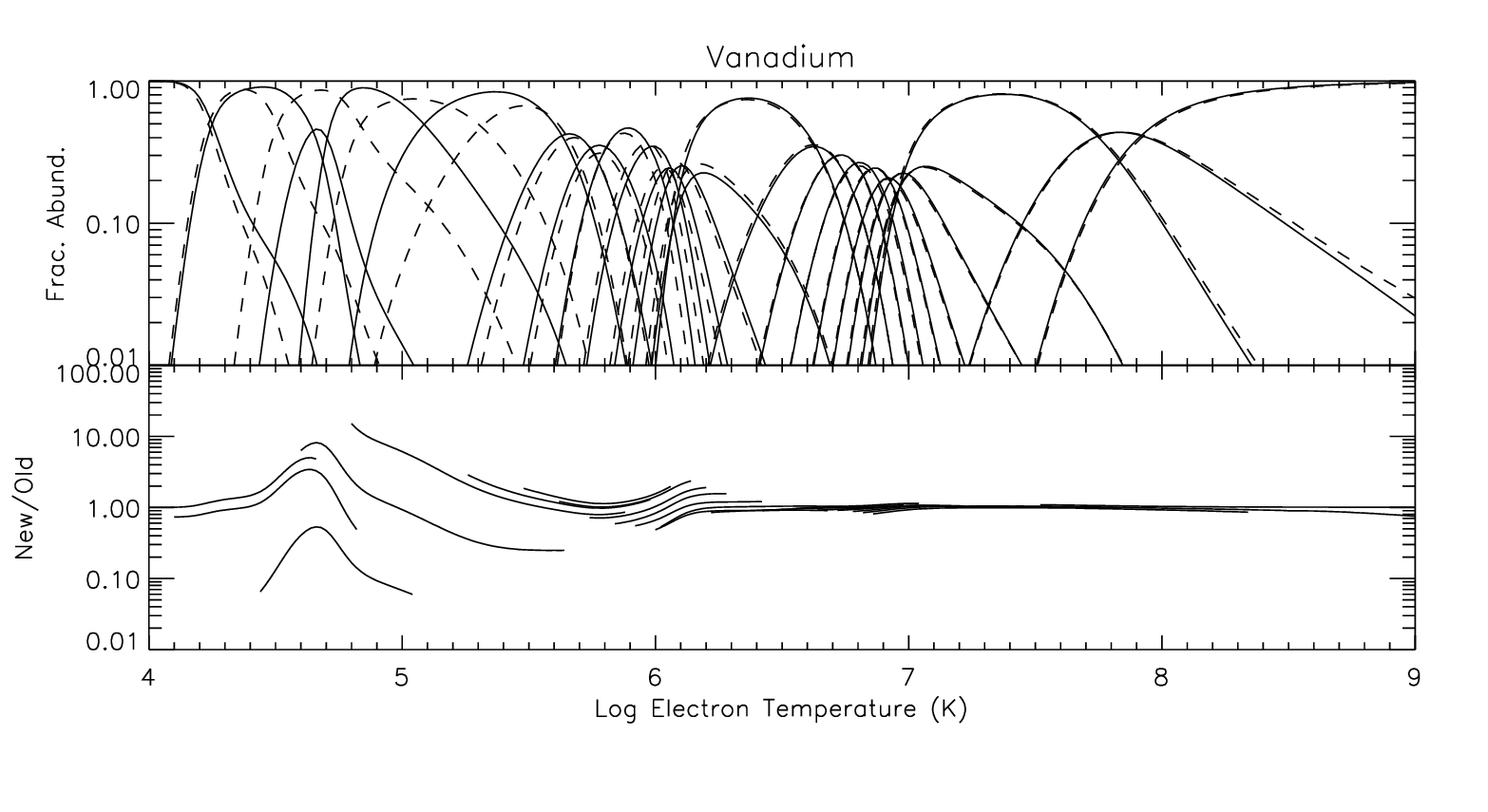}
  \caption[]{Same as Fig.~\protect\ref{fig:H Bryans} but for V.}
  \label{fig:V Bryans}
\end{figure}

\begin{figure}
  \centering
  \includegraphics[angle=90]{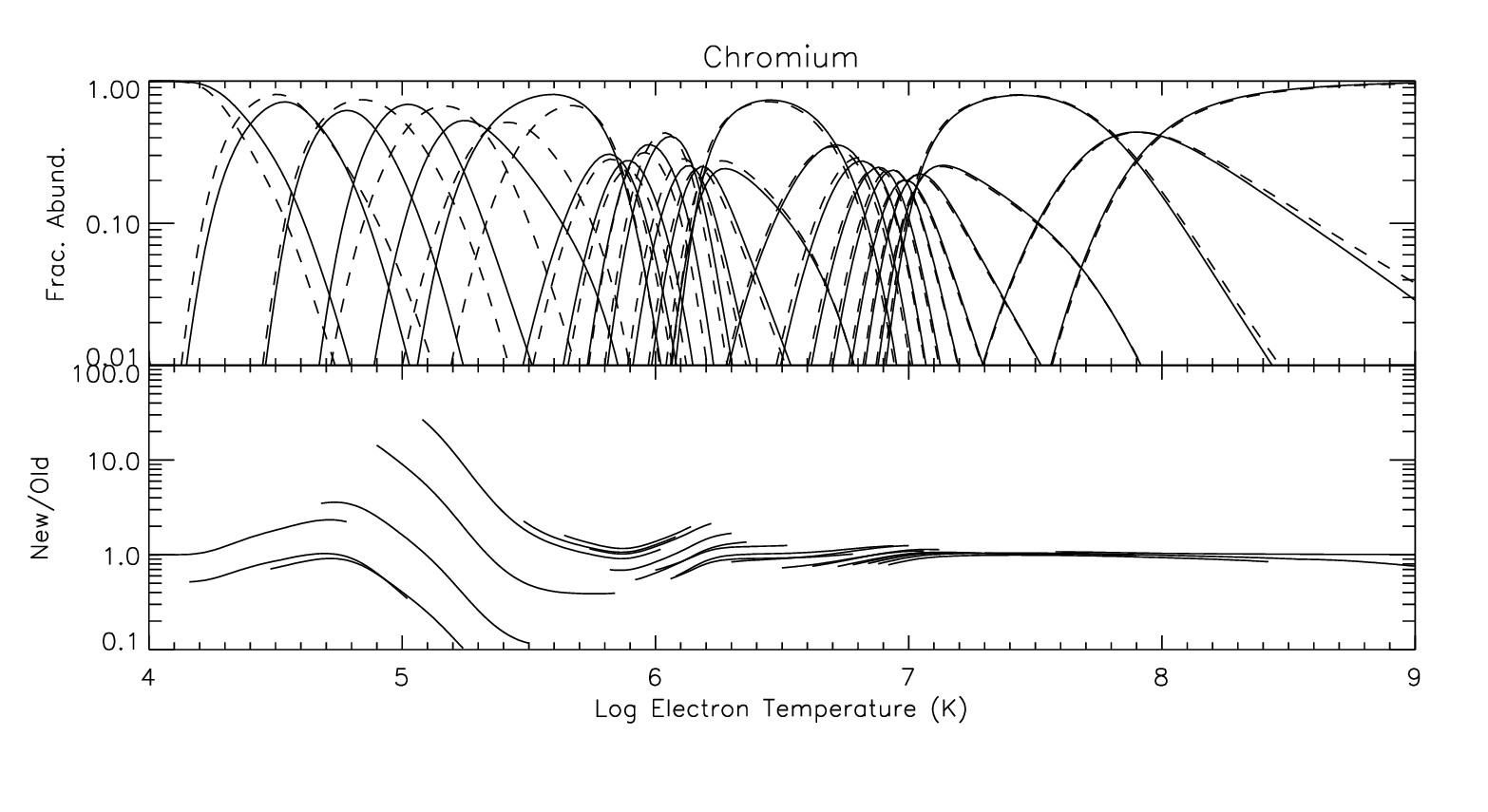}
  \caption[]{Same as Fig.~\protect\ref{fig:H Bryans} but for Cr.}
  \label{fig:Cr Bryans}
\end{figure}

\begin{figure}
  \centering
  \includegraphics[angle=90]{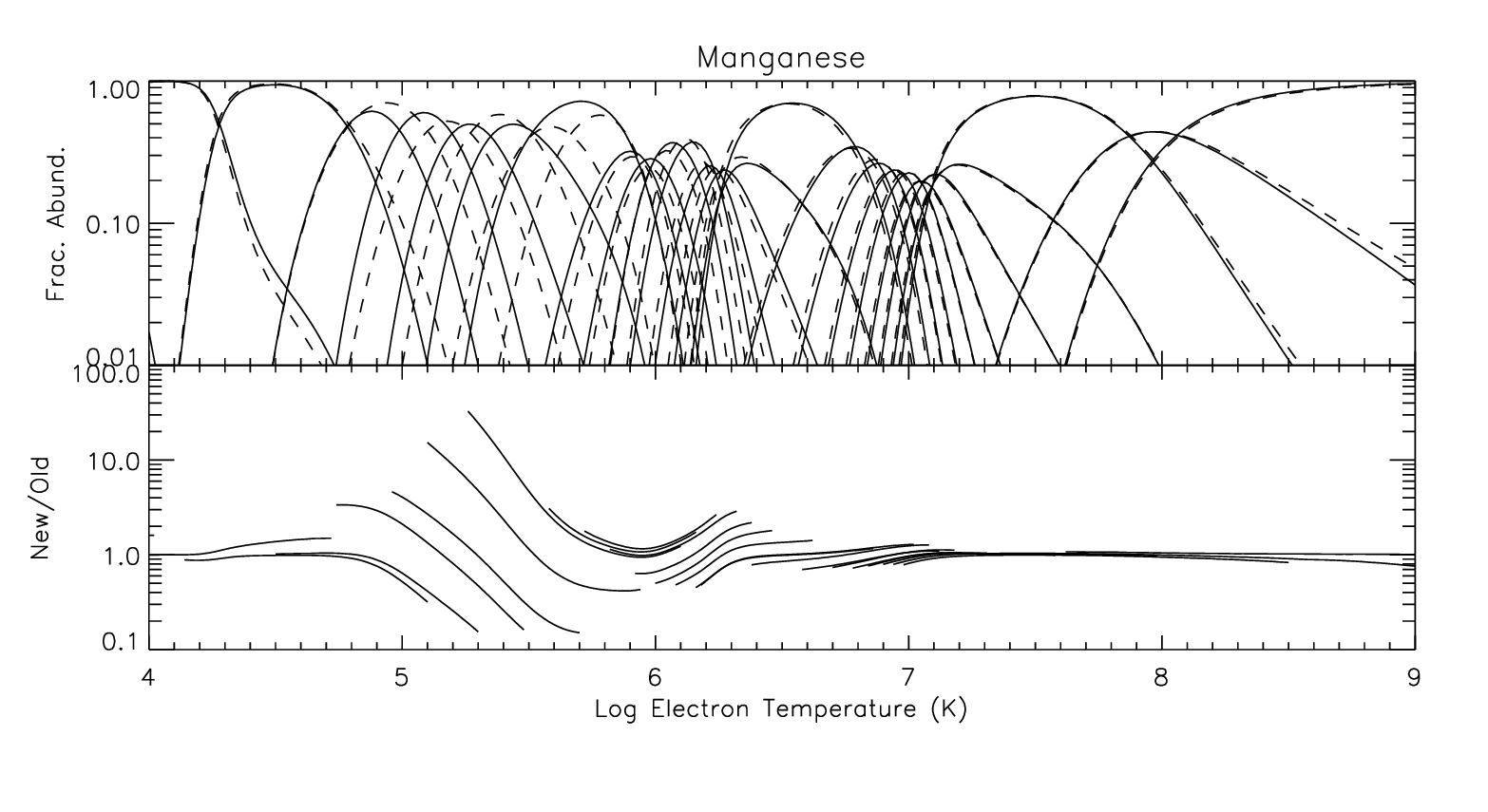}
  \caption[]{Same as Fig.~\protect\ref{fig:H Bryans} but for Mn.}
  \label{fig:Mn Bryans}
\end{figure}

\begin{figure}
  \centering
  \includegraphics[angle=90]{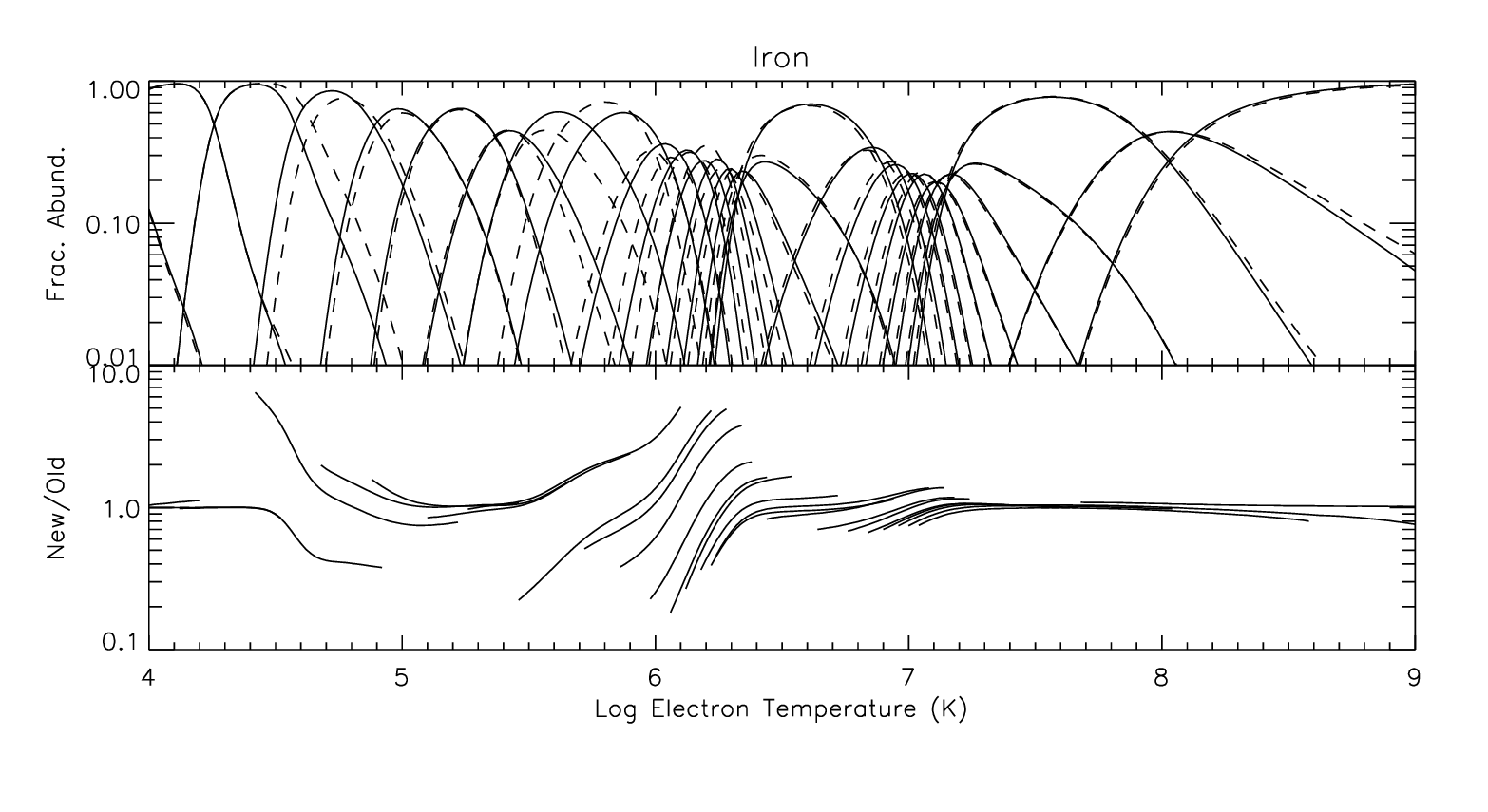}
  \caption[]{Same as Fig.~\protect\ref{fig:H Bryans} but for Fe.}
  \label{fig:Fe Bryans}
\end{figure}

\begin{figure}
  \centering
  \includegraphics[angle=90]{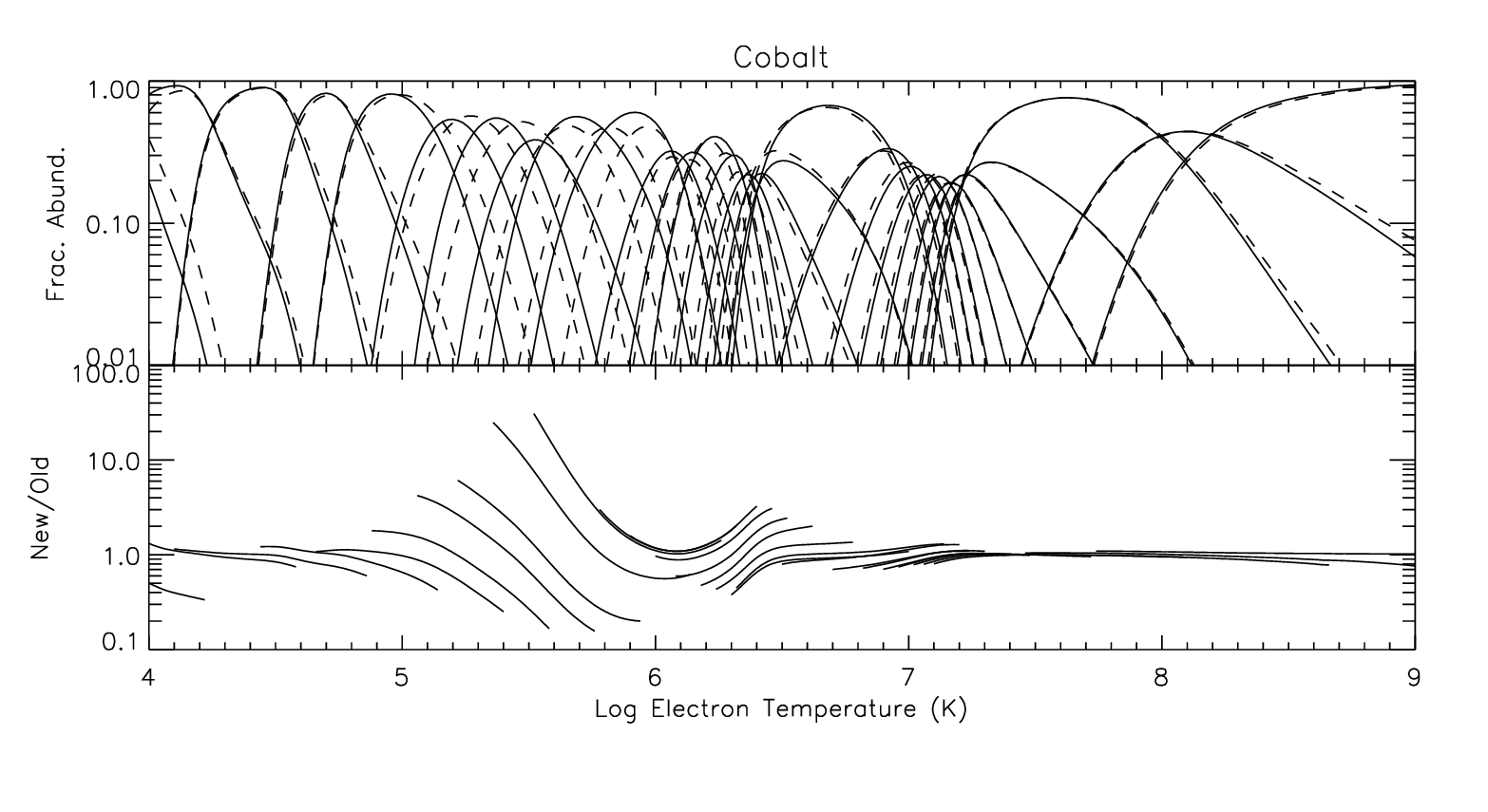}
  \caption[]{Same as Fig.~\protect\ref{fig:H Bryans} but for Co.}
  \label{fig:Co Bryans}
\end{figure}

\begin{figure}
  \centering
  \includegraphics[angle=90]{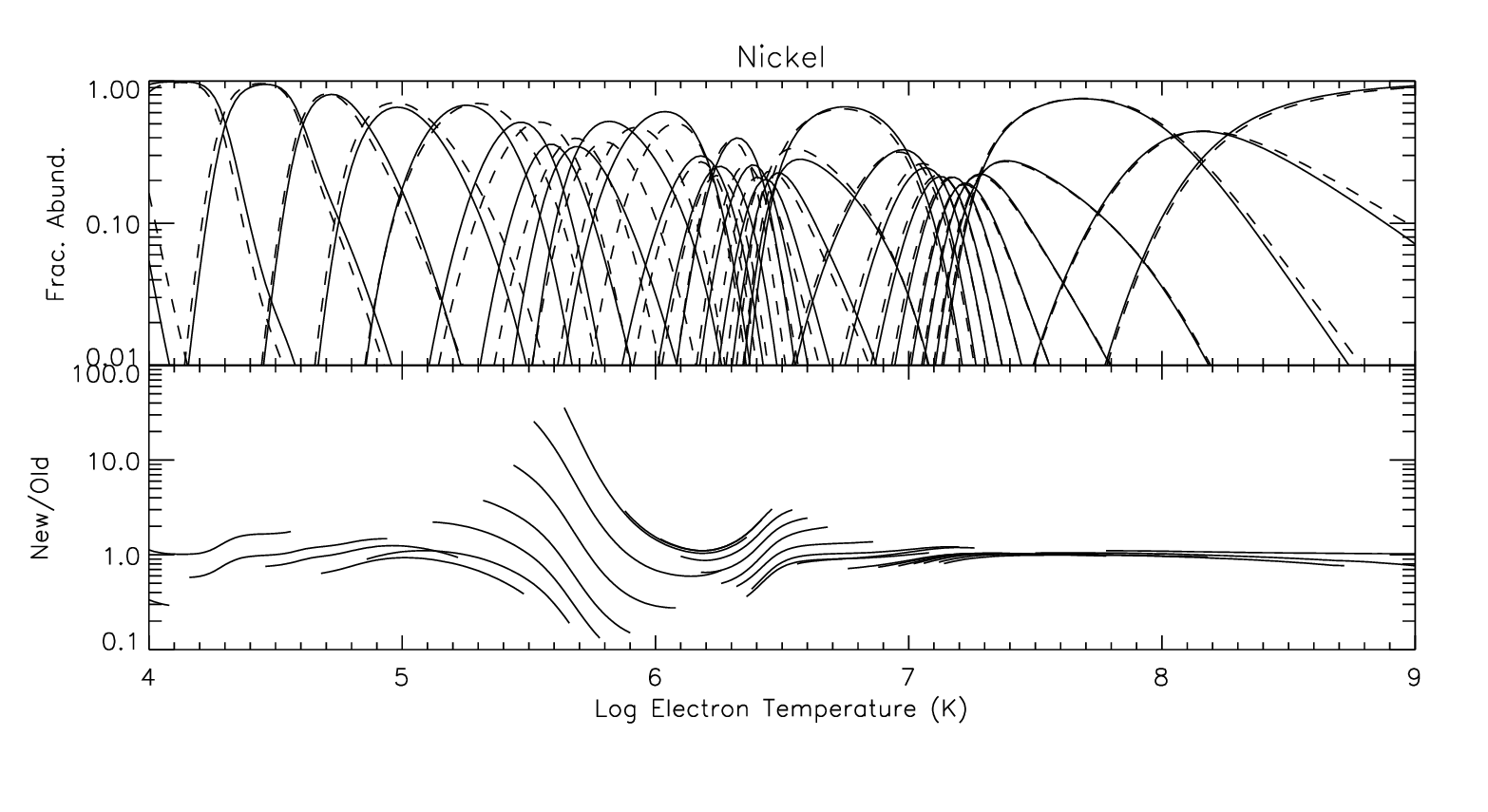}
  \caption[]{Same as Fig.~\protect\ref{fig:H Bryans} but for Ni.}
  \label{fig:Ni Bryans}
\end{figure}

\begin{figure}
  \centering
  \includegraphics[angle=90]{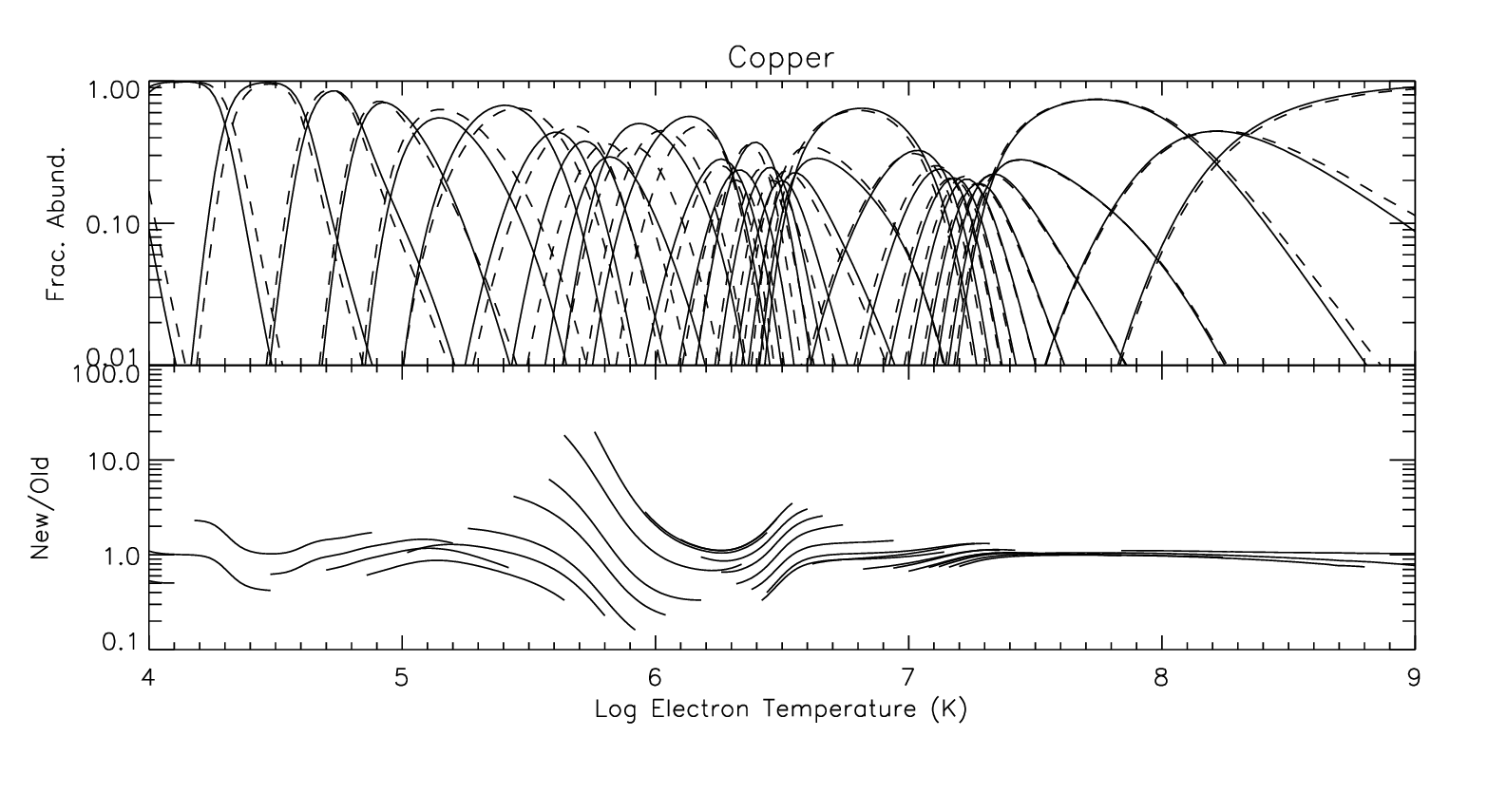}
  \caption[]{Same as Fig.~\protect\ref{fig:H Bryans} but for Cu.}
  \label{fig:Cu Bryans}
\end{figure}

\begin{figure}
  \centering
  \includegraphics[angle=90]{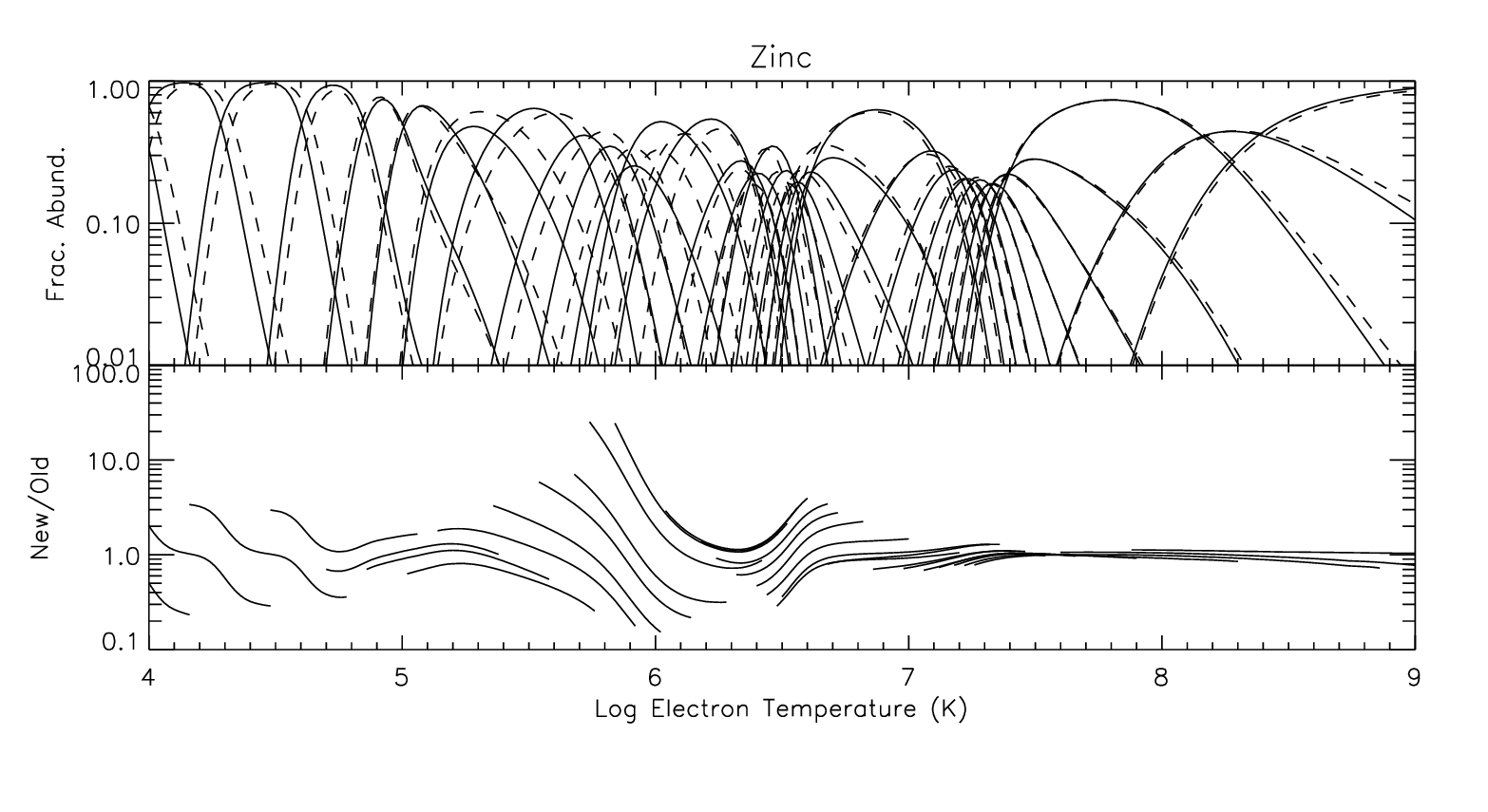}
  \caption[]{Same as Fig.~\protect\ref{fig:H Bryans} but for Zn.}
  \label{fig:Zn Bryans}
\end{figure}

Here we provide electronic tables of the CIE fractional
abundances  for all elements from H through Zn calculated using 
these data (Tables~\ref{tab:H}--\ref{tab:Zn}).
These tabulations
are provided for a $T_e$ range of $10^4$--$10^9$~K.
For ease of comparison with previous CIE fractional abundance calculations
we present figures showing the present results along with those of
\citet{Mazz98a} in Figs.~\ref{fig:H Mazz}--\ref{fig:Zn Mazz},
and the present results along with those of
\citet{Brya06a} in Figs.~\ref{fig:H Bryans}--\ref{fig:Zn Bryans}.

\section{A new approach to derive average emission measures and temperatures}
\label{sec:method}

Using the method described in Sec.~\ref{sec:em}, the assumption of constant
temperature and density, and our updated CIE results, we can
calculate the $EM$ curve for each of the observed spectral lines listed in 
Table~\ref{tab:linelist}.
Due to oversimplifications of the plasma model,
uncertainties in the observations, and
errors in the atomic data, there is no common intersection
of all $EM$ curves at a single $[T_c,EM_c]$. So one must calculate the most 
likely $EM$ and $T_e$ of the plasma
based on the range of values where the $EM$ curves cross one another.
To determine these values we have developed a
mathematically more rigorous approach than has been used in the past
for isothermal plasmas.
Here we use the emission lines from Si to illustrate this new method.
We calculate the $EM$ curves using a constant electron density
of $1.8\times 10^8$~cm$^{-3}$ as was reported by \citet{Feld99a}
for the same source region.

\begin{figure}
  \includegraphics{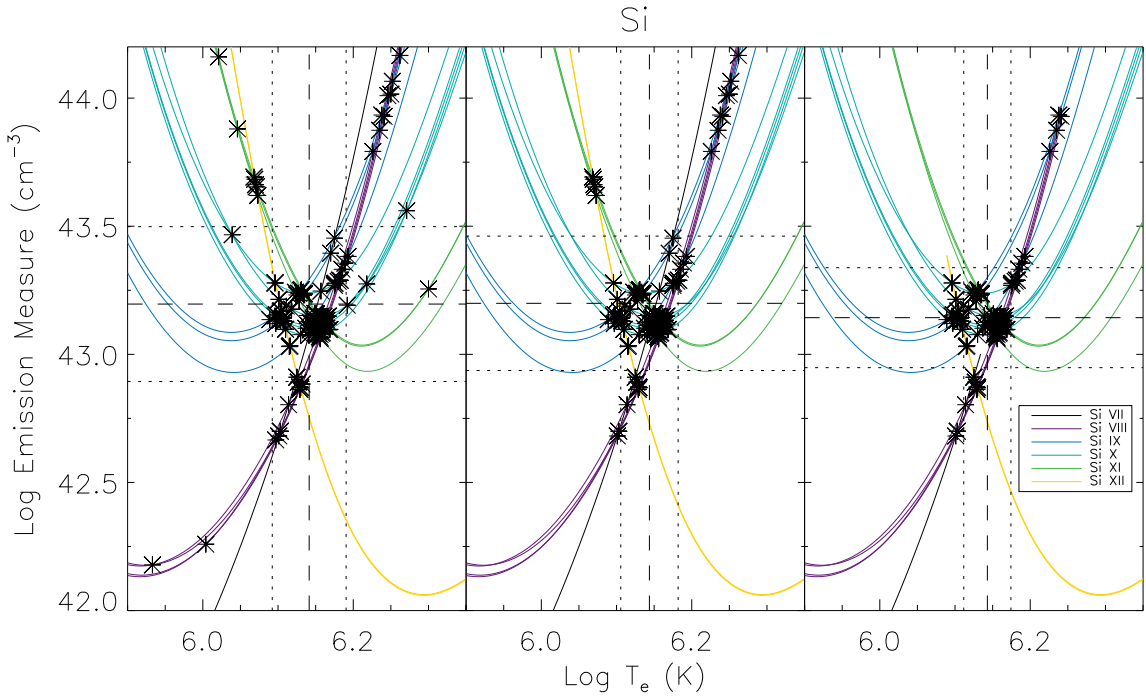}
  \caption[]{$EM$ versus $T_e$ curves of all  emission lines observed here from 
  Si.
  The dashed lines indicate the mean $\log_{10}EM$ and $\log_{10}T_e$ and the dotted lines show the 
  standard deviations of these values.  Asterisks indicate where the curves cross.
  The left panel shows the results after Step 1 of the analysis,
  the middle panel shows the results after Step 2, and
  the right panel shows the results after Step 3.
  See Sec.~\ref{sec:method} for a description of each step.}
  \label{fig:si}
\end{figure}

Step 1 of our approach is to take the mean of all  crossing points of the $EM$ curves
for a given group of lines.
This can be seen in the left panel of Fig.~\ref{fig:si}.  In this
panel we have marked with an asterisk
 every crossing point of the $EM$ curves shown.

The $EM$ vs.\ $T_e$ curves vary more slowly in log-log space than in
linear space.  
Also, because of the shape of the curves, any outlying crossings 
are far more likely to occur at a higher $EM$ than at a lower $EM$.
Thus, those crossings that fall far from the preponderance skew the average always
towards higher values of the $EM$. 
To avoid giving undue weight to these
points we calculate the mean in log space,
where $\langle \log_{10} EM\rangle \le \log_{10}\langle EM\rangle$.
This is equivalent to taking the geometric mean rather than the more
common arithmetic mean. The log of the geometric mean $EM$ is given by
\begin{equation}
  \langle\log_{10} EM\rangle=\log_{10}\left( \prod_{i=1}^{n} EM_i \right) ^{1/n}
  = \frac{1}{n}\sum_{i=1}^{n} \log_{10} EM_i
\end{equation}
and its standard deviation by
\begin{equation}
  \delta\langle \log_{10}EM\rangle=\sqrt{\frac{\sum_{i=1}^{n} (\log_{10} EM_i 
  -\langle\log_{10} EM\rangle)^2}{n}} 
\end{equation}
where $n$ is the number of crossing points over which the mean is being taken
and $EM_i$ is the value of $EM$ at each of these crossings.
By a similar argument the mean and standard deviation of $T_e$ are
 calculated in the same way.
From here on, unless otherwise stated, 
when we discuss the mean and standard deviation of the $EM$ and $T_e$ we are
referring to the geometric mean and geometric standard deviation.
In Fig.~\ref{fig:si} the mean $\log_{10}EM$ and mean $\log_{10}T_e$ are shown as dashed lines
 and the standard deviations by dotted lines.

Step 2  eliminates the less physically probable crossings when
two $EM$ curves  cross one another more than once.
For any two curves we select only the
crossing point that is closest, in the $\log EM$-$\log T_e$ plane,
 to the 
mean calculated values of the $EM$ and $T_e$
from Step 1. 
In Step 2 we also exclude some additional unphysical crossing points. 
In all cases where there
are multiple emission lines from a single ion, the $EM$ curves are nearly parallel.
Often, these curves nearly overlap with one another and can cross in one  or more
places. We attribute the crossings of these lines to errors in the effective line
emission rate coefficients and/or issues with the observed line intensities. For this
reason, we exclude these crossings from our calculation. 
For Si emission lines, such crossings are seen for Si~{\sc viii}, {\sc x}, and {\sc xi}
(Fig.~\ref{fig:si}).
Using this reduced set of crossings we
recalculate the mean and standard deviation of the $EM$ and temperature.
This plot is shown in the middle panel of  Fig.~\ref{fig:si}.

In Step 3 we further reduce the dataset by considering only $EM$ curves in the
temperature range where $f^m\ge0.01$. The reliability of all published  CIE calculations
is uncertain below this fractional abundance.
\citet{Brya06a}  compared the results of CIE calculations using 2 different compilations
of state-of-the-art DR and RR datasets.  Agreement at peak abundance was
found to be within 10\% and within 50\% when going to temperatures
where the fractional
abundance is 0.01.
Outside this temperature range, for values of $f^m<0.01$, the reliability of the CIE
calculations grows significantly worse.

In the right
panel of Fig.~\ref{fig:si} we show the same $EM$ curves as in the middle
panel but only for the temperature range where $f^m\ge0.01$.
It is this mean $EM$ and $T_e$ after Step 3 that we consider the most likely
$EM$ and $T_e$ for a given set of emission lines.
Henceforth, when discussing the results after all three steps
of our analysis, we refer to the $EM$ and $T_e$ as coming from
the Geometric mean Emission Measure (GEM) method.

\section{Coronal abundance enhancement factors}
\label{sec:fip}

Our first step in determining the coronal abundance of 
the observed elements is to assume
that the high-FIP
elements Ne and Ar have the same abundance in the corona as they do in the photosphere.
This follows the approach taken by \citet{Feld98a}.
Using the photospheric abundances of
Feldman \& Laming (2000; see Table~\ref{tab:abundances})
we have calculated the
geometric mean $EM$ from emission lines of Ne and Ar
using the GEM method outlined in Sec.~\ref{sec:method},
giving $\langle\log_{10} EM_{\rm high-FIP}\rangle$.

An objective of the present paper is to investigate the apparent abundance
discrepancy of Li- and Na-like ions.  Previous studies, such as those of
\citet{Dupr72a}, \citet{Feld98a}, and \citet{Land02a}, have found the abundance of these ions
to be greater than those of ions in other isoelectronic sequences.
In order that this discrepancy does not affect our calculation of the
FIP factors of each element we do not include any Li- or Na-like lines in the
calculations of the FIP factors detailed below.

\citet{Land02a} also reported a  difference in $T_e$
derived from N-like and non--N-like ions within Group~I.
However, unlike the Li- and Na-like abundance discrepancy, this has not been
reported in the literature previously.  
If we adopt the uniform FIP factor of 3.5 used by \citet{Land02a} 
for all high-FIP N-like and non--N-like ions
and implement our GEM method we find no discrepancy in the $T_e$
derived from N-like and non--N-like ions.
This is discussed in more detail in Sec.~\ref{sec:discuss}. 
For these reasons, in this
section we include both
N-like and non--N-like ions in our analysis.

\begin{figure}
  \includegraphics{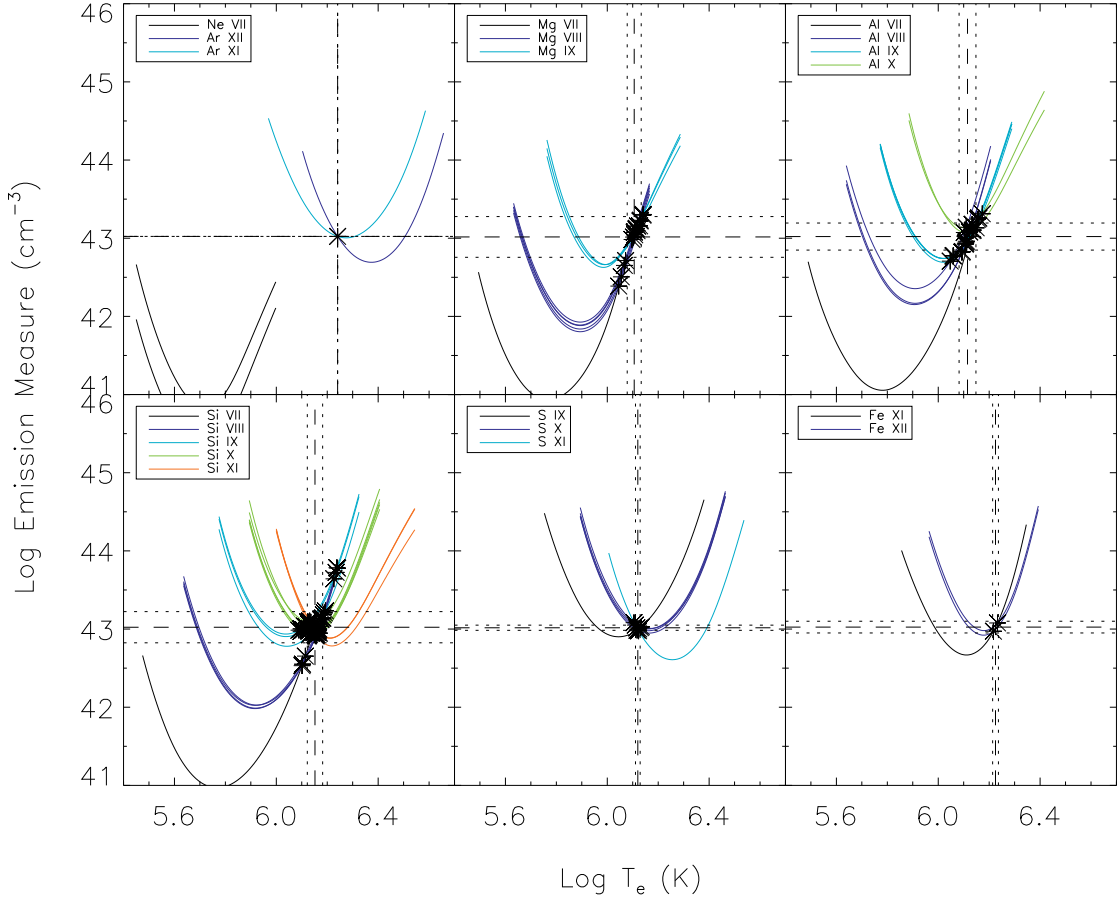}
  \caption[]{$EM$ versus $T_e$ curves of all the emission lines from each of the low- and moderate-FIP
  elements using the GEM method described in Sec.~\ref{sec:method}.  Na, K and Ca are excluded 
  as in this SUMER dataset there are not enough
  observed emission lines from these elements to determine a mean $EM$.
  The upper left panel shows the high-FIP elements Ne and Ar.}
  \label{fig:all elem}
\end{figure}

The results of the GEM analysis of the high-FIP elements
 can be seen in the upper left panel of Fig.~\ref{fig:all elem}.
The exclusion of the Li- and Na-like ions results in only a single crossing
remaining after the 3 steps---due to 2 emission
lines from Ar~{\sc xi} and Ar~{\sc xii}.  We use the value of the
$EM$ at this point as our reference value.
Restricting the temperature range to that where the  fractional
abundance of an ion is greater than 1\% limits us to this single crossing
since the 2 Ne~{\sc vii} $EM$ curves are below this 
limit at the $T_e$ values where they cross the $EM$ curves of Ar~{\sc xi} and Ar~{\sc xii}.
It is not ideal that we are left with only a single crossing
but we believe this represents an improvement over the work of
\citet{Feld98a}.  There they used only a single line, whereas
here we use two.  Additionally, the line they selected was Li-like
O~{\sc vi}.  As we have discussed above, and will also discuss
in Sec.~\ref{sec:discuss}, there are several reasons to treat this line with
suspicion.
It is also worth noting that the  crossing of the Ar lines
results in $\log_{10} T_e=6.24$.  This is $\sim 0.1$ in the dex higher than
the temperature derived from the other emission lines (see 
later in this section and Sec.~\ref{sec:groups}).  However, in the absence
of additional non--Li- and Na-like emission from other
high-FIP elements, we consider normalizing
to this crossing of Ar lines to be the best approach to analyzing this
particular observation.

We next separate all other emission lines 
by the element responsible for the emission and,
again using the GEM method,
calculate the mean $EM$ of each of the low- and moderate-FIP elements individually
using the photospheric elemental abundances as our starting value,
giving $\langle\log_{10} EM_{\rm X}\rangle$ for each element X.
For each of these low- to moderate-FIP elements, we determine an ``enhancement factor''
$f_{\rm X}$
for the elemental abundance that will result in the same derived $EM$ as
found for the high-FIP element Ar.
From Eqns.~\ref{eqn:cont func} and \ref{eqn:em} we see that the elemental abundance,
\begin{equation}
f(X)=\frac{n({\rm X})}{n({\rm H})},
\end{equation}
is inversely proportional to the $EM$ of the emitting plasma so the $f_{\rm X}$
values can be calculated as
\begin{equation}
  \label{eqn:fip}
  \log_{10}f_{\rm X} = \langle\log_{10}EM_{\rm X}\rangle
                      -\langle\log_{10}EM_{\rm high-FIP}\rangle 
\end{equation}
where
\begin{equation}
  f_{\rm X} = \frac{f(X)_{\rm corona}}{f(X)_{\rm photosphere}}.
\end{equation}

For the emission from the elements Mg, Al, Si, S, 
 and Fe we
show the $EM$ as a function of $T_e$ in Fig.~\ref{fig:all elem}
where we have used our derived coronal elemental abundances.
These are subject to the 3 steps of the GEM method
but in this case we only show the last step.
The derived elemental abundances are given in Table~\ref{tab:abundances}.

From Eq.~\ref{eqn:fip}, we estimate the absolute error in 
$\log_{10}f_{\rm X}$ as the 
quadrature sum of the standard deviations of the $EM$ from the
high-FIP elements and the $EM$ from the individual element X, i.e.,
\begin{equation}
  \delta\langle\log_{10}f_{\rm X}\rangle = 
  \sqrt{\delta\langle\log_{10}EM_{\rm X}\rangle^2  +
  \delta\langle\log_{10}EM_{\rm high-FIP}\rangle^2}
\end{equation}
However, since only a single crossing of Ar lines is used to determine
the high-FIP $EM$, there is no error associated with 
$\log_{10}EM_{\rm high-FIP}$ and the error in $\log_{10}f_{\rm X}$ 
reduces to $\delta\langle\log_{10}EM_{\rm X}\rangle$ and
is thus probably an underestimate.
Given our derived errors in $\langle\log_{10}EM\rangle$ presented in 
Sec.~\ref{sec:groups}
we estimate $\langle\log_{10}EM_{\rm high-FIP}\rangle$ 
is good to $\sim\pm 0.3$ in the dex.
However, due to insufficient data, we do not attempt to assign an error 
to $\delta\langle\log_{10}EM_{\rm high-FIP}\rangle$.
Instead, we leave the errors in the FIP factors
as they are, but note that they are likely underestimates.

With the Li- and Na-like lines omitted, we are left with only 2 emission
lines from Na and Ca and a single emission line from K
in this observation.
Both Na emission lines are from the same charge state 
and their associated $EM$ curves are therefore almost parallel.
The same is true for the 2 emission lines from Ca.
We thus have no crossing points of the $EM$
curves over the $T_e$ range considered for Na, K, and Ca.
Also, 
as discussed in Sec.~\ref{sec:discuss}, we believe the emission from 
Li-like N~{\sc v} and O~{\sc vi} ions
to be from a cooler region of plasma so 
 we do not determine an average $EM$ from the curves crossings of these elements.

\begin{figure}
  \includegraphics{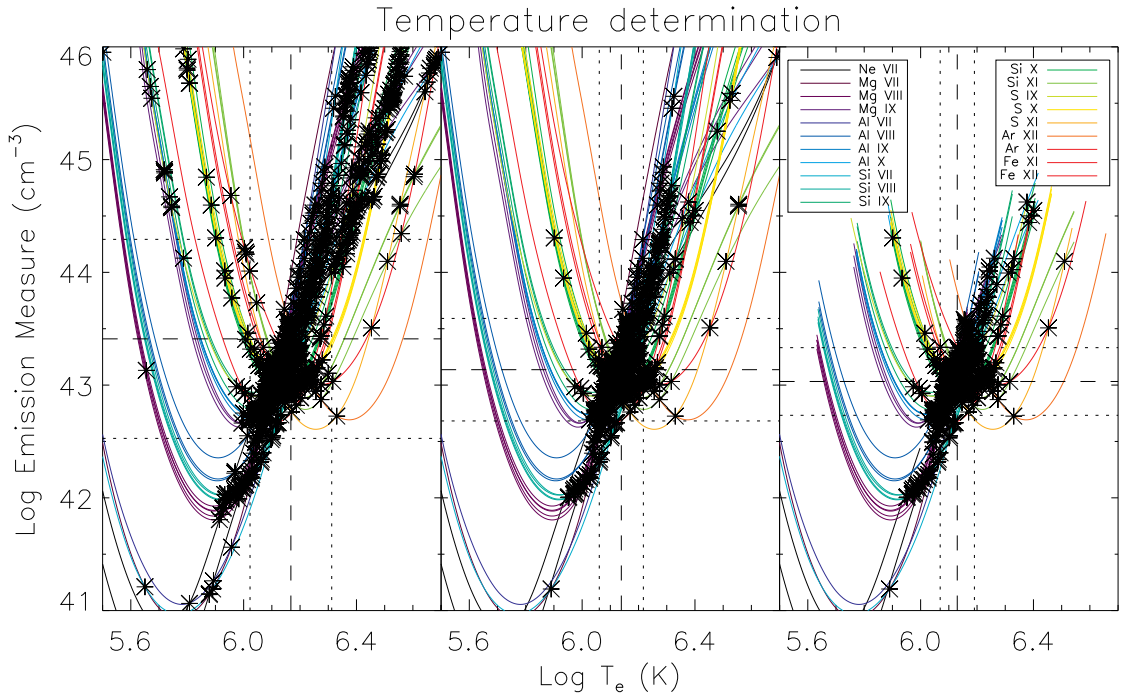}
  \caption[]{$EM$ versus $T_e$ curves of all the emission lines from each 
  of the elements Ne, Mg, Al, Si, S, Ar, and Fe
  (excluding Li- and Na-like ions).
  The $T_e$ derived from these elements were used to
  determine the FIP factors of  Na, K and Ca. Asterisks indicate where the curves cross.
  The three panels show the three steps of the GEM method as in Fig.~\ref{fig:si}.}
  \label{fig:te}
\end{figure}

To determine the FIP factors of Na, K and Ca we use their $EM$ values
at the $T_e$ determined from those emission lines for which we have already 
calculated FIP factors.  This mean $T_e$ determination is shown in 
Fig.~\ref{fig:te} where emission lines of Ne, Mg, Al, Si, S, 
Ar, and Fe have been considered (excluding Li- and Na-like ions).
This gives a value of $\log_{10} T_e=6.13\pm 0.06$, at which value we calculate
the $EM$ of each of the Na, K and Ca lines.
(For Na and Ca, for which we have 2 emission lines, we take the average value of
the 2 $EM$ values at $\log_{10} T_e=6.13$.)
  We then determine a FIP factor
for each of these elements that will give the same $EM$ as for the high-FIP
element Ar.
We estimate the absolute error in 
$\log_{10}f_{\rm X}$ of these 3 elements by calculating their $EM$ at the
values of the extremes of the errors associated with $T_e$,
i.e., the $EM$ at $\log_{10} T_e=6.07$ and $\log_{10} T_e=6.19$.

\begin{figure}
  \includegraphics{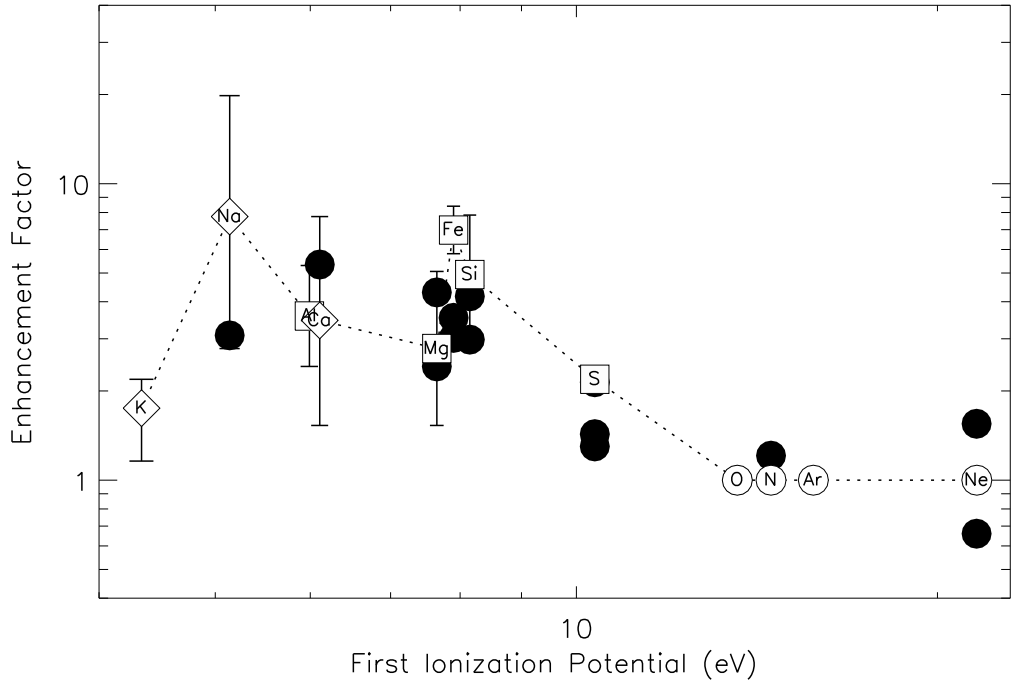}
  \caption[]{Coronal abundance enhancement factor (i.e., FIP factor) used for each of the elements
             versus their first ionization potential.  Open circles indicate
	     the high-FIP elements O, N, Ne and Ar where no enhancement was assumed.
	     The elements in squares indicate  those which had their enhancement
	     factor determined by matching their mean $EM$ with that of the
	     high-FIP element Ar.  Elements marked with diamonds are those that
	     did not have enough crossings to determine their $EM$ in this way
	     (see Sec.~\ref{sec:fip} for further details).
	     The dotted line is purely to guide the eye.
	     The solid circles are the results of \citet{Feld98a} for
	     N, O, Ne, Na, Mg, Si, S, Ca, and Fe.
	     Feldman et al.\ scaled their results to O and assumed
	     a FIP factor of 1 for this element.  We also set the
	     O FIP factor to 1, so their and our FIP factors for O
	     lie directly on top of one another in this plot. }

  \label{fig:fip}
\end{figure}

Table~\ref{tab:abundances} lists the enhancement factors,
which are often called FIP factors or FIP biases,
for all of the elements present in the observation.
We also give the resulting coronal elemental abundances.
The FIP factors are also shown in Fig.~\ref{fig:fip} alongside
the results of \citet{Feld98a}.
Note that Figs.~\ref{fig:si}--\ref{fig:fip} show the results when using the
CIE fractional abundances of the present paper.
We have repeated the analysis using the \citet{Mazz98a} CIE fractional
abundances.  We do not show figures of these results, but in
Table~\ref{tab:abundances} we list the FIP factors and coronal
abundances determined when using these older CIE data.

\section{Analysis by groups}
\label{sec:groups}
\begin{figure}
  \includegraphics{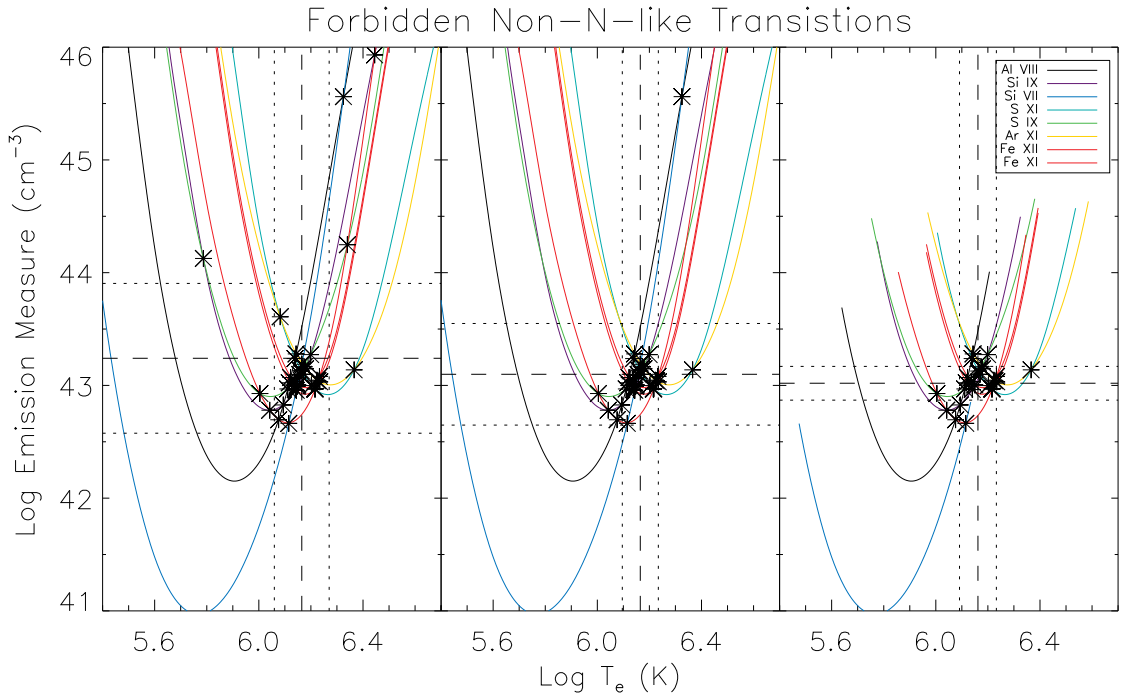}
  \caption[]{$EM$ versus $T_e$ curves for the emission lines of Group~Ia using our inferred 
  coronal abundances.  Asterisks indicate where the curves cross.
  The three panels show the three steps of the GEM method as in Fig.~\ref{fig:si}.}
  \label{fig:1a}
\end{figure}

\begin{figure}
  \includegraphics{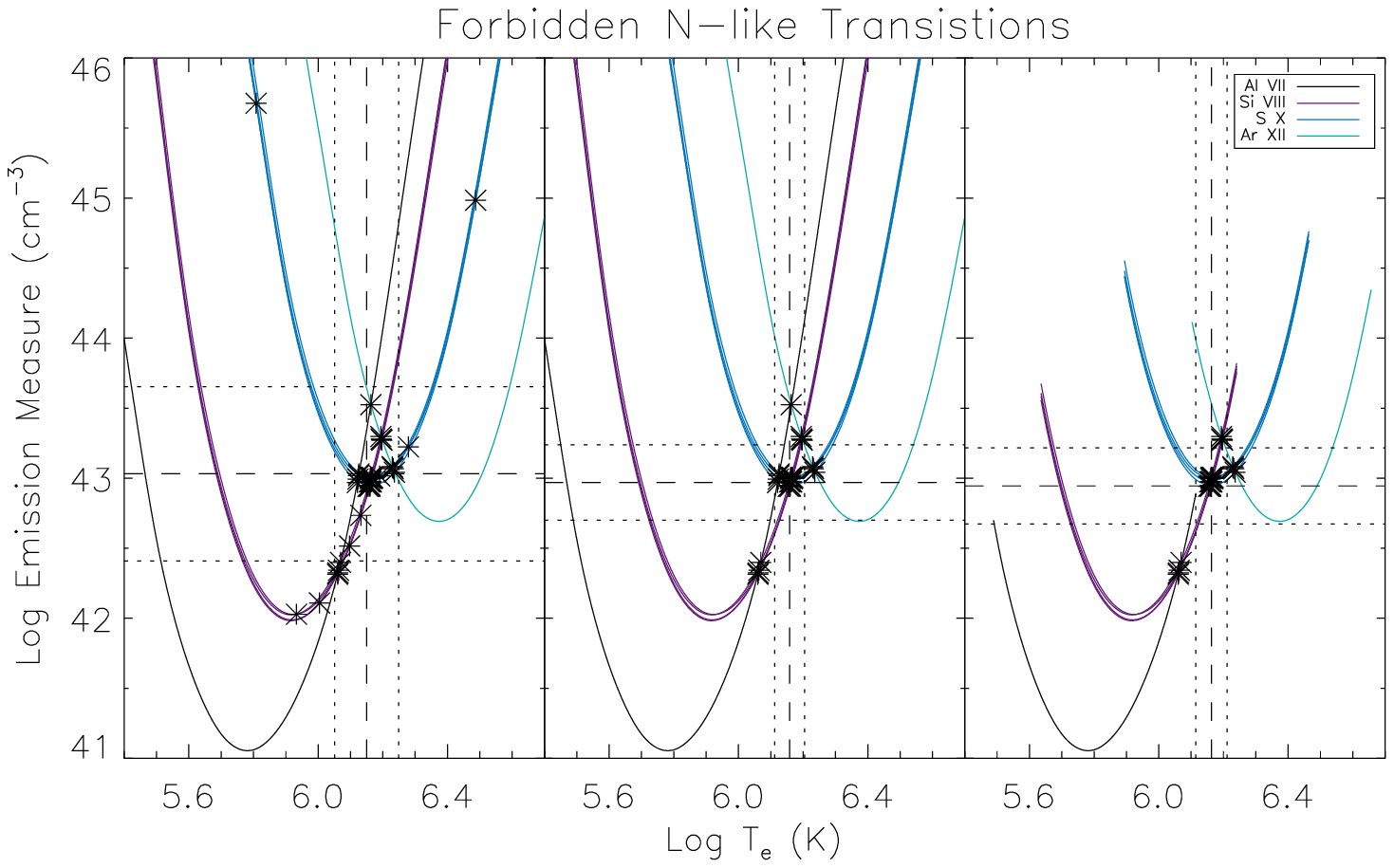}
  \caption[]{Same as Fig.~\ref{fig:1a} but for Group~Ib.}
  \label{fig:1b}
\end{figure}

\begin{figure}
  \includegraphics{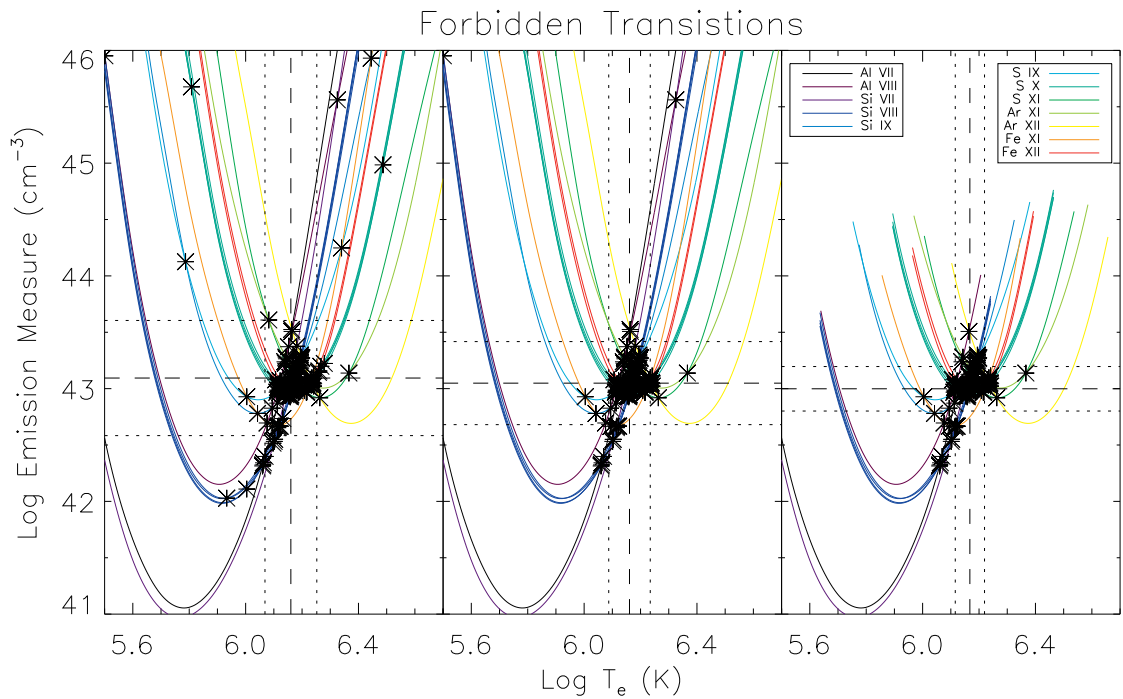}
  \caption[]{Same as Fig.~\ref{fig:1a} but for Group~I as a whole.}
  \label{fig:1}
\end{figure}

\begin{figure}
  \includegraphics{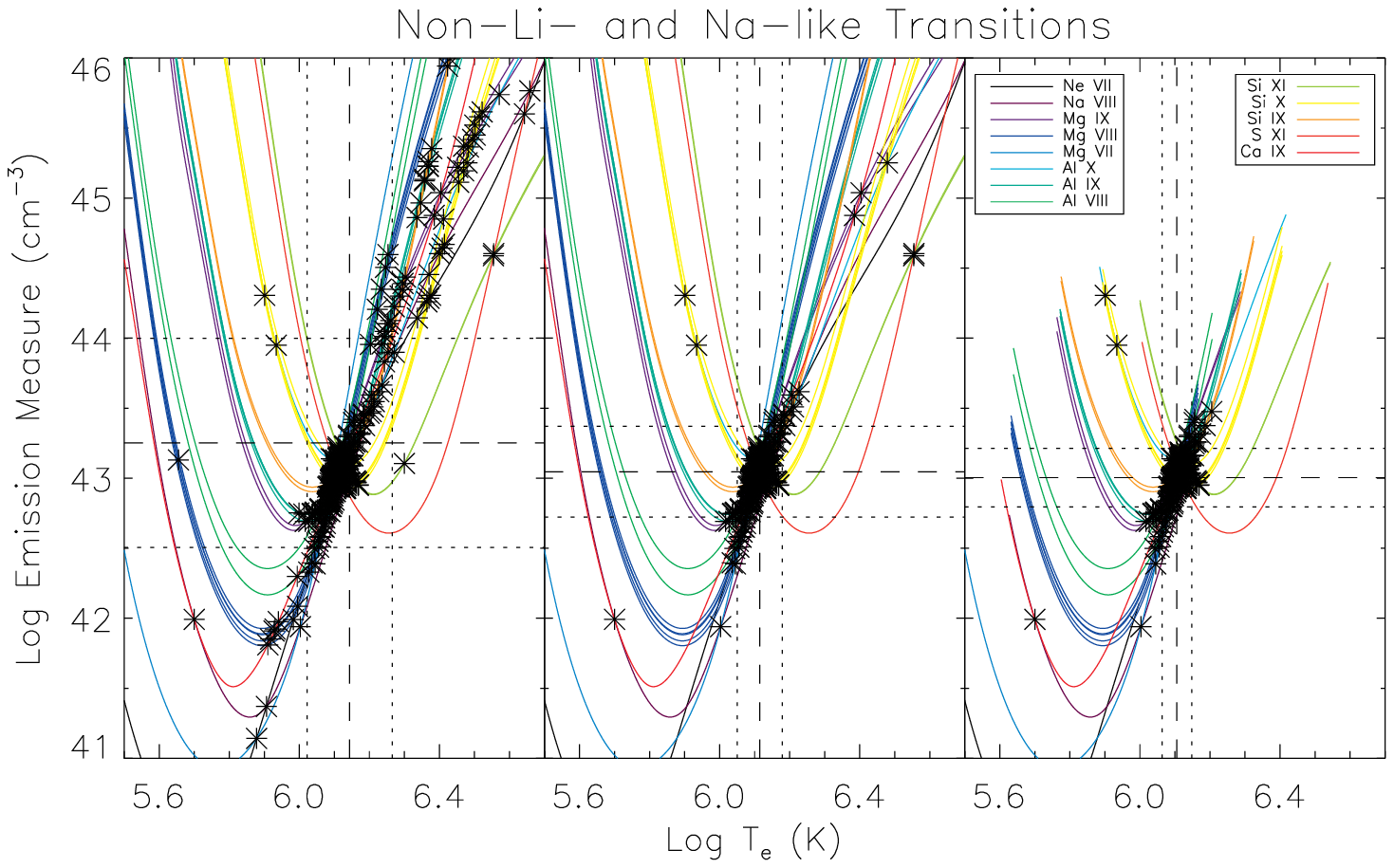}
  \caption[]{Same as Fig.~\ref{fig:1a} but for Group~IIb.}
  \label{fig:2b}
\end{figure}

\begin{figure}
  \includegraphics{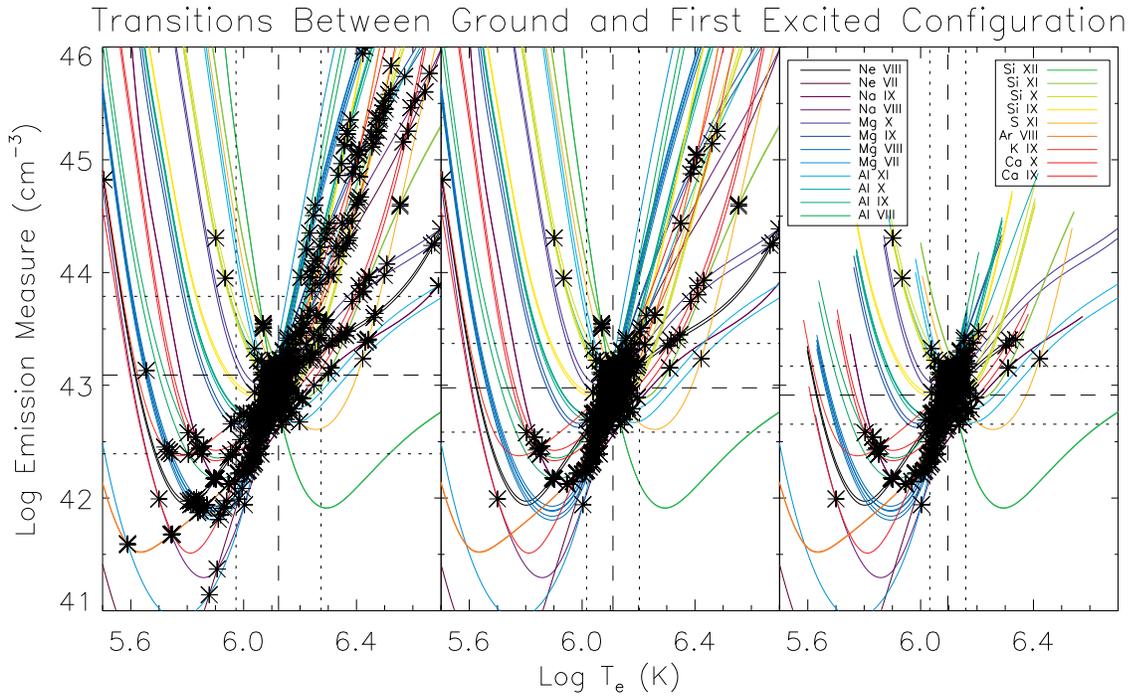}
  \caption[]{Same as Fig.~\ref{fig:1a} but for Group~II as a whole, excluding emission lines from N~{\sc v}
  and O~{\sc vi}.}
  \label{fig:2}
\end{figure}

\begin{figure}
  \includegraphics{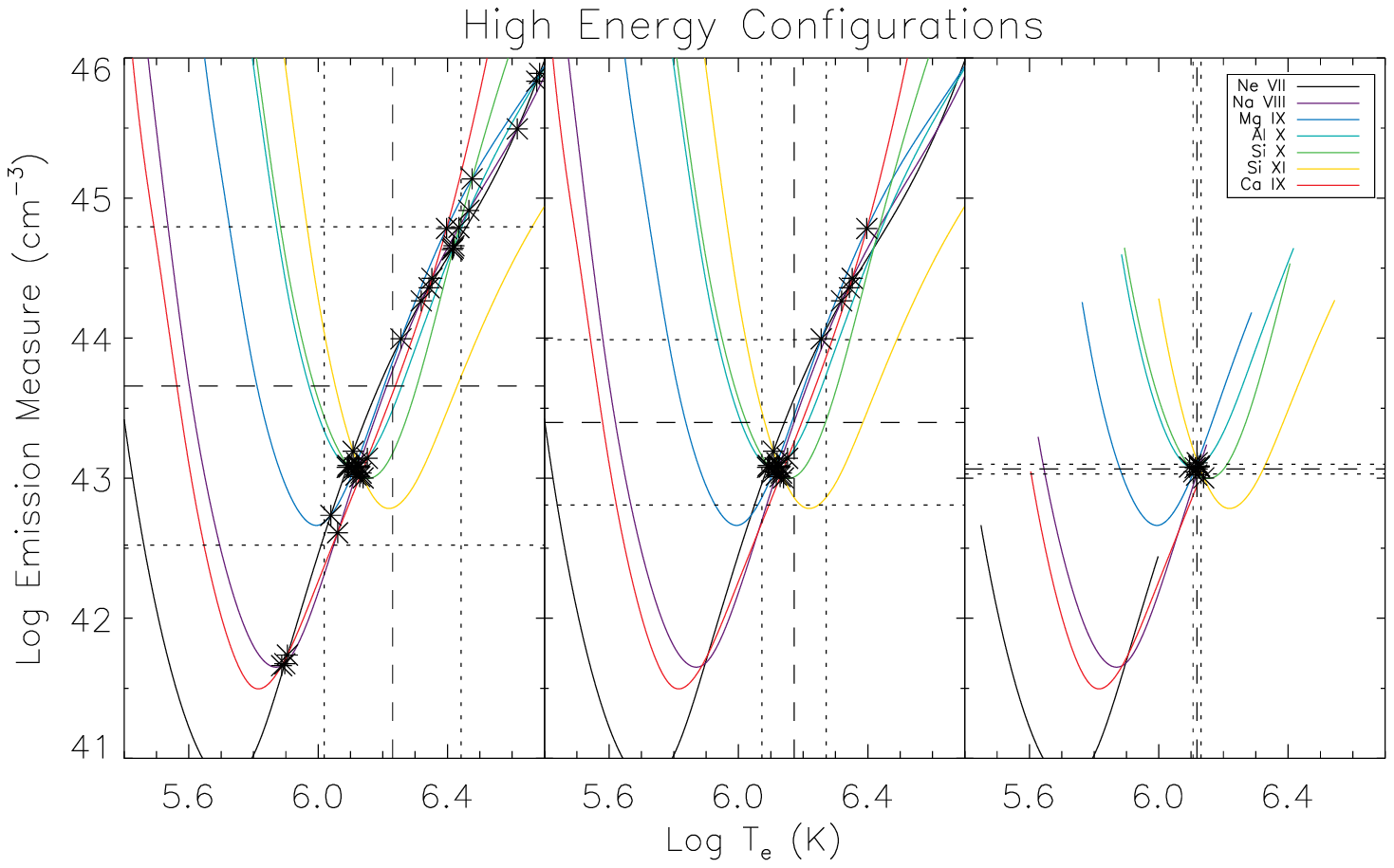}
  \caption[]{Same as Fig.~\ref{fig:1a} but for Group~III.}
  \label{fig:3}
\end{figure}

Using our derived coronal abundances  we calculate
the $EM$ and $T_e$ of each of the line categorizations
given in Sec.~\ref{sec:observations}.
Figures~\ref{fig:1a}--\ref{fig:3} show the GEM approach
 as applied to each of these groups.
We also give the results in Table~\ref{tab:averages}
listing the geometric mean and standard
deviation of the $EM$ and $T_e$  after each step
of the GEM method.

For the Group~I and II categorizations we show their individual subcategorizations
as well as the groups as a whole.
In the case of Group~IIb, the
emission lines have been
further subdivided by separating out the N~{\sc v} and O~{\sc vi} lines. This is 
because the $EM$ curves
from these lines do not match well with the others in this group.  
We elaborate on the possible
reasons for this in Sec.~\ref{sec:discuss}. 
When we consider Group~II as a whole, these lines are also excluded.

\begin{figure}
  \includegraphics{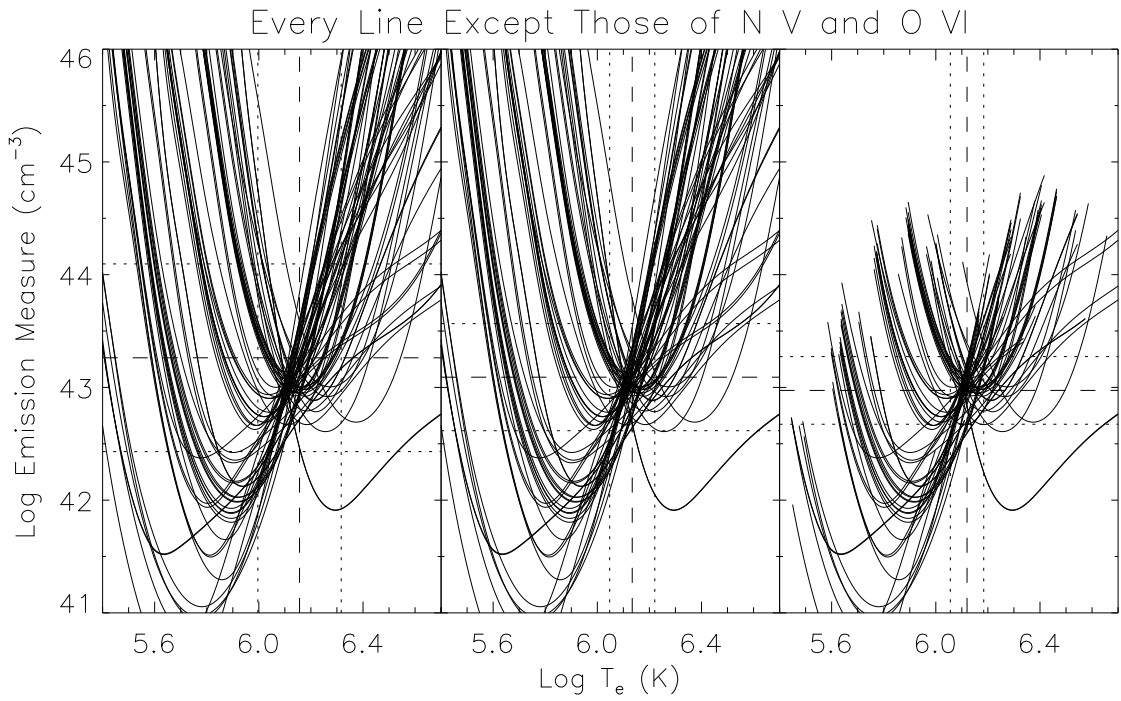}
  \caption[]{Same as Fig.~\ref{fig:1a} but for all emission lines in the observation except
  those from N~{\sc v} and O~{\sc vi}.  Due to the large number of crossings we exclude the asterisks
  for clarity.}
  \label{fig:all}
\end{figure}

\begin{figure}
  \includegraphics{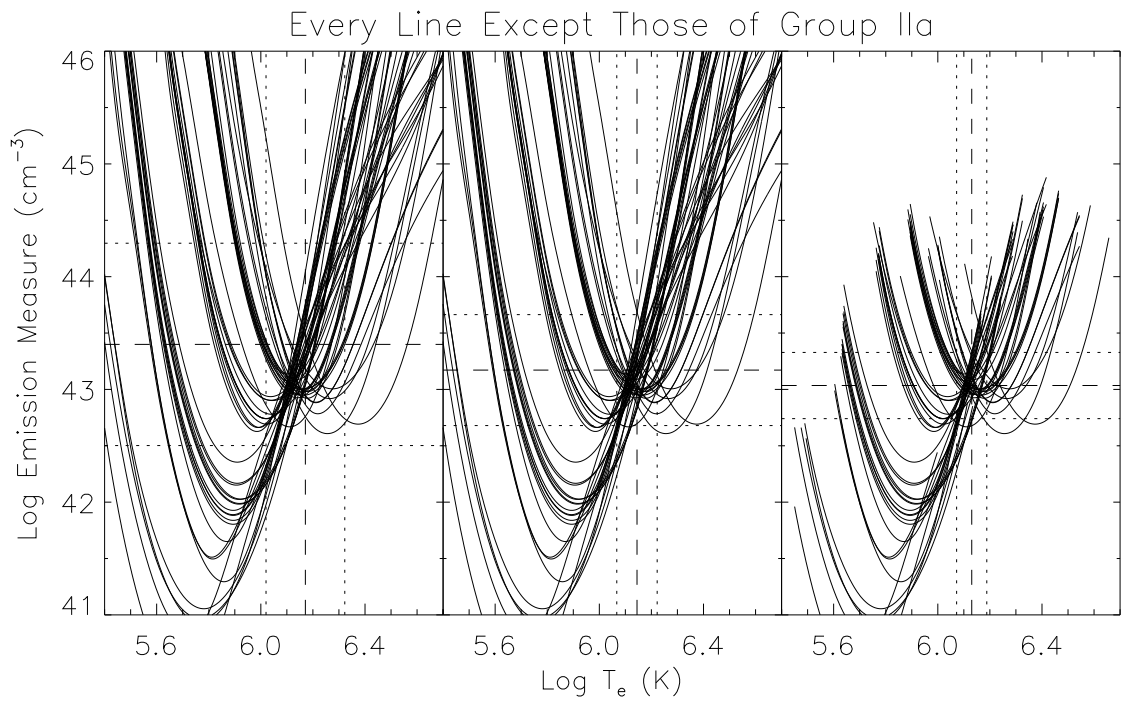}
  \caption[]{Same as Fig.~\ref{fig:all} but excluding emission lines 
  from all Li- and Na-like ions.}
  \label{fig:allx}
\end{figure}

In addition to the division by groups we calculate the mean $EM$ and $T_e$
from every emission line (but again excluding the Li-like N~{\sc v} and O~{\sc vi} lines).
This is done both  including and excluding Li- and Na-like ions with
the results shown in Figs.~\ref{fig:all} and \ref{fig:allx},
respectively, and listed in Table~\ref{tab:averages}.
It should be noted that this is not simply the sum of all the crossings from
the individual groups.  It also includes crossings between lines from different
groups and results in a total of 1428 and 872 crossings (including and
excluding Li- and Na-like lines, respectively).

\begin{figure}
  \includegraphics{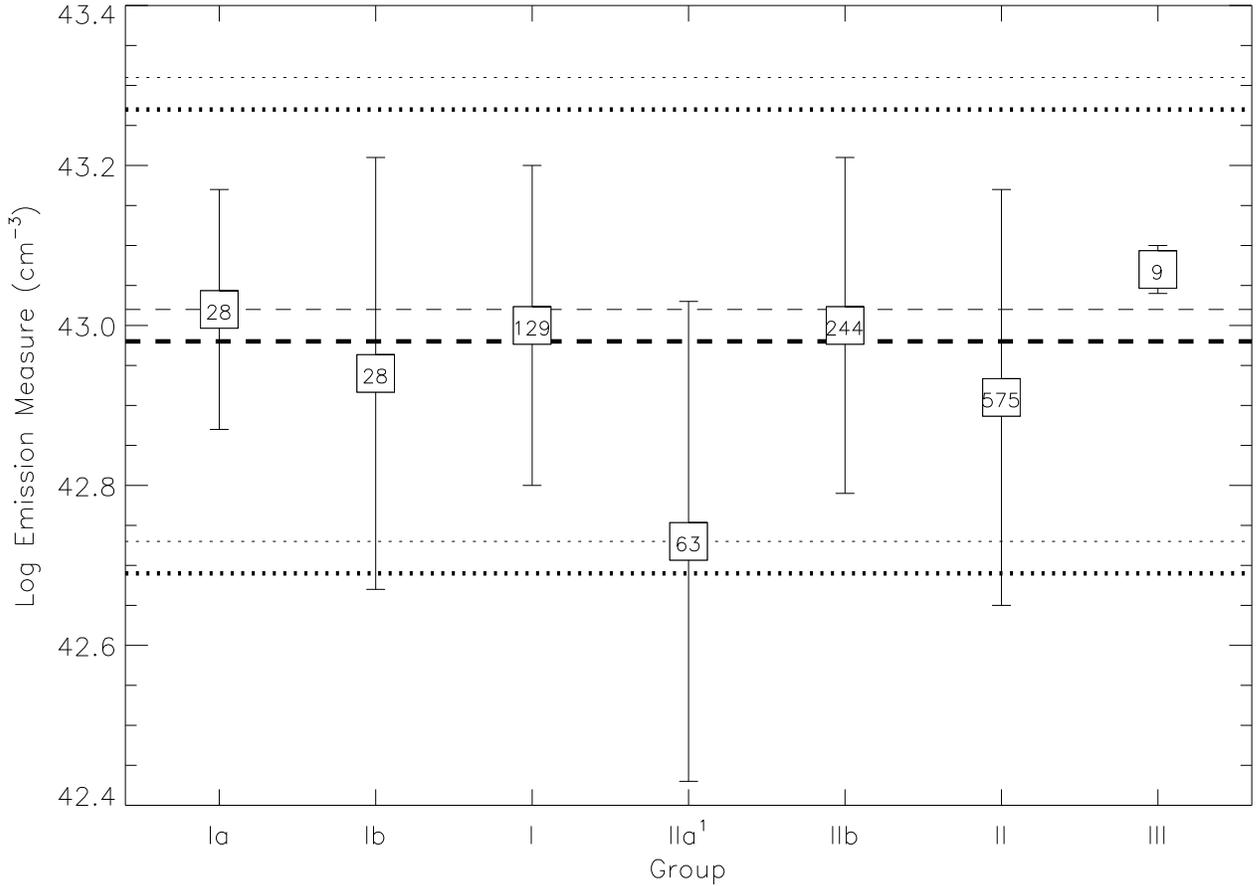}
  \caption[]{Mean $\log_{10}EM$ values for each of the groups using the GEM method (as listed in
  Table~\ref{tab:averages}).  The numbers in the data points represent the number of
  $EM$ curve crossings that were used to derive the mean $EM$.  The error bars on the points
  are $\pm \delta\langle\log_{10}EM\rangle$.  Group~IIa$^1$ excludes the  O~{\sc vi} and 
  N~{\sc v} lines.
  The dashed and dotted lines indicate the mean and standard deviation, respectively,
  when every emission line, except N~{\sc v} and O~{\sc vi}, is considered.
  The thick lines include emission lines from Li- and Na-like ions
  (Fig.~\ref{fig:all}) and the thin lines
  exclude emission lines from these ions (Fig.~\ref{fig:allx}). }
  \label{fig:em figs}
\end{figure}

\begin{figure}
  \includegraphics{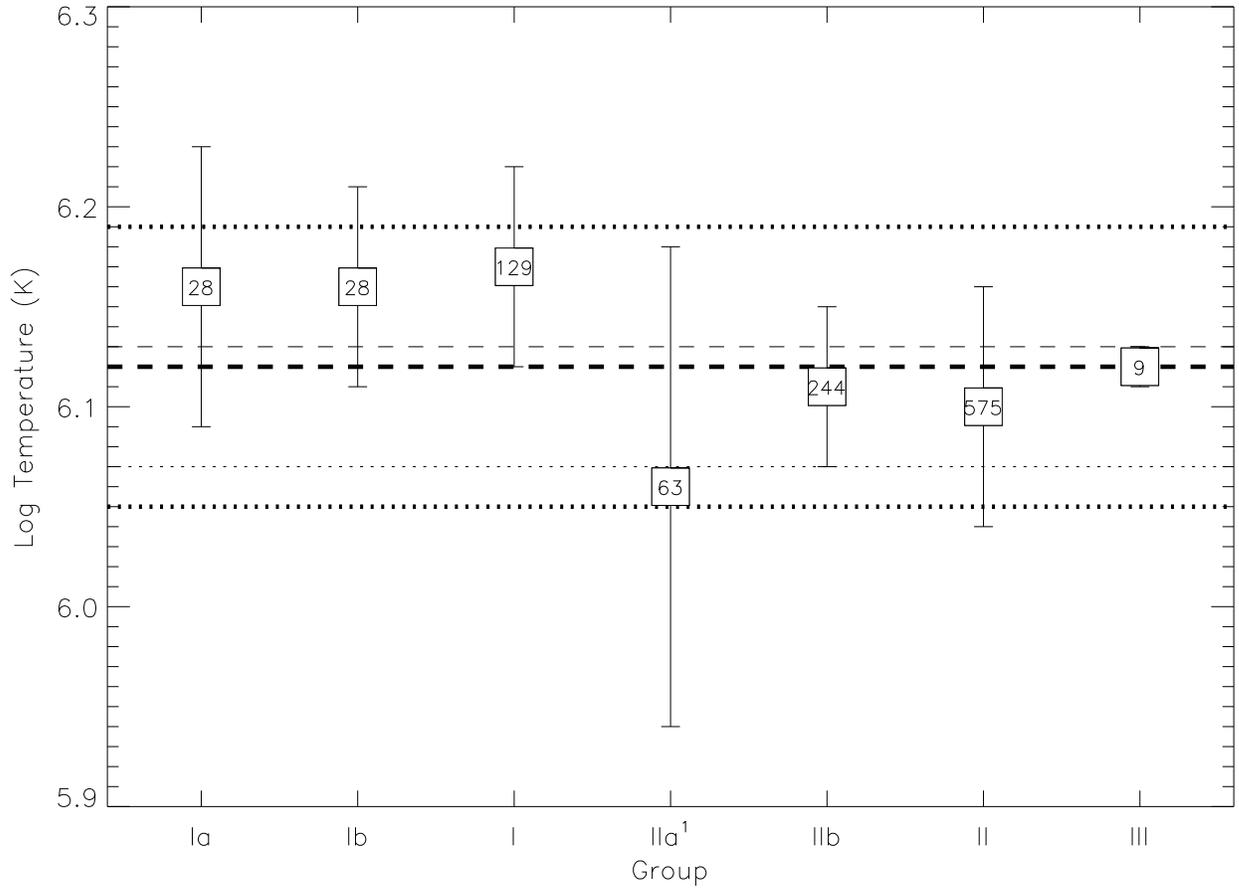}
  \caption[]{Same as Fig.~\ref{fig:em figs} but for $\log_{10}T_e$.
  The upper values of the standard deviation for the thick and thin lines
  lie on top of one another.}
  \label{fig:temp figs}
\end{figure}

The results of our analysis, as given in Table~\ref{tab:averages}, are shown 
in graphical
form in Fig.~\ref{fig:em figs} for the variation of $\log_{10}EM$ versus group and 
Fig.~\ref{fig:temp figs} for the
variation of $\log_{10}T_e$ with group. The numbers in the data points in these 
figures are the number of
crossings that were used to determine the average value and the errors 
shown are $\pm \delta\langle\log_{10}EM\rangle$  of the
mean.  The average and standard deviation of $\log_{10}EM$ and $\log_{10}T_e$ 
as determined from every
emission line are shown for comparison as dashed and dotted lines, respectively.
We show these values with and without Li- and Na-like ions included
in the $EM$ calculation (i.e., Figs.~\ref{fig:all} and \ref{fig:allx}, respectively).
The thick lines  are the average and standard deviations
when Li- and Na-like ions are included (excluding N~{\sc v} and O~{\sc vi})
and the thin are when they are
excluded.

\section{Discussion}
\label{sec:discuss}

\subsection{Updated CIE fractional abundances}
\label{sec:discuss cie}

One of the aims of the present paper is to investigate the effect of our new CIE fractional
abundances on the $EM$ analysis.  But first we look at how the updated
recombination and ionization data impact the fractional abundances themselves.
Perhaps the most widely used recommended CIE fractional abundances are
those of \citet{Mazz98a}.
Comparison of the current CIE fractional abundances with these
are shown in Figs.~\ref{fig:H Mazz}--\ref{fig:Zn Mazz}.
We also compare with the  recently recommended CIE fractional
abundances of \citet{Brya06a} in Figs.~\ref{fig:H Bryans}--\ref{fig:Zn Bryans}.
A comparison of the works of \citet{Mazz98a} and \citet{Brya06a}
was discussed in \citet{Brya06a},
showing the effects of the new DR and RR data  on the
\citet{Mazz98a} results.

We compare the current CIE results with those of
\citet{Mazz98a} for temperatures where 
$f^m \ge 0.01$.  As discussed in
\citet{Brya06a} and in Sec.~\ref{sec:method}, the reliability of the atomic data is
uncertain below this abundance.
Differences between the current CIE results and those of
\citet{Mazz98a} are large for all elements other than H, He, and Li.
Factors of typically at least 2 difference in abundance are found for at
least one ionization stage of each of these elements.
Differences are often much larger.
We draw particular attention to the extremely large differences in
abundance and peak formation temperature of Sc, Ti, V, Cr, Mn,
Co, Ni, Cu, and Zn in the $T_e$ range of $10^4$--$10^6$~K.
Differences for these elements can be up to a factor of 30.
Such variation between the current results and those of \citet{Mazz98a}
is a result of the new recombination and ionization
rate coefficients being used here.

We also compare our present results
with the more recent recommended CIE fractional abundances of
\citet{Brya06a}.  The DR and RR rate coefficients used in this work
are largely the same as those used by Bryans et al.\ with the exception
of the Mg-like ions of H through Zn, the Al- through Ar-like ions of Fe,
and some corrections to the fitting of other ions.
The most significant changes in atomic data bewteen this work and
\citet{Brya06a} is the introduction of the \citet{Dere07a} EII
rate coefficients.
As expected, differences between the present results and those of
\citet{Brya06a} are not as large as those found between the present
results and those of \citet{Mazz98a}.
However, large differences do remain.
The differences highlighted above for Sc, Ti, V, Cr, Mn,
Co, Ni, Cu, and Zn in the $T_e$ range of $10^4$--$10^6$~K
are also present in the comparison with Bryans et al. 
For other elements, abundance differences of a factor a few are not uncommon.
 We 
attribute all these differences primarily to the EII rate coefficients.

In Secs.~\ref{sec:discuss fip obs} and \ref{sec:discuss groups}
we discuss the impact of these updated CIE calculations on the
analysis of the present SUMER observation.
However, only a selection of the ions discussed in this section are present
in the SUMER observation.
We recommend that the CIE fractional abundances provided
here be used in all future analysis of astrophysical spectra
until the next revision of the CIE fractional abundances is published.

\subsection{Comparison with FIP factor observations}
\label{sec:discuss fip obs}
For this same SUMER observation, FIP factors were also determined 
by \citet{Feld98a}.
The present results are shown in comparison to those of
Feldman et al.\ in Fig.~\ref{fig:fip}.
They recommend a FIP factor of 1 for the high-FIP elements,
a factor of 4 for the low-FIP elements, and a factor of somewhere  between
1 and 2 for S.
These results are the basis of the approach taken by \citet{Land02a}
who assumed a FIP factor of unity for the moderate- and high-FIP elements,
S, O, N, Ar, and Ne, and a uniform factor of 3.5 for all of the low-FIP elements,
K, Na, Al, Ca, Mg, Fe, and Si.

We believe our present results are more robust than those of \citet{Feld98a}.
Firstly, our reference $EM$ value is taken from the crossing
of 2 Ar $EM$ curves whereas \citet{Feld98a}
use the emission from a single Li-like O~{\sc vi} line as their reference value.
This O~{\sc vi} line had an order of magnitude more counts than any other line in
the dataset used by Feldman et al.\ and thus seems a natural reference emission
line.  However, given the apparent systematic abundance discrepancy
of Li-like ions (which the authors acknowledge),
the O~{\sc vi} line may not be the most reliable to use
as an $EM$ reference value.

Furthermore, in determining the FIP factor for each element we generally use
more emission lines than \citet{Feld98a}.  
For most elements we have multiple emission lines, ranging from
3 lines for Fe to as many as 18  for Si.
The only exceptions are the elements K, Na, and Ca as have already
been discussed in Sec.~\ref{sec:fip}.
For K we only have 1 emission line and for Na and Ca we have 2.
\citet{Feld98a}, however, use only 1 or 2 emission
lines to determine the FIP factors for each of the elements they consider.

An additional source of unreliability in the \citet{Feld98a} results lies in
the method they use to
estimate the plasma temperature.
They use the crossing points of curves of
FIP factors vs.\ $T_e$
from different elements.  They estimate $\log_{10}T_e=6.13$ (the same
value at which we ultimately arrive) but only calculate these FIP factor
vs.\ $T_e$ curves on a temperature grid of 0.1 in the log.  From their figures
it is reasonable to conclude that any value in the range
of $\log_{10}T_e=6.1$ to 6.2 would fit their data points.
In which case, their reported FIP factors could range from $\sim 1.5$ to 11.
However, \citet{Feld00a} estimate the error in these FIP factors to
be of the order of 25\% which seems to be a significant underestimate.

Of the low-FIP elements, we find rough agreement between our results and those
of Feldman et al.\ in the sense that the abundance of the
low-FIP elements is enhanced over the high-FIP elements,
though one should note that Feldman et al.\ did not
ascribe errors to their results.
The largest differences between our results and those of
Feldman et al.\ occur for Na and Ca, where we find differences
of a factor of 2.5 and 1.5 respectively.
However, our results for these elements should be considered 
with some care since they are not determined from an average
of crossing points but from the $EM$ at a given $T_e$.
The error bars on our results for Na and Ca are also relatively large and 
the Feldman et al.\ results lie within these errors.
Our result for K (a FIP factor of 1.75) is in disagreement with
the Feldman et al.\ conclusion that the low-FIP elements are best
fitted with an enhancement factor of 4.  However, it should be noted that
Feldman et al.\ did not calculate the FIP factor for K itself
and that our analysis uses only one line of K.

Finally, we compare the FIP factor results of our GEM method
when we utilize the CIE fractional abundances of the present paper
and those of \citet{Mazz98a}.  These results are given in Table~\ref{tab:abundances}.
We find that the effect of our new CIE fractional abundances is
largest for K, Ca, and Fe.  In the case of K and Fe the differences in
FIP factor are not within the estimated errors using our new CIE results.
Naturally, these differences are also seen in the log of the inferred coronal
abundances.

\subsection{Comparison with FIP factor model}
\label{sec:discuss fip model}

The FIP effect model of \citet{Lami04a,Lami08a} allows an opportunity to 
quantitatively compare our
derived coronal elemental abundances with those of theory.
The Laming model builds on that of \citet{Schw99a} by explaining
the FIP effect in terms of Alfv\'{e}n waves in the chromosphere.
These Alfv\'{e}n waves drive a pondermotive force on their reflection
or transmission at the chromosphere-corona boundary which results
in the elemental fractionation.

\begin{figure}
  \includegraphics{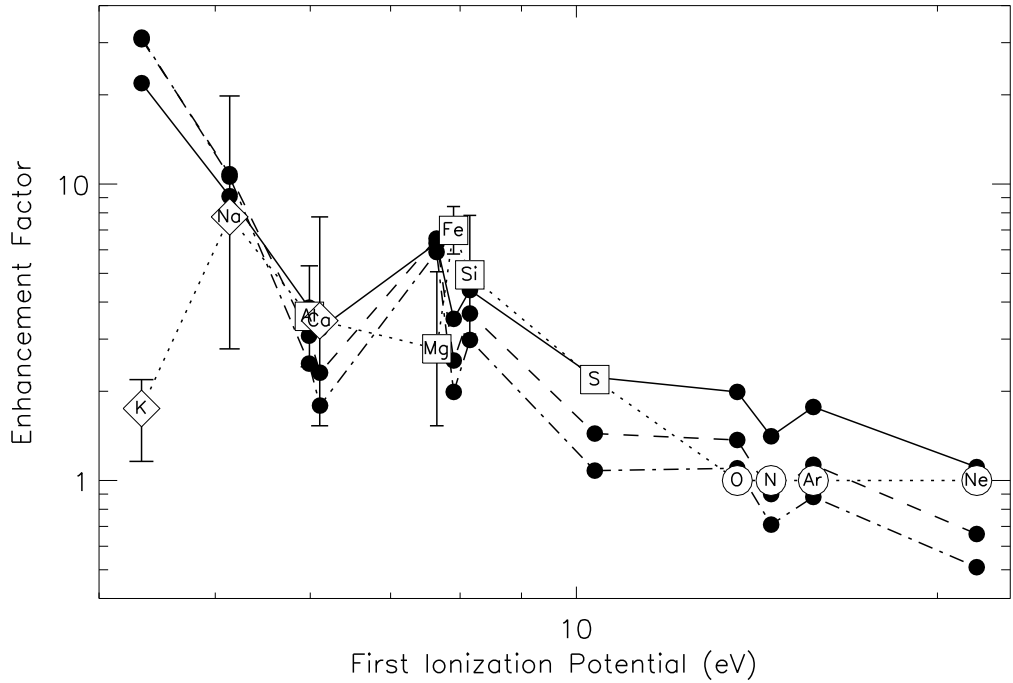}
  \caption[]{Coronal abundance enhancement factor (i.e., FIP factor) used for each of the elements
             versus their first ionization potential.  Open symbols
	     represent the present results; refer to Fig.~\ref{fig:fip}
	     for details.
	     The solid circles are the results of the model of \citet{Lami08a} for
	     for upward Alfv\'{e}n wave energy fluxes of 2, 8, and
	     32 ({\it solid}, {\it dashed}, and
	     {\it dot-dashed lines}, respectively) 
	     in units of $10^6$~ergs~cm$^{-2}$~s$^{-1}$.
	     Lines have been drawn between points only to guide the eye.}
  \label{fig:fip model}
\end{figure}

The extent of the FIP effect on each species is dependent on the upward energy
flux of the Alfv\'{e}n waves.  \citet{Lami08a} gives results for a
number of wave energy fluxes and we compare these results with ours
for wave energy fluxes of 2, 8 and 32 in units of $10^6$~ergs~cm$^{-2}$~s$^{-1}$.
We show these comparisons in Fig.~\ref{fig:fip model}.
Our results suggest that upward wave
energy fluxes in this range best describe the solar
conditions at the time of this particular SUMER observation.
Our data generally fit the model well, with the exception of K.
However, as has already been discussed in
Sec.~\ref{sec:fip}, our result for K should be considered less reliable than
the other elements since we were limited to only a single K emission line
in the SUMER observation.  

It should also be noted that the low-FIP results of the present work were
calculated relative  to a high-FIP enhancement of 1, while in the
Laming model the high-FIP elements do show a slight abundance
variation dependent on their FIP value.
If we were to normalize to the Ar FIP factor of the \citet{Lami08a} model
this would introduce a shift in our FIP factors of somewhere between a
factor of 0.88 to a factor of 1.77.

\subsection{Groups}
\label{sec:discuss groups}

We have used the same group splitting as that used by \citet{Land02a} and thus can
compare directly with their results.
Table~\ref{tab:Landi} shows their results for
the mean and ``error'' of $\log_{10}EM$ and $\log_{10}T_e$  for the various groups.
However, unlike the present work, Landi et al.\ quotes
the mean and the error as judged by eye as opposed to our GEM method.
As our results demonstrate, they have considerably underestimated
the uncertainty of their results.

The results of \citet{Land02a} suggest a difference in the temperature derived from the
subsets of Group~I, with $\log_{10} T_e=6.13\pm 0.01$ and $6.17\pm 0.01$ for Groups~Ia and Ib,
respectively. We do not see this difference in our analysis. In the present work,
Groups~Ia and Ib give $\log_{10} T_e=6.16\pm 0.07$ and $6.16\pm 0.05$, respectively.  Within our
error bars, we see no distinction between the N-like and non--N-like ions in this group.
Using the same uniform low-FIP factor of 3.5 used by \citet{Land02a}
and the GEM method,
the distinction remains unobserved as we find values of
$\log_{10} T_e=6.16\pm 0.05$ and $6.17\pm 0.03$ for Groups~Ia and Ib,
respectively.
The temperatures derived from Groups~IIa\footnote[1]{Excluding the Li-like 
N~{\sc v} and O~{\sc vi} lines from Group~IIa}, 
IIb, and III agree reasonably well with those of
\citet{Land02a}. We note that Landi et al.\ excluded the N~{\sc v} and O~{\sc vi} lines from
their calculation of the Group~IIa lines. When comparing with their results we also exclude
these lines.

\begin{figure}
  \includegraphics{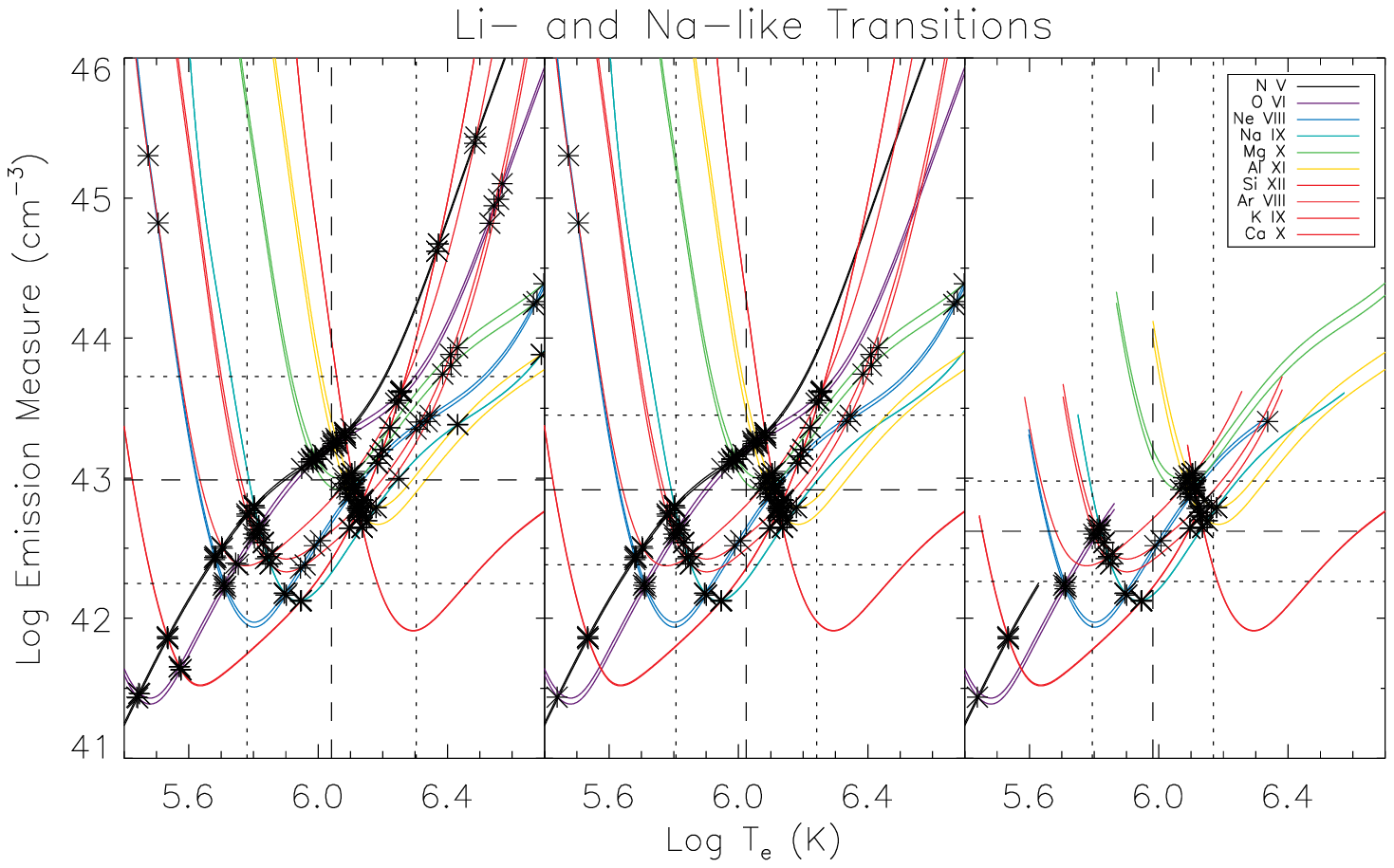}
  \caption[]{Same as Fig.~\ref{fig:1a} but for Group~IIa.}
  \label{fig:2a}
\end{figure}

\begin{figure}
  \includegraphics{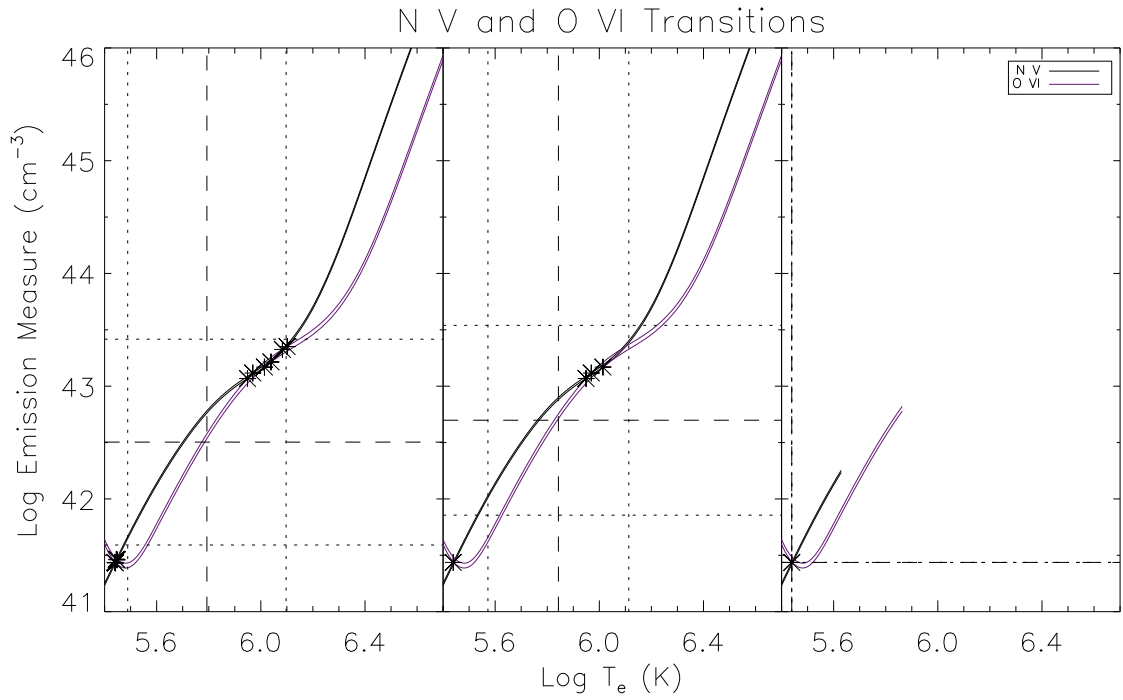}
  \caption[]{Same as Fig.~\ref{fig:2a} but showing only emission lines from N~{\sc v}
  and O~{\sc vi}.}
  \label{fig:2a_on}
\end{figure}

Figure~\ref{fig:2a} shows the $EM$ curves for the lines in Group~IIa.
The largest
discrepancies from the other lines in this group
can be seen to come from the 2 pairs of N~{\sc v}
and O~{\sc vi} lines.  It is interesting to note
that these lines are the lowest in temperature of peak formation 
of all the ions considered here (see Figs.~\ref{fig:H Bryans}--\ref{fig:Zn Bryans}
and Tables~\ref{tab:H}--\ref{tab:Zn}) and as a result
the majority of the crossings from these lines are excluded in the right panel
of Fig.~\ref{fig:2a} when we ignore fractional abundances below 0.01. This
perhaps goes some way to highlight the need for care when using fractional
abundances of such low values.

\begin{figure}
  \includegraphics{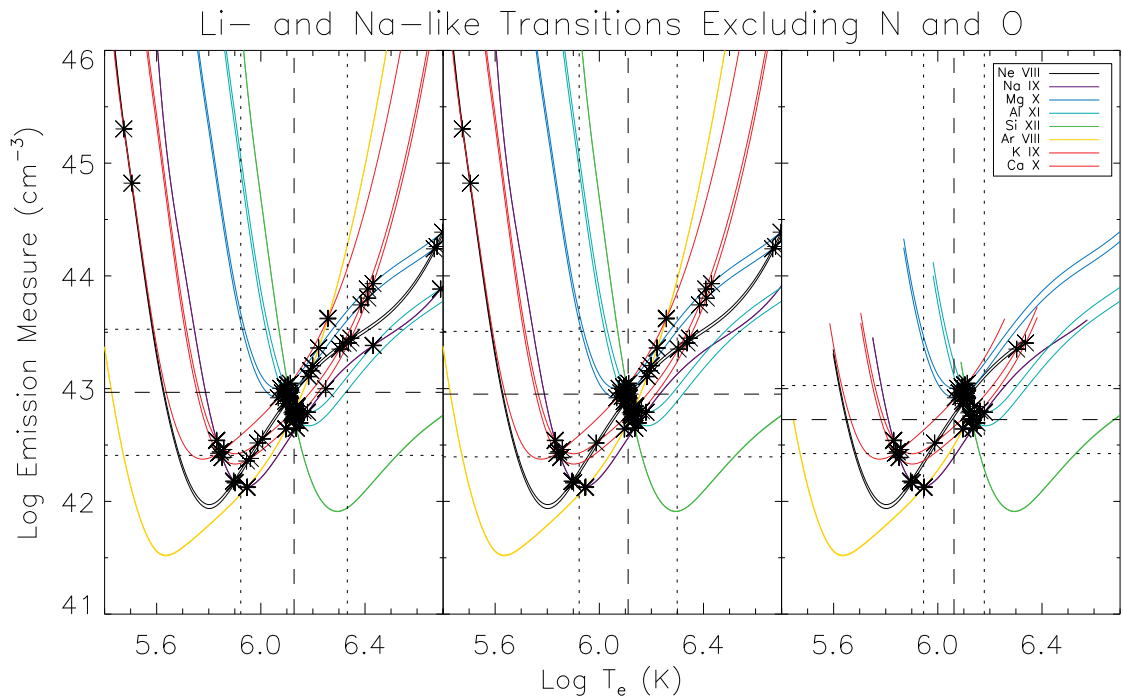}
  \caption[]{Same as Fig.~\ref{fig:2a} but excluding emission lines from N~{\sc v}
  and O~{\sc vi}.}
  \label{fig:2ax}
\end{figure}

Given the disagreement with the other lines in Group~IIa, and the lower formation 
temperature of N~{\sc v}
and O~{\sc vi} compared to the other ions in the group, it is possible
that the emission lines from these two ions originate from a different region of plasma.
Thus, we have excluded the O~{\sc vi}
and N~{\sc v} lines and recalculated the $EM$ curves.
Figure~\ref{fig:2ax} shows this reduced set of $EM$ curves.
We have also done the same for the O and N lines on their own in Fig.~\ref{fig:2a_on}.
A much lower average temperature of $\log_{10} T_e=5.44$ is derived from 
these curves.  We do not give an estimated error on this value since
the final result comes from a single crossing point and has no
standard deviation.
Given that the N~{\sc v} and O~{\sc vi} ions have
lower formation temperatures than the other ions of this observation, this
suggests that the source of emission from these ions is from a different region of plasma
with a lower temperature than that emitting the lines from other elements.

One of the questions that this paper seeks to address is the
apparent discrepancy between the $EM$s derived from Li- and Na-like and
that of all other ions.  This has been identified previously 
\citep[e.g.,][]{Dupr72a,Feld98a}.
All Li- and Na-like lines in this observation come from transitions between
the ground and first excited configuration, i.e., our Group~II.
So we first investigate the difference between Li- and Na-like ions and
all other ions within this group.
Our results for the Li- and Na-like ions are shown in
Fig.~\ref{fig:2ax} (excluding N~{\sc v} and O~{\sc vi})
 and can be compared to the $EM$
from the other Group~II ions shown in Fig.~\ref{fig:2b}.
Results are also given in Table~\ref{tab:averages}.
A comparison of our present results and those of \citet{Land02a}
can be seen in Table~\ref{tab:Landi}.
The combination of using the most up-to-date atomic data, an improved method
of arriving at the most likely $EM$ and $T_e$, and a reanalysis of the FIP factors,
has led to us finding no sign of any
discrepancy between the emission of Li- and Na-like
ions and all the other Group~II ions.
In the present case, the difference between the $EM$ from
Groups~IIa\footnotemark[1] (excluding 
N~{\sc v} and O~{\sc vi} lines) and IIb is within
 the error bars on the $EM$ of Group~IIa\footnotemark[1].

In addition to the comparison of $EM$ within Group~II, we also
compare the $EM$ from the Li- and Na-like ions (Group~IIa$^1$)
with the $EM$ derived
from every other ion in the observation (i.e., those from
Groups~I, IIb, and III).
The comparison between the $EM$s from the
Li- and Na-like ions and all other ions is shown
 in Fig.~\ref{fig:em figs}.  One can see that
 the $EM$ from Li- and Na-like lines alone
(point IIa$^1$) overlaps, within the errors, with the average determined
excluding Li- and Na-like ions (thin dashed line).
Thus, we find no statistically meaningful difference in the $EM$
derived from Li- and Na-like ions and that
from every other ion in the observation.

Given that we find fairly good agreement 
in $EM$ and $T_e$ between each of the group categories, our
best estimate of the $EM$ and $T_e$ of the emitting plasma is found by applying
our analysis method  to every emission line
(excluding the discrepant N~{\sc v} and O~{\sc vi} lines).
These results are shown in Fig.~\ref{fig:all} and give $\log_{10} EM=42.98\pm 0.29$
and $\log_{10} T_e=6.12\pm 0.07$.
If, in addition to excluding the N~{\sc v} and O~{\sc vi} lines, we also
exclude the Li- and Na-like lines then the calculated values become
$\log_{10} EM=43.02\pm 0.29$ and $\log_{10} T_e=6.13\pm 0.06$ (see
Fig.~\ref{fig:allx}).
Landi et al.\ estimate $\log_{10} EM=43.20\pm 0.15$
and $\log_{10} T_e=6.13$ (no error given) for the plasma by combining results from 
Groups~I and IIb.
Our results agree, within our errors, with those of \citet{Land02a}.
Our results have larger errors, which we believe to be
more realistic due to our more rigorous method of calculating
the mean and standard deviation of $EM$ and $T_e$.

To investigate the effects of the updated CIE data on these results,
we compare the $EM$ and $T_e$ derived for each group when utilizing
the recommended CIE fractional abundances of the present paper
and those of \citet{Mazz98a}.  These results are given in Table~\ref{tab:Landi}.
While differences are found, they are all within the errors.
It is interesting to note that the large differences found in the
FIP factors do not translate into differences on the same scale
for the derived $EM$ and $T_e$.  Nonetheless, this would not necessarily
be the case when applied to other observations, so we recommend the future
use of the CIE fractional abundances presented here.

\subsection{Other Issues}

Despite an overall general agreement in the $EM$ and $T_e$ of each of the groups,
there are a number of indications that the observed emission does not come
from an isothermal plasma.
We have already discussed the possibility that the Li-like
N~{\sc v} and O~{\sc vi} lines come from a cooler region of gas.
Even when these lines are removed from the Group~IIa categorization, there
remains a large scatter in the crossing points of the emission from
 Li- and Na-like ions (Fig.~\ref{fig:2ax}) suggesting the isothermal assumption is not
 entirely accurate.
There is also some evidence of a low temperature component from Groups~I and 
IIb (Figs.~\ref{fig:1} and \ref{fig:2b}, respectively).
It is also possible that the relatively large errors in $EM$ and $T_e$
are suggestive of a non-isothermal plasma.
To determine whether the crossings that fall away from the average are indeed a
product of the non-isothermal nature of the plasma, and not some error in the atomic data,
one would have to perform a differential emission measure (DEM) analysis, which is
beyond the scope of this paper.

A further possible source of error in our analysis is that the ionization
balance was calculated using the zero-density approximation.
This issue has been raised by \citet{Feld00a} in reference to Fe$^{8+}$
emission.  They claim that over half of the population can be
in metastable levels at coronal densities,
but there are no emission lines from Fe$^{8+}$ in the observation
analyzed in the present paper.
The sensitivity of emission from Li-like ions has been investigated
by \citet{Doyl05a}.  These authors found that  
the contribution function of emission from
Li-like lines only becomes significantly affected on reaching densities 
$\ge 10^{11}$~cm$^{-3}$, orders of magnitude higher than the density
of $1.8\times 10^8$~cm$^{-3}$ inferred by \citet{Feld99a} for
the observation analyzed here.
We thus expect the zero-density approximation to be valid
in the present case, but a full density-dependent analysis 
of every emission line in the observation would be required
before one could answer this issue with complete certainty.
Again, such a study is beyond the scope of this paper.

\section{Proposals for future observations}
\label{sec:future}

Our work shows that SUMER observations can go a long way towards constraining FIP
models such as those of \citet{Lami04a,Lami08a}.  Even better constraints can be achieved
through the simultaneous observation of lines from a number of additional charge
states. More lines from high-FIP elements such as N, O, Ne, and Ar are required
to better determine the $EM$ for these high-FIP elements, which can
then be used to normalize the
low-FIP elements.  For N, O, and Ne, emission lines from H- and
He-like stages need to be observed to avoid using Li-like ions.
These charge states are predicted to be abundant for these elements
at coronal temperatures.
This may require simultaneous observations using separate, 
cross-calibrated spectrometers.
For Ar, emission lines from the Ne-, F-, O-, and N-like ions would lie
in the $6.0\le\log T_e\le 6.2$ range typical of the corona.
Additional line
 observations from elements such as
Na, K, and Ca, for which we have few lines in the present observation,
are also needed to better constrain their FIP factors.

\section{Summary}
\label{sec:conclusion}

This work has reanalyzed data from a SUMER coronal observation in an attempt to
improve upon previous methods of analysis.
We have given a brief review of and implemented 
state-of-the-art electron-ion recombination and ionization data.
We have updated the CIE results of \citet{Brya06a}
by using recently published DR data for Mg-like ions of the elements 
from H through Zn and for
Al- through Ar-like Fe ions,
and have updated the EII data to those of \citet{Dere07a} for all ions
of H through Zn.
We have also set out a new, mathematically rigorous, approach for
determining the $EM$ and $T_e$ of an emitting
plasma within the isothermal approximation.
Using these new CIE data and our approach for determining
the $EM$, we calculated the FIP factors of the observed elements.

Our assessment is generally in reasonable agreement with a previous study
of the FIP factors \citep{Feld98a}.
Also, we are in reasonable agreement with the FIP-effect model
of \citet{Lami08a} using an Alfv\'{e}n wave energy flux in the range
$\sim 2$--$32\times10^6$~erg~cm$^{-2}$~s$^{-1}$.
However, our results differ from those of \citet{Land02a} in certain respects. 
The difference between the temperature derived using lines from non--N- and N-like
ions is not evident when we apply out analysis technique.
Also, the previously reported
discrepancy between the $EM$ derived from Li- and Na-like lines and
the $EM$ from all other lines (Groups~I, IIb, and III)
is not  supported by our results, rather the two agree at
 the $1\sigma$ level.

Our best estimate of the $EM$ and $T_e$ of the emitting plasma of this observation is $\log_{10}
EM=42.98\pm 0.29$ and $\log_{10} T_e=6.12\pm 0.07$ when we include all lines
except those from N~{\sc v} and O~{\sc vi}, and $\log_{10}
EM=43.02\pm 0.29$ and $\log_{10} T_e=6.13\pm 0.06$ when we additionally exclude all
Li- and Na-like lines.  There remains variation in the crossing
points of the $EM$ vs.\ $T_e$ curves that are suggestive of errors in the atomic data,
the observations, or the
solar physics model used.  However, 
from the results of the present work it is not 
possible to say where the source of these
errors lie.  Further improvements to the atomic
database and new insight into the physical conditions of the upper solar atmosphere are needed
before these questions can be answered.
Given the evidence for regions of differing temperature,
a DEM analysis might go some way to
resolving the discrepancies found in the present paper.

\acknowledgments 
We thank H.\ Bruhns, H.\ Kreckel, J.\ M.\ Laming, and M.\ Lestinsky
for stimulating discussions. 
We also thank K.\ P.\ Dere for providing corrected versions of his
EII rate coefficients and for discussions thereon.
CHIANTI is a collaborative project involving the NRL
(USA), RAL (UK), MSSL (UK), the Universities of Florence (Italy) and Cambridge (UK),
and George Mason University (USA). P.B.\ and D.W.S.\ were supported in part by the
NASA Solar and  Heliospheric Physics Supporting Research and Technology program and
the NASA Astronomy and Physics Research and Analysis Program.
E.L.\ was supported by NASA grants NNG06EA14I and NNH06CD24 as well as other NASA grants.

\begin{deluxetable}{cccccc}
\tablewidth{0pt}
\tablecaption{Emission lines and intensities used in the present study
\label{tab:linelist}}
\tablehead{\colhead{Group} & \colhead{Ion} & \colhead{Sequence} & \colhead{Wavelength} & 
\colhead{Transition} & \colhead{Intensity} \\
\colhead{} & \colhead{} & \colhead{} & \colhead{(\AA)} & 
\colhead{} & \colhead{(ergs~cm$^{-2}$~s$^{-1}$~sr$^{-1}$)}}
\startdata
IIa & N {\sc v}	 &Li & 1238.82 		& $2s\:^2S_{1/2}-2p\:^2P_{3/2}$			&2.520     \\
IIa & N {\sc v}	 &Li & 1242.80 		& $2s\:^2S_{1/2}-2p\:^2P_{1/2}$			&1.420     \\
IIa & O {\sc vi} 	 &Li & 1031.91 	& $2s\:^2S_{1/2}-2p\:^2P_{3/2}$			&63.000    \\
IIa & O {\sc vi} 	 &Li & 1037.62 	& $2s\:^2S_{1/2}-2p\:^2P_{1/2}$			&28.500    \\
IIa & Ne {\sc viii} 	 &Li & 770.41  	& $2s\:^2S_{1/2}-2p\:^2P_{3/2}$			&30.700    \\
IIa & Ne {\sc viii} 	 &Li & 780.32  	& $2s\:^2S_{1/2}-2p\:^2P_{1/2}$ 		       &14.200    \\
IIb & Ne {\sc vii} 	 &Be & 895.17  	& $2s^2\:^1S_{0}-2s2p\:^3P_{1}$ 		       &0.132	  \\
III &  Ne {\sc vii} 	 &Be & 973.33  	& $2s2p\:^1P_{1}-2p^2\:^1D_{2}$ 		       &0.070	  \\
\enddata
\tablecomments{We list here the all the emission lines of the SUMER 1996 November 21 21:16~UT
to 1996 November 22 02:28~UT observation that are used in the analysis here.
This table is reproduced from \citet{Land02a} with known transition 
assignment errors corrected.
Lines that are blended or have uncertain intensities have also been omitted.}
\end{deluxetable}
\clearpage

\begin{deluxetable}{ccc}
\tabletypesize{\tiny}
\tablecaption{Hydrogen CIE Fractional Abundances
\label{tab:H}}
\tablewidth{0pt}
\tablehead{
\colhead{log(T)} 
& \colhead{H{$^{0+}$}}
& \colhead{H{$^{1+}$}}}
\startdata
   4.00 &    0.001 &    2.751  \\
   4.10 &    0.026 &    1.240  \\
   4.20 &    0.318 &    0.285  \\
   4.30 &    1.089 &    0.037  \\
   4.40 &    1.895 &    0.006  \\
   4.50 &    2.584 &    0.001  \\
   4.60 &    3.162 &    0.000  \\
   4.70 &    3.649 &    0.000  \\
\enddata
\tablecomments{Calculated $-\log_{10}$ of the fractional abundance for all
ionization stages of hydrogen. We use the RR 
rate coefficients of \protect\citet{Badn06c} and
the EII rate coefficients of \protect\citet{Dere07a}.}
\end{deluxetable}

\clearpage

\begin{deluxetable}{cccc}
\tabletypesize{\tiny}
\tablecaption{Helium CIE Fractional Abundances
\label{tab:He}}
\tablewidth{0pt}
\tablehead{
\colhead{log(T)} 
& \colhead{He{$^{0+}$}}
& \colhead{He{$^{1+}$}}
& \colhead{He{$^{2+}$}}}
\startdata
   4.00 &    0.000 &    8.813 &   15.000  \\
   4.10 &    0.000 &    6.135 &   15.000  \\
   4.20 &    0.000 &    3.980 &   15.000  \\
   4.30 &    0.003 &    2.240 &   13.153  \\
   4.40 &    0.061 &    0.885 &    8.981  \\
   4.50 &    0.496 &    0.167 &    5.896  \\
   4.60 &    1.295 &    0.023 &    3.845  \\
   4.70 &    2.043 &    0.006 &    2.288  \\
\enddata
\tablecomments{Same as Table~\protect\ref{tab:H} but for helium and
additionally using the DR rate 
coefficients of \protect\citet{Badn06b}.
Fractional abundances are cut off at 
$10^{-15}$. For ease of machine readability, values less than
$10^{-15}$ are given $-\log_{10}$ values of 15.}
\end{deluxetable}

\clearpage

\begin{deluxetable}{ccccc}
\tabletypesize{\tiny}
\tablecaption{Lithium CIE Fractional Abundances
\label{tab:Li}}
\tablewidth{0pt}
\tablehead{
\colhead{log(T)} 
& \colhead{Li{$^{0+}$}}
& \colhead{Li{$^{1+}$}}
& \colhead{Li{$^{2+}$}}
& \colhead{Li{$^{3+}$}}}
\startdata
   4.00 &    2.408 &    0.002 &   15.000 &   15.000  \\
   4.10 &    3.095 &    0.000 &   15.000 &   15.000  \\
   4.20 &    3.667 &    0.000 &   15.000 &   15.000  \\
   4.30 &    4.147 &    0.000 &   14.841 &   15.000  \\
   4.40 &    4.553 &    0.000 &   12.035 &   15.000  \\
   4.50 &    4.897 &    0.000 &    9.228 &   15.000  \\
   4.60 &    5.192 &    0.000 &    6.625 &   15.000  \\
   4.70 &    5.445 &    0.000 &    4.531 &   14.616  \\
\enddata
\tablecomments{Same as Table~\protect\ref{tab:He} but for lithium.}
\end{deluxetable}

\clearpage

\begin{deluxetable}{cccccc}
\tabletypesize{\tiny}
\tablecaption{Beryllium CIE Fractional Abundances
\label{tab:Be}}
\tablewidth{0pt}
\tablehead{
\colhead{log(T)}
& \colhead{Be{$^{0+}$}} 
& \colhead{Be{$^{1+}$}}
& \colhead{Be{$^{2+}$}}
& \colhead{Be{$^{3+}$}}
& \colhead{Be{$^{4+}$}}}
\startdata
   4.00 &    0.021 &    1.335 &    6.887 &   15.000 &   15.000  \\
   4.10 &    0.114 &    0.637 &    4.177 &   15.000 &   15.000  \\
   4.20 &    0.379 &    0.240 &    2.156 &   15.000 &   15.000  \\
   4.30 &    0.865 &    0.161 &    0.762 &   15.000 &   15.000  \\
   4.40 &    1.797 &    0.603 &    0.134 &   15.000 &   15.000  \\
   4.50 &    2.987 &    1.363 &    0.020 &   15.000 &   15.000  \\
   4.60 &    4.072 &    2.065 &    0.004 &   15.000 &   15.000  \\
   4.70 &    5.007 &    2.656 &    0.001 &   12.940 &   15.000  \\
\enddata
\tablecomments{Same as Table~\protect\ref{tab:He} but for beryllium.}
\end{deluxetable}

\clearpage

\begin{deluxetable}{ccccccc}
\tabletypesize{\tiny}
\tablecaption{Boron CIE Fractional Abundances
\label{tab:B}}
\tablewidth{0pt}
\tablehead{
\colhead{log(T)} 
& \colhead{B{$^{0+}$}}
& \colhead{B{$^{1+}$}}
& \colhead{B{$^{2+}$}}
& \colhead{B{$^{3+}$}}
& \colhead{B{$^{4+}$}}
& \colhead{B{$^{5+}$}}}
\startdata
   4.00 &    0.382 &    0.233 &    9.957 &   15.000 &   15.000 &   15.000  \\
   4.10 &    1.029 &    0.043 &    7.367 &   15.000 &   15.000 &   15.000  \\
   4.20 &    1.516 &    0.013 &    5.445 &   14.843 &   15.000 &   15.000  \\
   4.30 &    1.822 &    0.007 &    3.913 &   10.708 &   15.000 &   15.000  \\
   4.40 &    2.052 &    0.005 &    2.668 &    7.368 &   15.000 &   15.000  \\
   4.50 &    2.265 &    0.012 &    1.653 &    4.664 &   15.000 &   15.000  \\
   4.60 &    2.518 &    0.068 &    0.859 &    2.503 &   15.000 &   15.000  \\
   4.70 &    2.964 &    0.318 &    0.397 &    0.930 &   15.000 &   15.000  \\
\enddata
\tablecomments{Same as Table~\protect\ref{tab:He} but for boron.}
\end{deluxetable}

\clearpage

\begin{deluxetable}{cccccccc}
\tabletypesize{\tiny}
\tablecaption{Carbon CIE Fractional Abundances
\label{tab:C}}
\tablewidth{0pt}
\tablehead{
\colhead{log(T)}
& \colhead{C{$^{0+}$}}
& \colhead{C{$^{1+}$}}
& \colhead{C{$^{2+}$}}
& \colhead{C{$^{3+}$}}
& \colhead{C{$^{4+}$}}
& \colhead{C{$^{5+}$}}
& \colhead{C{$^{6+}$}}}
\startdata
   4.00 &    0.014 &    1.487 &   10.947 &   15.000 &   15.000 &   15.000 &   15.000  \\
   4.10 &    0.203 &    0.428 &    7.258 &   15.000 &   15.000 &   15.000 &   15.000  \\
   4.20 &    0.795 &    0.076 &    4.809 &   15.000 &   15.000 &   15.000 &   15.000  \\
   4.30 &    1.369 &    0.019 &    3.125 &   13.250 &   15.000 &   15.000 &   15.000  \\
   4.40 &    1.768 &    0.013 &    1.918 &    9.615 &   15.000 &   15.000 &   15.000  \\
   4.50 &    2.099 &    0.043 &    1.068 &    6.842 &   14.727 &   15.000 &   15.000  \\
   4.60 &    2.476 &    0.164 &    0.508 &    4.733 &   10.382 &   15.000 &   15.000  \\
   4.70 &    2.970 &    0.423 &    0.207 &    3.170 &    7.020 &   15.000 &   15.000  \\
\enddata
\tablecomments{Same as Table~\protect\ref{tab:He} but for carbon.}
\end{deluxetable}

\clearpage

\begin{deluxetable}{ccccccccc}
\tabletypesize{\tiny}
\tablecaption{Nitrogen CIE Fractional Abundances
\label{tab:N}}
\tablewidth{0pt}
\tablehead{
\colhead{log(T)} 
& \colhead{N{$^{0+}$}}
& \colhead{N{$^{1+}$}}
& \colhead{N{$^{2+}$}}
& \colhead{N{$^{3+}$}}
& \colhead{N{$^{4+}$}}
& \colhead{N{$^{5+}$}}
& \colhead{N{$^{6+}$}}
& \colhead{N{$^{7+}$}}}
\startdata
   4.00 &    0.000 &    3.390 &   14.613 &   15.000 &   15.000 &   15.000 &   15.000 &   15.000  \\
   4.10 &    0.008 &    1.746 &   10.223 &   15.000 &   15.000 &   15.000 &   15.000 &   15.000  \\
   4.20 &    0.140 &    0.559 &    6.507 &   15.000 &   15.000 &   15.000 &   15.000 &   15.000  \\
   4.30 &    0.680 &    0.102 &    4.089 &   13.747 &   15.000 &   15.000 &   15.000 &   15.000  \\
   4.40 &    1.290 &    0.024 &    2.554 &    9.663 &   15.000 &   15.000 &   15.000 &   15.000  \\
   4.50 &    1.758 &    0.023 &    1.476 &    6.596 &   15.000 &   15.000 &   15.000 &   15.000  \\
   4.60 &    2.180 &    0.095 &    0.722 &    4.316 &   12.687 &   15.000 &   15.000 &   15.000  \\
   4.70 &    2.690 &    0.318 &    0.288 &    2.706 &    9.065 &   15.000 &   15.000 &   15.000  \\
\enddata
\tablecomments{Same as Table~\protect\ref{tab:He} but for nitrogen.}
\end{deluxetable}

\clearpage

\begin{deluxetable}{cccccccccc}
\tabletypesize{\tiny}
\tablecaption{Oxygen CIE Fractional Abundances
\label{tab:O}}
\tablewidth{0pt}
\tablehead{
\colhead{log(T)} 
& \colhead{O{$^{0+}$}}
& \colhead{O{$^{1+}$}}
& \colhead{O{$^{2+}$}}
& \colhead{O{$^{3+}$}}
& \colhead{O{$^{4+}$}}
& \colhead{O{$^{5+}$}}
& \colhead{O{$^{6+}$}}
& \colhead{O{$^{7+}$}}
& \colhead{O{$^{8+}$}}}
\startdata
   4.00 &    0.000 &    2.959 &   15.000 &   15.000 &   15.000 &   15.000 &   15.000 &   15.000 &   15.000  \\
   4.10 &    0.017 &    1.409 &   12.192 &   15.000 &   15.000 &   15.000 &   15.000 &   15.000 &   15.000  \\
   4.20 &    0.244 &    0.367 &    8.276 &   15.000 &   15.000 &   15.000 &   15.000 &   15.000 &   15.000  \\
   4.30 &    0.910 &    0.057 &    5.558 &   15.000 &   15.000 &   15.000 &   15.000 &   15.000 &   15.000  \\
   4.40 &    1.510 &    0.014 &    3.616 &   11.944 &   15.000 &   15.000 &   15.000 &   15.000 &   15.000  \\
   4.50 &    1.890 &    0.009 &    2.183 &    8.206 &   15.000 &   15.000 &   15.000 &   15.000 &   15.000  \\
   4.60 &    2.188 &    0.033 &    1.173 &    5.441 &   13.325 &   15.000 &   15.000 &   15.000 &   15.000  \\
   4.70 &    2.550 &    0.158 &    0.521 &    3.452 &    9.310 &   15.000 &   15.000 &   15.000 &   15.000  \\
\enddata
\tablecomments{Same as Table~\protect\ref{tab:He} but for oxygen.}
\end{deluxetable}

\clearpage

\begin{deluxetable}{ccccccccccc}
\tabletypesize{\tiny}
\tablecaption{Fluorine CIE Fractional Abundances
\label{tab:F}}
\tablewidth{0pt}
\tablehead{
\colhead{log(T)} 
& \colhead{F{$^{0+}$}}
& \colhead{F{$^{1+}$}}
& \colhead{F{$^{2+}$}}
& \colhead{F{$^{3+}$}}
& \colhead{F{$^{4+}$}}
& \colhead{F{$^{5+}$}}
& \colhead{F{$^{6+}$}}
& \colhead{F{$^{7+}$}}
& \colhead{F{$^{8+}$}}
& \colhead{F{$^{9+}$}}}
\startdata
   4.00 &    0.000 &    4.765 &   15.000 &   15.000 &   15.000 &   15.000 &   15.000 &   15.000 &   15.000 &   15.000  \\
   4.10 &    0.001 &    2.820 &   13.944 &   15.000 &   15.000 &   15.000 &   15.000 &   15.000 &   15.000 &   15.000  \\
   4.20 &    0.024 &    1.268 &    9.524 &   15.000 &   15.000 &   15.000 &   15.000 &   15.000 &   15.000 &   15.000  \\
   4.30 &    0.315 &    0.288 &    6.129 &   15.000 &   15.000 &   15.000 &   15.000 &   15.000 &   15.000 &   15.000  \\
   4.40 &    1.052 &    0.040 &    3.951 &   13.757 &   15.000 &   15.000 &   15.000 &   15.000 &   15.000 &   15.000  \\
   4.50 &    1.673 &    0.011 &    2.429 &    9.568 &   15.000 &   15.000 &   15.000 &   15.000 &   15.000 &   15.000  \\
   4.60 &    2.046 &    0.024 &    1.354 &    6.422 &   15.000 &   15.000 &   15.000 &   15.000 &   15.000 &   15.000  \\
   4.70 &    2.364 &    0.112 &    0.653 &    4.134 &   10.748 &   15.000 &   15.000 &   15.000 &   15.000 &   15.000  \\
\enddata
\tablecomments{Same as Table~\protect\ref{tab:He} but for fluorine.}
\end{deluxetable}

\clearpage

\begin{deluxetable}{cccccccccccc}
\tabletypesize{\tiny}
\tablecaption{Neon CIE Fractional Abundances
\label{tab:Ne}}
\tablewidth{0pt}
\tablehead{
\colhead{log(T)} 
& \colhead{Ne{$^{0+}$}}
& \colhead{Ne{$^{1+}$}}
& \colhead{Ne{$^{2+}$}}
& \colhead{Ne{$^{3+}$}}
& \colhead{Ne{$^{4+}$}}
& \colhead{Ne{$^{5+}$}}
& \colhead{Ne{$^{6+}$}}
& \colhead{Ne{$^{7+}$}}
& \colhead{Ne{$^{8+}$}}
& \colhead{Ne{$^{9+}$}}
& \colhead{Ne{$^{10+}$}}}
\startdata
   4.00 &    0.000 &    7.324 &   15.000 &   15.000 &   15.000 &   15.000 &   15.000 &   15.000 &   15.000 &   15.000 &   15.000  \\
   4.10 &    0.000 &    4.891 &   15.000 &   15.000 &   15.000 &   15.000 &   15.000 &   15.000 &   15.000 &   15.000 &   15.000  \\
   4.20 &    0.001 &    2.918 &   12.845 &   15.000 &   15.000 &   15.000 &   15.000 &   15.000 &   15.000 &   15.000 &   15.000  \\
   4.30 &    0.021 &    1.330 &    8.442 &   15.000 &   15.000 &   15.000 &   15.000 &   15.000 &   15.000 &   15.000 &   15.000  \\
   4.40 &    0.303 &    0.299 &    5.146 &   15.000 &   15.000 &   15.000 &   15.000 &   15.000 &   15.000 &   15.000 &   15.000  \\
   4.50 &    1.090 &    0.037 &    3.079 &   10.508 &   15.000 &   15.000 &   15.000 &   15.000 &   15.000 &   15.000 &   15.000  \\
   4.60 &    1.800 &    0.016 &    1.686 &    6.961 &   15.000 &   15.000 &   15.000 &   15.000 &   15.000 &   15.000 &   15.000  \\
   4.70 &    2.261 &    0.081 &    0.784 &    4.411 &   11.839 &   15.000 &   15.000 &   15.000 &   15.000 &   15.000 &   15.000  \\
\enddata
\tablecomments{Same as Table~\protect\ref{tab:He} but for neon.}
\end{deluxetable}

\clearpage

\begin{deluxetable}{ccccccccccccc}
\tabletypesize{\tiny}
\tablecaption{Sodium CIE Fractional Abundances
\label{tab:Na}}
\tablewidth{0pt}
\tablehead{
\colhead{log(T)} 
& \colhead{Na{$^{0+}$}}
& \colhead{Na{$^{1+}$}}
& \colhead{Na{$^{2+}$}}
& \colhead{Na{$^{3+}$}}
& \colhead{Na{$^{4+}$}}
& \colhead{Na{$^{5+}$}}
& \colhead{Na{$^{6+}$}}
& \colhead{Na{$^{7+}$}}
& \colhead{Na{$^{8+}$}}
& \colhead{Na{$^{9+}$}}
& \colhead{Na{$^{10+}$}}
& \colhead{Na{$^{11+}$}}}
\startdata
   4.00 &    3.201 &    0.000 &   15.000 &   15.000 &   15.000 &   15.000 &   15.000 &   15.000 &   15.000 &   15.000 &   15.000 &   15.000  \\
   4.10 &    3.873 &    0.000 &   14.221 &   15.000 &   15.000 &   15.000 &   15.000 &   15.000 &   15.000 &   15.000 &   15.000 &   15.000  \\
   4.20 &    4.435 &    0.000 &   11.417 &   15.000 &   15.000 &   15.000 &   15.000 &   15.000 &   15.000 &   15.000 &   15.000 &   15.000  \\
   4.30 &    4.908 &    0.000 &    8.613 &   15.000 &   15.000 &   15.000 &   15.000 &   15.000 &   15.000 &   15.000 &   15.000 &   15.000  \\
   4.40 &    5.310 &    0.000 &    6.012 &   15.000 &   15.000 &   15.000 &   15.000 &   15.000 &   15.000 &   15.000 &   15.000 &   15.000  \\
   4.50 &    5.652 &    0.000 &    3.918 &   12.315 &   15.000 &   15.000 &   15.000 &   15.000 &   15.000 &   15.000 &   15.000 &   15.000  \\
   4.60 &    5.931 &    0.002 &    2.243 &    8.194 &   15.000 &   15.000 &   15.000 &   15.000 &   15.000 &   15.000 &   15.000 &   15.000  \\
   4.70 &    6.126 &    0.046 &    1.000 &    5.061 &   13.104 &   15.000 &   15.000 &   15.000 &   15.000 &   15.000 &   15.000 &   15.000  \\
\enddata
\tablecomments{Same as Table~\protect\ref{tab:He} but for sodium.}
\end{deluxetable}

\clearpage

\begin{deluxetable}{cccccccccccccc}
\tabletypesize{\tiny}
\tablecaption{Magnesium CIE Fractional Abundances
\label{tab:Mg}}
\tablewidth{0pt}
\tablehead{
\colhead{log(T)} 
& \colhead{Mg{$^{0+}$}}
& \colhead{Mg{$^{1+}$}}
& \colhead{Mg{$^{2+}$}}
& \colhead{Mg{$^{3+}$}}
& \colhead{Mg{$^{4+}$}}
& \colhead{Mg{$^{5+}$}}
& \colhead{Mg{$^{6+}$}}
& \colhead{Mg{$^{7+}$}}
& \colhead{Mg{$^{8+}$}}
& \colhead{Mg{$^{9+}$}}
& \colhead{Mg{$^{10+}$}}
& \colhead{Mg{$^{11+}$}}
& \colhead{Mg{$^{12+}$}}}
\startdata
   4.00 &    0.272 &    0.333 &    4.014 &   15.000 &   15.000 &   15.000 &   15.000 &   15.000 &   15.000 &   15.000 &   15.000 &   15.000 &   15.000  \\
   4.10 &    0.616 &    0.125 &    2.118 &   15.000 &   15.000 &   15.000 &   15.000 &   15.000 &   15.000 &   15.000 &   15.000 &   15.000 &   15.000  \\
   4.20 &    1.093 &    0.129 &    0.754 &   15.000 &   15.000 &   15.000 &   15.000 &   15.000 &   15.000 &   15.000 &   15.000 &   15.000 &   15.000  \\
   4.30 &    1.991 &    0.614 &    0.127 &   15.000 &   15.000 &   15.000 &   15.000 &   15.000 &   15.000 &   15.000 &   15.000 &   15.000 &   15.000  \\
   4.40 &    3.156 &    1.413 &    0.017 &   12.784 &   15.000 &   15.000 &   15.000 &   15.000 &   15.000 &   15.000 &   15.000 &   15.000 &   15.000  \\
   4.50 &    4.215 &    2.146 &    0.003 &    9.795 &   15.000 &   15.000 &   15.000 &   15.000 &   15.000 &   15.000 &   15.000 &   15.000 &   15.000  \\
   4.60 &    5.125 &    2.761 &    0.001 &    7.040 &   15.000 &   15.000 &   15.000 &   15.000 &   15.000 &   15.000 &   15.000 &   15.000 &   15.000  \\
   4.70 &    5.904 &    3.273 &    0.000 &    4.844 &   13.296 &   15.000 &   15.000 &   15.000 &   15.000 &   15.000 &   15.000 &   15.000 &   15.000  \\
\enddata
\tablecomments{Same as Table~\protect\ref{tab:He} but for magnesium.}
\end{deluxetable}

\clearpage

\begin{deluxetable}{ccccccccccccccc}
\setlength{\tabcolsep}{0.06in}
\tabletypesize{\tiny}
\tablecaption{Aluminum CIE Fractional Abundances
\label{tab:Al}}
\tablewidth{0pt}
\tablehead{
\colhead{log(T)}
& \colhead{Al{$^{0+}$}}
& \colhead{Al{$^{1+}$}}
& \colhead{Al{$^{2+}$}}
& \colhead{Al{$^{3+}$}}
& \colhead{Al{$^{4+}$}}
& \colhead{Al{$^{5+}$}}
& \colhead{Al{$^{6+}$}}
& \colhead{Al{$^{7+}$}}
& \colhead{Al{$^{8+}$}}
& \colhead{Al{$^{9+}$}}
& \colhead{Al{$^{10+}$}}
& \colhead{Al{$^{11+}$}}
& \colhead{Al{$^{12+}$}}
& \colhead{Al{$^{13+}$}}}
\startdata
   4.00 &    1.422 &    0.017 &    6.249 &   15.000 &   15.000 &   15.000 &   15.000 &   15.000 &   15.000 &   15.000 &   15.000 &   15.000 &   15.000 &   15.000  \\
   4.10 &    1.765 &    0.008 &    4.422 &   12.816 &   15.000 &   15.000 &   15.000 &   15.000 &   15.000 &   15.000 &   15.000 &   15.000 &   15.000 &   15.000  \\
   4.20 &    1.969 &    0.005 &    3.074 &    9.001 &   15.000 &   15.000 &   15.000 &   15.000 &   15.000 &   15.000 &   15.000 &   15.000 &   15.000 &   15.000  \\
   4.30 &    2.145 &    0.007 &    2.030 &    5.972 &   15.000 &   15.000 &   15.000 &   15.000 &   15.000 &   15.000 &   15.000 &   15.000 &   15.000 &   15.000  \\
   4.40 &    2.338 &    0.030 &    1.206 &    3.545 &   15.000 &   15.000 &   15.000 &   15.000 &   15.000 &   15.000 &   15.000 &   15.000 &   15.000 &   15.000  \\
   4.50 &    2.616 &    0.137 &    0.610 &    1.653 &   15.000 &   15.000 &   15.000 &   15.000 &   15.000 &   15.000 &   15.000 &   15.000 &   15.000 &   15.000  \\
   4.60 &    3.214 &    0.562 &    0.446 &    0.434 &   12.598 &   15.000 &   15.000 &   15.000 &   15.000 &   15.000 &   15.000 &   15.000 &   15.000 &   15.000  \\
   4.70 &    4.382 &    1.558 &    0.940 &    0.067 &    9.420 &   15.000 &   15.000 &   15.000 &   15.000 &   15.000 &   15.000 &   15.000 &   15.000 &   15.000  \\
\enddata
\tablecomments{Same as Table~\protect\ref{tab:He} but for aluminum.}
\end{deluxetable}

\clearpage

\begin{deluxetable}{cccccccccccccccc}
\setlength{\tabcolsep}{0.06in}
\tabletypesize{\tiny}
\tablecaption{Silicon CIE Fractional Abundances
\label{tab:Si}}
\tablewidth{0pt}
\tablehead{
\colhead{log(T)} 
& \colhead{Si{$^{0+}$}}
& \colhead{Si{$^{1+}$}}
& \colhead{Si{$^{2+}$}}
& \colhead{Si{$^{3+}$}}
& \colhead{Si{$^{4+}$}}
& \colhead{Si{$^{5+}$}}
& \colhead{Si{$^{6+}$}}
& \colhead{Si{$^{7+}$}}
& \colhead{Si{$^{8+}$}}
& \colhead{Si{$^{9+}$}}
& \colhead{Si{$^{10+}$}}
& \colhead{Si{$^{11+}$}}
& \colhead{Si{$^{12+}$}}
& \colhead{Si{$^{13+}$}}
& \colhead{Si{$^{14+}$}}}
\startdata
   4.00 &    0.638 &    0.113 &    4.786 &   15.000 &   15.000 &   15.000 &   15.000 &   15.000 &   15.000 &   15.000 &   15.000 &   15.000 &   15.000 &   15.000 &   15.000  \\
   4.10 &    1.435 &    0.017 &    2.955 &   13.766 &   15.000 &   15.000 &   15.000 &   15.000 &   15.000 &   15.000 &   15.000 &   15.000 &   15.000 &   15.000 &   15.000  \\
   4.20 &    2.109 &    0.014 &    1.627 &    9.882 &   15.000 &   15.000 &   15.000 &   15.000 &   15.000 &   15.000 &   15.000 &   15.000 &   15.000 &   15.000 &   15.000  \\
   4.30 &    2.645 &    0.089 &    0.739 &    6.960 &   15.000 &   15.000 &   15.000 &   15.000 &   15.000 &   15.000 &   15.000 &   15.000 &   15.000 &   15.000 &   15.000  \\
   4.40 &    3.180 &    0.323 &    0.281 &    4.929 &   11.328 &   15.000 &   15.000 &   15.000 &   15.000 &   15.000 &   15.000 &   15.000 &   15.000 &   15.000 &   15.000  \\
   4.50 &    3.749 &    0.670 &    0.105 &    3.499 &    7.915 &   15.000 &   15.000 &   15.000 &   15.000 &   15.000 &   15.000 &   15.000 &   15.000 &   15.000 &   15.000  \\
   4.60 &    4.299 &    1.030 &    0.044 &    2.420 &    5.235 &   15.000 &   15.000 &   15.000 &   15.000 &   15.000 &   15.000 &   15.000 &   15.000 &   15.000 &   15.000  \\
   4.70 &    4.820 &    1.372 &    0.032 &    1.567 &    3.088 &   15.000 &   15.000 &   15.000 &   15.000 &   15.000 &   15.000 &   15.000 &   15.000 &   15.000 &   15.000  \\
\enddata
\tablecomments{Same as Table~\protect\ref{tab:He} but for silicon and
using the DR and RR rate coefficients of \protect\citet{Mazz98a} for
ions not calculated by \protect\citet{Badn06b,Badn06c}.}
\end{deluxetable}

\clearpage

\begin{deluxetable}{ccccccccccccccccc}
\setlength{\tabcolsep}{0.05in}
\tabletypesize{\tiny}
\tablecaption{Phosphorus CIE Fractional Abundances
\label{tab:P}}
\tablewidth{0pt}
\tablehead{
\colhead{log(T)} 
& \colhead{P{$^{0+}$}}
& \colhead{P{$^{1+}$}}
& \colhead{P{$^{2+}$}}
& \colhead{P{$^{3+}$}}
& \colhead{P{$^{4+}$}}
& \colhead{P{$^{5+}$}}
& \colhead{P{$^{6+}$}}
& \colhead{P{$^{7+}$}}
& \colhead{P{$^{8+}$}}
& \colhead{P{$^{9+}$}}
& \colhead{P{$^{10+}$}}
& \colhead{P{$^{11+}$}}
& \colhead{P{$^{12+}$}}
& \colhead{P{$^{13+}$}}
& \colhead{P{$^{14+}$}}
& \colhead{P{$^{15+}$}}}
\startdata
   4.00 &    0.089 &    0.733 &    6.302 &   15.000 &   15.000 &   15.000 &   15.000 &   15.000 &   15.000 &   15.000 &   15.000 &   15.000 &   15.000 &   15.000 &   15.000 &   15.000  \\
   4.10 &    0.670 &    0.105 &    3.573 &   13.045 &   15.000 &   15.000 &   15.000 &   15.000 &   15.000 &   15.000 &   15.000 &   15.000 &   15.000 &   15.000 &   15.000 &   15.000  \\
   4.20 &    1.541 &    0.017 &    2.014 &    8.887 &   15.000 &   15.000 &   15.000 &   15.000 &   15.000 &   15.000 &   15.000 &   15.000 &   15.000 &   15.000 &   15.000 &   15.000  \\
   4.30 &    2.213 &    0.040 &    1.091 &    5.922 &   15.000 &   15.000 &   15.000 &   15.000 &   15.000 &   15.000 &   15.000 &   15.000 &   15.000 &   15.000 &   15.000 &   15.000  \\
   4.40 &    2.647 &    0.157 &    0.522 &    3.792 &   12.363 &   15.000 &   15.000 &   15.000 &   15.000 &   15.000 &   15.000 &   15.000 &   15.000 &   15.000 &   15.000 &   15.000  \\
   4.50 &    3.050 &    0.409 &    0.219 &    2.322 &    8.920 &   15.000 &   15.000 &   15.000 &   15.000 &   15.000 &   15.000 &   15.000 &   15.000 &   15.000 &   15.000 &   15.000  \\
   4.60 &    3.524 &    0.767 &    0.107 &    1.329 &    6.353 &   12.260 &   15.000 &   15.000 &   15.000 &   15.000 &   15.000 &   15.000 &   15.000 &   15.000 &   15.000 &   15.000  \\
   4.70 &    4.082 &    1.207 &    0.140 &    0.671 &    4.421 &    8.516 &   15.000 &   15.000 &   15.000 &   15.000 &   15.000 &   15.000 &   15.000 &   15.000 &   15.000 &   15.000  \\
\enddata
\tablecomments{Same as Table~\protect\ref{tab:Si} but for phosphorus.}
\end{deluxetable}

\clearpage

\begin{deluxetable}{cccccccccccccccccc}
\setlength{\tabcolsep}{0.04in}
\tabletypesize{\tiny}
\tablecaption{Sulfur CIE Fractional Abundances
\label{tab:S}}
\tablewidth{0pt}
\tablehead{
\colhead{log(T)} 
& \colhead{S{$^{0+}$}}
& \colhead{S{$^{1+}$}}
& \colhead{S{$^{2+}$}}
& \colhead{S{$^{3+}$}}
& \colhead{S{$^{4+}$}}
& \colhead{S{$^{5+}$}}
& \colhead{S{$^{6+}$}}
& \colhead{S{$^{7+}$}}
& \colhead{S{$^{8+}$}}
& \colhead{S{$^{9+}$}}
& \colhead{S{$^{10+}$}}
& \colhead{S{$^{11+}$}}
& \colhead{S{$^{12+}$}}
& \colhead{S{$^{13+}$}}
& \colhead{S{$^{14+}$}}
& \colhead{S{$^{15+}$}}
& \colhead{S{$^{16+}$}}}
\startdata
   4.00 &    0.181 &    0.468 &    8.954 &   15.000 &   15.000 &   15.000 &   15.000 &   15.000 &   15.000 &   15.000 &   15.000 &   15.000 &   15.000 &   15.000 &   15.000 &   15.000 &   15.000  \\
   4.10 &    0.949 &    0.052 &    5.894 &   15.000 &   15.000 &   15.000 &   15.000 &   15.000 &   15.000 &   15.000 &   15.000 &   15.000 &   15.000 &   15.000 &   15.000 &   15.000 &   15.000  \\
   4.20 &    1.850 &    0.006 &    3.766 &   11.913 &   15.000 &   15.000 &   15.000 &   15.000 &   15.000 &   15.000 &   15.000 &   15.000 &   15.000 &   15.000 &   15.000 &   15.000 &   15.000  \\
   4.30 &    2.521 &    0.004 &    2.163 &    7.967 &   15.000 &   15.000 &   15.000 &   15.000 &   15.000 &   15.000 &   15.000 &   15.000 &   15.000 &   15.000 &   15.000 &   15.000 &   15.000  \\
   4.40 &    2.899 &    0.044 &    1.023 &    5.016 &   12.472 &   15.000 &   15.000 &   15.000 &   15.000 &   15.000 &   15.000 &   15.000 &   15.000 &   15.000 &   15.000 &   15.000 &   15.000  \\
   4.50 &    3.228 &    0.232 &    0.384 &    3.068 &    8.554 &   15.000 &   15.000 &   15.000 &   15.000 &   15.000 &   15.000 &   15.000 &   15.000 &   15.000 &   15.000 &   15.000 &   15.000  \\
   4.60 &    3.671 &    0.580 &    0.141 &    1.884 &    5.855 &   13.801 &   15.000 &   15.000 &   15.000 &   15.000 &   15.000 &   15.000 &   15.000 &   15.000 &   15.000 &   15.000 &   15.000  \\
   4.70 &    4.176 &    0.983 &    0.087 &    1.108 &    3.903 &   10.045 &   15.000 &   15.000 &   15.000 &   15.000 &   15.000 &   15.000 &   15.000 &   15.000 &   15.000 &   15.000 &   15.000  \\
\enddata
\tablecomments{Same as Table~\protect\ref{tab:Si} but for sulfur.}
\end{deluxetable}

\clearpage

\begin{deluxetable}{ccccccccccccccccccc}
\setlength{\tabcolsep}{0.03in}
\tabletypesize{\tiny}
\tablecaption{Chlorine CIE Fractional Abundances
\label{tab:Cl}}
\tablewidth{0pt}
\tablehead{
\colhead{log(T)}
& \colhead{Cl{$^{0+}$}}
& \colhead{Cl{$^{1+}$}}
& \colhead{Cl{$^{2+}$}}
& \colhead{Cl{$^{3+}$}}
& \colhead{Cl{$^{4+}$}}
& \colhead{Cl{$^{5+}$}}
& \colhead{Cl{$^{6+}$}}
& \colhead{Cl{$^{7+}$}}
& \colhead{Cl{$^{8+}$}}
& \colhead{Cl{$^{9+}$}}
& \colhead{Cl{$^{10+}$}}
& \colhead{Cl{$^{11+}$}}
& \colhead{Cl{$^{12+}$}}
& \colhead{Cl{$^{13+}$}}
& \colhead{Cl{$^{14+}$}}
& \colhead{Cl{$^{15+}$}}
& \colhead{Cl{$^{16+}$}}
& \colhead{Cl{$^{17+}$}}}
\startdata
   4.00 &    0.003 &    2.230 &    9.867 &   15.000 &   15.000 &   15.000 &   15.000 &   15.000 &   15.000 &   15.000 &   15.000 &   15.000 &   15.000 &   15.000 &   15.000 &   15.000 &   15.000 &   15.000  \\
   4.10 &    0.074 &    0.808 &    5.856 &   15.000 &   15.000 &   15.000 &   15.000 &   15.000 &   15.000 &   15.000 &   15.000 &   15.000 &   15.000 &   15.000 &   15.000 &   15.000 &   15.000 &   15.000  \\
   4.20 &    0.606 &    0.124 &    3.103 &   11.838 &   15.000 &   15.000 &   15.000 &   15.000 &   15.000 &   15.000 &   15.000 &   15.000 &   15.000 &   15.000 &   15.000 &   15.000 &   15.000 &   15.000  \\
   4.30 &    1.505 &    0.031 &    1.414 &    7.445 &   15.000 &   15.000 &   15.000 &   15.000 &   15.000 &   15.000 &   15.000 &   15.000 &   15.000 &   15.000 &   15.000 &   15.000 &   15.000 &   15.000  \\
   4.40 &    2.455 &    0.184 &    0.466 &    4.346 &   12.168 &   15.000 &   15.000 &   15.000 &   15.000 &   15.000 &   15.000 &   15.000 &   15.000 &   15.000 &   15.000 &   15.000 &   15.000 &   15.000  \\
   4.50 &    3.415 &    0.572 &    0.138 &    2.415 &    8.341 &   15.000 &   15.000 &   15.000 &   15.000 &   15.000 &   15.000 &   15.000 &   15.000 &   15.000 &   15.000 &   15.000 &   15.000 &   15.000  \\
   4.60 &    4.178 &    1.021 &    0.067 &    1.318 &    5.854 &   12.960 &   15.000 &   15.000 &   15.000 &   15.000 &   15.000 &   15.000 &   15.000 &   15.000 &   15.000 &   15.000 &   15.000 &   15.000  \\
   4.70 &    4.794 &    1.489 &    0.114 &    0.701 &    4.134 &    9.499 &   15.000 &   15.000 &   15.000 &   15.000 &   15.000 &   15.000 &   15.000 &   15.000 &   15.000 &   15.000 &   15.000 &   15.000  \\
\enddata
\tablecomments{Same as Table~\protect\ref{tab:Si} but for chlorine.}
\end{deluxetable}

\clearpage

\begin{deluxetable}{cccccccccccccccccccc}
\setlength{\tabcolsep}{0.02in}
\tabletypesize{\tiny}
\tablecaption{Argon CIE Fractional Abundances
\label{tab:Ar}}
\tablewidth{0pt}
\tablehead{
\colhead{log(T)} 
& \colhead{Ar{$^{0+}$}}
& \colhead{Ar{$^{1+}$}}
& \colhead{Ar{$^{2+}$}}
& \colhead{Ar{$^{3+}$}}
& \colhead{Ar{$^{4+}$}}
& \colhead{Ar{$^{5+}$}}
& \colhead{Ar{$^{6+}$}}
& \colhead{Ar{$^{7+}$}}
& \colhead{Ar{$^{8+}$}}
& \colhead{Ar{$^{9+}$}}
& \colhead{Ar{$^{10+}$}}
& \colhead{Ar{$^{11+}$}}
& \colhead{Ar{$^{12+}$}}
& \colhead{Ar{$^{13+}$}}
& \colhead{Ar{$^{14+}$}}
& \colhead{Ar{$^{15+}$}}
& \colhead{Ar{$^{16+}$}}
& \colhead{Ar{$^{17+}$}}
& \colhead{Ar{$^{18+}$}}}
\startdata
   4.00 &    0.000 &    3.459 &   13.305 &   15.000 &   15.000 &   15.000 &   15.000 &   15.000 &   15.000 &   15.000 &   15.000 &   15.000 &   15.000 &   15.000 &   15.000 &   15.000 &   15.000 &   15.000 &   15.000  \\
   4.10 &    0.009 &    1.693 &    8.681 &   15.000 &   15.000 &   15.000 &   15.000 &   15.000 &   15.000 &   15.000 &   15.000 &   15.000 &   15.000 &   15.000 &   15.000 &   15.000 &   15.000 &   15.000 &   15.000  \\
   4.20 &    0.197 &    0.439 &    5.025 &   14.395 &   15.000 &   15.000 &   15.000 &   15.000 &   15.000 &   15.000 &   15.000 &   15.000 &   15.000 &   15.000 &   15.000 &   15.000 &   15.000 &   15.000 &   15.000  \\
   4.30 &    0.984 &    0.049 &    2.700 &    9.243 &   15.000 &   15.000 &   15.000 &   15.000 &   15.000 &   15.000 &   15.000 &   15.000 &   15.000 &   15.000 &   15.000 &   15.000 &   15.000 &   15.000 &   15.000  \\
   4.40 &    1.939 &    0.039 &    1.133 &    5.405 &   14.578 &   15.000 &   15.000 &   15.000 &   15.000 &   15.000 &   15.000 &   15.000 &   15.000 &   15.000 &   15.000 &   15.000 &   15.000 &   15.000 &   15.000  \\
   4.50 &    3.053 &    0.365 &    0.247 &    2.759 &    9.717 &   15.000 &   15.000 &   15.000 &   15.000 &   15.000 &   15.000 &   15.000 &   15.000 &   15.000 &   15.000 &   15.000 &   15.000 &   15.000 &   15.000  \\
   4.60 &    4.288 &    1.005 &    0.065 &    1.400 &    6.732 &   14.248 &   15.000 &   15.000 &   15.000 &   15.000 &   15.000 &   15.000 &   15.000 &   15.000 &   15.000 &   15.000 &   15.000 &   15.000 &   15.000  \\
   4.70 &    5.172 &    1.527 &    0.109 &    0.718 &    4.780 &   10.658 &   15.000 &   15.000 &   15.000 &   15.000 &   15.000 &   15.000 &   15.000 &   15.000 &   15.000 &   15.000 &   15.000 &   15.000 &   15.000  \\
\enddata
\tablecomments{Same as Table~\protect\ref{tab:Si} but for argon.}
\end{deluxetable}

\clearpage

\begin{deluxetable}{ccccccccccc}
\tabletypesize{\tiny}
\tablecaption{Potassium CIE Fractional Abundances
\label{tab:K}}
\tablewidth{0pt}
\tablehead{
\colhead{log(T)} 
& \colhead{K{$^{0+}$}}
& \colhead{K{$^{1+}$}}
& \colhead{K{$^{2+}$}}
& \colhead{K{$^{3+}$}}
& \colhead{K{$^{4+}$}}
& \colhead{K{$^{5+}$}}
& \colhead{K{$^{6+}$}}
& \colhead{K{$^{7+}$}}
& \colhead{K{$^{8+}$}}
& \colhead{K{$^{9+}$}}}
\startdata
   4.00 &    3.425 &    0.000 &   11.456 &   15.000 &   15.000 &   15.000 &   15.000 &   15.000 &   15.000 &   15.000 \\
   4.10 &    3.999 &    0.000 &    8.520 &   15.000 &   15.000 &   15.000 &   15.000 &   15.000 &   15.000 &   15.000 \\
   4.20 &    4.478 &    0.000 &    5.799 &   15.000 &   15.000 &   15.000 &   15.000 &   15.000 &   15.000 &   15.000 \\
   4.30 &    4.879 &    0.000 &    3.614 &   11.687 &   15.000 &   15.000 &   15.000 &   15.000 &   15.000 &   15.000 \\
   4.40 &    5.224 &    0.006 &    1.861 &    7.434 &   15.000 &   15.000 &   15.000 &   15.000 &   15.000 &   15.000 \\
   4.50 &    5.639 &    0.134 &    0.577 &    4.155 &   10.438 &   15.000 &   15.000 &   15.000 &   15.000 &   15.000 \\
   4.60 &    6.457 &    0.733 &    0.092 &    2.176 &    6.539 &   14.837 &   15.000 &   15.000 &   15.000 &   15.000 \\
   4.70 &    7.060 &    1.376 &    0.051 &    1.169 &    4.218 &   10.723 &   15.000 &   15.000 &   15.000 &   15.000 \\
\enddata
\tablecomments{Same as Table~\protect\ref{tab:Si} but for potassium.}
\end{deluxetable}

\clearpage

\begin{deluxetable}{cccccccccccc}
\tabletypesize{\tiny}
\tablecaption{Calcium CIE Fractional Abundances
\label{tab:Ca}}
\tablewidth{0pt}
\tablehead{
\colhead{log(T)} 
& \colhead{Ca{$^{0+}$}}
& \colhead{Ca{$^{1+}$}}
& \colhead{Ca{$^{2+}$}}
& \colhead{Ca{$^{3+}$}}
& \colhead{Ca{$^{4+}$}}
& \colhead{Ca{$^{5+}$}}
& \colhead{Ca{$^{6+}$}}
& \colhead{Ca{$^{7+}$}}
& \colhead{Ca{$^{8+}$}}
& \colhead{Ca{$^{9+}$}}
& \colhead{Ca{$^{10+}$}}}
\startdata
   4.00 &    1.131 &    0.040 &    1.884 &   15.000 &   15.000 &   15.000 &   15.000 &   15.000 &   15.000 &   15.000 &   15.000\\
   4.10 &    1.734 &    0.129 &    0.624 &   15.000 &   15.000 &   15.000 &   15.000 &   15.000 &   15.000 &   15.000 &   15.000\\
   4.20 &    2.747 &    0.697 &    0.098 &   12.421 &   15.000 &   15.000 &   15.000 &   15.000 &   15.000 &   15.000 &   15.000\\
   4.30 &    3.945 &    1.505 &    0.014 &    9.346 &   15.000 &   15.000 &   15.000 &   15.000 &   15.000 &   15.000 &   15.000\\
   4.40 &    5.015 &    2.230 &    0.003 &    6.564 &   15.000 &   15.000 &   15.000 &   15.000 &   15.000 &   15.000 &   15.000\\
   4.50 &    5.931 &    2.835 &    0.001 &    4.342 &   11.865 &   15.000 &   15.000 &   15.000 &   15.000 &   15.000 &   15.000\\
   4.60 &    6.613 &    3.234 &    0.001 &    2.634 &    7.897 &   15.000 &   15.000 &   15.000 &   15.000 &   15.000 &   15.000\\
   4.70 &    6.908 &    3.269 &    0.013 &    1.534 &    5.230 &   10.863 &   15.000 &   15.000 &   15.000 &   15.000 &   15.000\\
\enddata
\tablecomments{Same as Table~\protect\ref{tab:Si} but for calcium.}
\end{deluxetable}

\clearpage

\begin{deluxetable}{cccccccccccc}
\tabletypesize{\tiny}
\tablecaption{Scandium CIE Fractional Abundances
\label{tab:Sc}}
\tablewidth{0pt}
\tablehead{
\colhead{log(T)} 
& \colhead{Sc{$^{0+}$}}
& \colhead{Sc{$^{1+}$}}
& \colhead{Sc{$^{2+}$}}
& \colhead{Sc{$^{3+}$}}
& \colhead{Sc{$^{4+}$}}
& \colhead{Sc{$^{5+}$}}
& \colhead{Sc{$^{6+}$}}
& \colhead{Sc{$^{7+}$}}
& \colhead{Sc{$^{8+}$}}
& \colhead{Sc{$^{9+}$}}
& \colhead{Sc{$^{10+}$}}}
\startdata
   4.00 &    1.265 &    0.028 &    2.130 &   10.843 &   15.000 &   15.000 &   15.000 &   15.000 &   15.000 &   15.000 &   15.000\\
   4.10 &    1.650 &    0.050 &    1.060 &    7.050 &   15.000 &   15.000 &   15.000 &   15.000 &   15.000 &   15.000 &   15.000\\
   4.20 &    2.069 &    0.144 &    0.563 &    4.354 &   15.000 &   15.000 &   15.000 &   15.000 &   15.000 &   15.000 &   15.000\\
   4.30 &    2.530 &    0.307 &    0.302 &    2.313 &   15.000 &   15.000 &   15.000 &   15.000 &   15.000 &   15.000 &   15.000\\
   4.40 &    3.097 &    0.599 &    0.231 &    0.797 &   12.234 &   15.000 &   15.000 &   15.000 &   15.000 &   15.000 &   15.000\\
   4.50 &    4.172 &    1.417 &    0.721 &    0.113 &    8.814 &   15.000 &   15.000 &   15.000 &   15.000 &   15.000 &   15.000\\
   4.60 &    5.562 &    2.567 &    1.566 &    0.013 &    6.185 &   15.000 &   15.000 &   15.000 &   15.000 &   15.000 &   15.000\\
   4.70 &    6.727 &    3.505 &    2.217 &    0.003 &    4.198 &   10.998 &   15.000 &   15.000 &   15.000 &   15.000 &   15.000\\
\enddata
\tablecomments{Same as Table~\protect\ref{tab:Si} but for scandium.}
\end{deluxetable}

\clearpage

\begin{deluxetable}{ccccccccccccc}
\tabletypesize{\tiny}
\tablecaption{Titanium CIE Fractional Abundances
\label{tab:Ti}}
\tablewidth{0pt}
\tablehead{
\colhead{log(T)} 
& \colhead{Ti{$^{0+}$}}
& \colhead{Ti{$^{1+}$}}
& \colhead{Ti{$^{2+}$}}
& \colhead{Ti{$^{3+}$}}
& \colhead{Ti{$^{4+}$}}
& \colhead{Ti{$^{5+}$}}
& \colhead{Ti{$^{6+}$}}
& \colhead{Ti{$^{7+}$}}
& \colhead{Ti{$^{8+}$}}
& \colhead{Ti{$^{9+}$}}
& \colhead{Ti{$^{10+}$}}
& \colhead{Ti{$^{11+}$}}}
\startdata
4.00 &    2.503 &    0.002 &	3.397 &   13.345 &   15.000 &	15.000 &   15.000 &   15.000 &   15.000 &   15.000 &   15.000 &   15.000\\
4.10 &    2.796 &    0.007 &	1.859 &    8.946 &   15.000 &	15.000 &   15.000 &   15.000 &   15.000 &   15.000 &   15.000 &   15.000\\
4.20 &    2.907 &    0.098 &	0.698 &    5.553 &   15.000 &	15.000 &   15.000 &   15.000 &   15.000 &   15.000 &   15.000 &   15.000\\
4.30 &    3.342 &    0.530 &	0.152 &    3.457 &   10.413 &	15.000 &   15.000 &   15.000 &   15.000 &   15.000 &   15.000 &   15.000\\
4.40 &    3.880 &    1.042 &	0.044 &    2.205 &    6.796 &	15.000 &   15.000 &   15.000 &   15.000 &   15.000 &   15.000 &   15.000\\
4.50 &    4.223 &    1.331 &	0.044 &    1.296 &    3.988 &	15.000 &   15.000 &   15.000 &   15.000 &   15.000 &   15.000 &   15.000\\
4.60 &    4.554 &    1.587 &	0.137 &    0.640 &    1.805 &	12.271 &   15.000 &   15.000 &   15.000 &   15.000 &   15.000 &   15.000\\
4.70 &    5.233 &    2.173 &	0.578 &    0.452 &    0.425 &	 8.688 &   15.000 &   15.000 &   15.000 &   15.000 &   15.000 &   15.000\\
\enddata
\tablecomments{Same as Table~\protect\ref{tab:Si} but for titanium.}
\end{deluxetable}

\clearpage

\begin{deluxetable}{ccccccccccccc}
\tabletypesize{\tiny}
\tablecaption{Vanadium CIE Fractional Abundances
\label{tab:V}}
\tablewidth{0pt}
\tablehead{
\colhead{log(T)} 
& \colhead{V{$^{0+}$}}
& \colhead{V{$^{1+}$}}
& \colhead{V{$^{2+}$}}
& \colhead{V{$^{3+}$}}
& \colhead{V{$^{4+}$}}
& \colhead{V{$^{5+}$}}
& \colhead{V{$^{6+}$}}
& \colhead{V{$^{7+}$}}
& \colhead{V{$^{8+}$}}
& \colhead{V{$^{9+}$}}
& \colhead{V{$^{10+}$}}
& \colhead{V{$^{11+}$}}}
\startdata
   4.00 &    2.694 &    0.001 &    3.460 &   14.920 &   15.000 &   15.000 &   15.000 &   15.000 &   15.000 &   15.000 &   15.000 &   15.000\\
   4.10 &    3.381 &    0.007 &    1.820 &   10.534 &   15.000 &   15.000 &   15.000 &   15.000 &   15.000 &   15.000 &   15.000 &   15.000\\
   4.20 &    3.666 &    0.117 &    0.626 &    6.793 &   15.000 &   15.000 &   15.000 &   15.000 &   15.000 &   15.000 &   15.000 &   15.000\\
   4.30 &    4.056 &    0.553 &    0.143 &    4.255 &   11.807 &   15.000 &   15.000 &   15.000 &   15.000 &   15.000 &   15.000 &   15.000\\
   4.40 &    4.432 &    0.972 &    0.051 &    2.508 &    7.532 &   15.000 &   15.000 &   15.000 &   15.000 &   15.000 &   15.000 &   15.000\\
   4.50 &    4.725 &    1.267 &    0.052 &    1.227 &    4.220 &   11.611 &   15.000 &   15.000 &   15.000 &   15.000 &   15.000 &   15.000\\
   4.60 &    5.127 &    1.635 &    0.194 &    0.491 &    1.845 &    6.896 &   15.000 &   15.000 &   15.000 &   15.000 &   15.000 &   15.000\\
   4.70 &    5.860 &    2.305 &    0.658 &    0.392 &    0.432 &    3.626 &   14.037 &   15.000 &   15.000 &   15.000 &   15.000 &   15.000\\
\enddata
\tablecomments{Same as Table~\protect\ref{tab:Si} but for vanadium.}
\end{deluxetable}

\clearpage

\begin{deluxetable}{cccccccccccccc}
\tabletypesize{\tiny}
\tablecaption{Chromium CIE Fractional Abundances
\label{tab:Cr}}
\tablewidth{0pt}
\tablehead{
\colhead{log(T)} 
& \colhead{Cr{$^{0+}$}}
& \colhead{Cr{$^{1+}$}}
& \colhead{Cr{$^{2+}$}}
& \colhead{Cr{$^{3+}$}}
& \colhead{Cr{$^{4+}$}}
& \colhead{Cr{$^{5+}$}}
& \colhead{Cr{$^{6+}$}}
& \colhead{Cr{$^{7+}$}}
& \colhead{Cr{$^{8+}$}}
& \colhead{Cr{$^{9+}$}}
& \colhead{Cr{$^{10+}$}}
& \colhead{Cr{$^{11+}$}}
& \colhead{Cr{$^{12+}$}}}
\startdata
4.00 &    1.948 &    0.005 &	4.512 &   15.000 &   15.000 &	15.000 &   15.000 &   15.000 &   15.000 &   15.000 &   15.000 &   15.000 &   15.000\\
4.10 &    2.814 &    0.002 &	2.700 &   11.913 &   15.000 &	15.000 &   15.000 &   15.000 &   15.000 &   15.000 &   15.000 &   15.000 &   15.000\\
4.20 &    3.542 &    0.018 &	1.385 &    7.922 &   15.000 &	15.000 &   15.000 &   15.000 &   15.000 &   15.000 &   15.000 &   15.000 &   15.000\\
4.30 &    4.088 &    0.111 &	0.646 &    5.032 &   14.278 &	15.000 &   15.000 &   15.000 &   15.000 &   15.000 &   15.000 &   15.000 &   15.000\\
4.40 &    4.318 &    0.300 &	0.303 &    2.963 &    9.446 &	15.000 &   15.000 &   15.000 &   15.000 &   15.000 &   15.000 &   15.000 &   15.000\\
4.50 &    4.460 &    0.564 &	0.158 &    1.489 &    5.770 &	13.316 &   15.000 &   15.000 &   15.000 &   15.000 &   15.000 &   15.000 &   15.000\\
4.60 &    4.739 &    0.938 &	0.186 &    0.634 &    3.207 &	 8.447 &   15.000 &   15.000 &   15.000 &   15.000 &   15.000 &   15.000 &   15.000\\
4.70 &    5.195 &    1.436 &	0.376 &    0.284 &    1.648 &	 5.203 &   11.647 &   15.000 &   15.000 &   15.000 &   15.000 &   15.000 &   15.000\\
\enddata
\tablecomments{Same as Table~\protect\ref{tab:Si} but for chromium.}
\end{deluxetable}

\clearpage

\begin{deluxetable}{cccccccccccccc}
\setlength{\tabcolsep}{0.06in}
\tabletypesize{\tiny}
\tablecaption{Manganese CIE Fractional Abundances
\label{tab:Mn}}
\tablewidth{0pt}
\tablehead{
\colhead{log(T)} 
& \colhead{Mn{$^{0+}$}}
& \colhead{Mn{$^{1+}$}}
& \colhead{Mn{$^{2+}$}}
& \colhead{Mn{$^{3+}$}}
& \colhead{Mn{$^{4+}$}}
& \colhead{Mn{$^{5+}$}}
& \colhead{Mn{$^{6+}$}}
& \colhead{Mn{$^{7+}$}}
& \colhead{Mn{$^{8+}$}}
& \colhead{Mn{$^{9+}$}}
& \colhead{Mn{$^{10+}$}}
& \colhead{Mn{$^{11+}$}}
& \colhead{Mn{$^{12+}$}}}
\startdata
4.00 &    1.767 &    0.008 &	4.065 &   15.000 &   15.000 &	15.000 &   15.000 &   15.000 &   15.000 &   15.000 &   15.000 &   15.000 &   15.000\\
4.10 &    2.611 &    0.003 &	2.314 &   12.339 &   15.000 &	15.000 &   15.000 &   15.000 &   15.000 &   15.000 &   15.000 &   15.000 &   15.000\\
4.20 &    3.110 &    0.052 &	0.953 &    8.212 &   15.000 &	15.000 &   15.000 &   15.000 &   15.000 &   15.000 &   15.000 &   15.000 &   15.000\\
4.30 &    3.591 &    0.415 &	0.211 &    5.152 &   14.884 &	15.000 &   15.000 &   15.000 &   15.000 &   15.000 &   15.000 &   15.000 &   15.000\\
4.40 &    4.200 &    0.987 &	0.048 &    3.176 &   10.057 &	15.000 &   15.000 &   15.000 &   15.000 &   15.000 &   15.000 &   15.000 &   15.000\\
4.50 &    4.607 &    1.345 &	0.026 &    1.876 &    6.516 &	15.000 &   15.000 &   15.000 &   15.000 &   15.000 &   15.000 &   15.000 &   15.000\\
4.60 &    4.934 &    1.597 &	0.051 &    1.071 &    4.079 &	10.443 &   15.000 &   15.000 &   15.000 &   15.000 &   15.000 &   15.000 &   15.000\\
4.70 &    5.321 &    1.887 &	0.148 &    0.565 &    2.470 &	 6.819 &   13.200 &   15.000 &   15.000 &   15.000 &   15.000 &   15.000 &   15.000\\
\enddata
\tablecomments{Same as Table~\protect\ref{tab:Si} but for manganese.}
\end{deluxetable}

\clearpage

\begin{deluxetable}{ccccccccccccccc}
\setlength{\tabcolsep}{0.06in}
\tabletypesize{\tiny}
\tablecaption{Iron CIE Fractional Abundances
\label{tab:Fe}}
\tablewidth{0pt}
\tablehead{
\colhead{log(T)} 
& \colhead{Fe{$^{0+}$}}
& \colhead{Fe{$^{1+}$}}
& \colhead{Fe{$^{2+}$}}
& \colhead{Fe{$^{3+}$}}
& \colhead{Fe{$^{4+}$}}
& \colhead{Fe{$^{5+}$}}
& \colhead{Fe{$^{6+}$}}
& \colhead{Fe{$^{7+}$}}
& \colhead{Fe{$^{8+}$}}
& \colhead{Fe{$^{9+}$}}
& \colhead{Fe{$^{10+}$}}
& \colhead{Fe{$^{11+}$}}
& \colhead{Fe{$^{12+}$}}
& \colhead{Fe{$^{13+}$}}}
\startdata
4.00 &    0.901 &    0.058 &	4.016 &   15.000 &   15.000 &	15.000 &   15.000 &   15.000 &   15.000 &   15.000 &   15.000 &   15.000 &   15.000 &	15.000\\
4.10 &    1.416 &    0.020 &	2.172 &   10.829 &   15.000 &	15.000 &   15.000 &   15.000 &   15.000 &   15.000 &   15.000 &   15.000 &   15.000 &	15.000\\
4.20 &    1.932 &    0.085 &	0.781 &    6.792 &   15.000 &	15.000 &   15.000 &   15.000 &   15.000 &   15.000 &   15.000 &   15.000 &   15.000 &	15.000\\
4.30 &    2.825 &    0.587 &	0.131 &    4.017 &   14.447 &	15.000 &   15.000 &   15.000 &   15.000 &   15.000 &   15.000 &   15.000 &   15.000 &	15.000\\
4.40 &    3.877 &    1.291 &	0.026 &    2.200 &    9.715 &	15.000 &   15.000 &   15.000 &   15.000 &   15.000 &   15.000 &   15.000 &   15.000 &	15.000\\
4.50 &    4.692 &    1.794 &	0.070 &    0.879 &    5.950 &	14.727 &   15.000 &   15.000 &   15.000 &   15.000 &   15.000 &   15.000 &   15.000 &	15.000\\
4.60 &    5.580 &    2.399 &	0.412 &    0.216 &    3.337 &	 9.594 &   15.000 &   15.000 &   15.000 &   15.000 &   15.000 &   15.000 &   15.000 &	15.000\\
4.70 &    6.502 &    3.061 &	0.881 &    0.072 &    1.699 &	 5.924 &   13.063 &   15.000 &   15.000 &   15.000 &   15.000 &   15.000 &   15.000 &	15.000\\
\enddata
\tablecomments{Same as Table~\protect\ref{tab:Si} but for iron.}
\end{deluxetable}

\clearpage

\begin{deluxetable}{ccccccccccccccc}
\setlength{\tabcolsep}{0.05in}
\tabletypesize{\tiny}
\tablecaption{Cobalt CIE Fractional Abundances
\label{tab:Co}}
\tablewidth{0pt}
\tablehead{
\colhead{log(T)} 
& \colhead{Co{$^{0+}$}}
& \colhead{Co{$^{1+}$}}
& \colhead{Co{$^{2+}$}}
& \colhead{Co{$^{3+}$}}
& \colhead{Co{$^{4+}$}}
& \colhead{Co{$^{5+}$}}
& \colhead{Co{$^{6+}$}}
& \colhead{Co{$^{7+}$}}
& \colhead{Co{$^{8+}$}}
& \colhead{Co{$^{9+}$}}
& \colhead{Co{$^{10+}$}}
& \colhead{Co{$^{11+}$}}
& \colhead{Co{$^{12+}$}}
& \colhead{Co{$^{13+}$}}}
\startdata
4.00 &    0.710 &    0.094 &	3.841 &   15.000 &   15.000 &	15.000 &   15.000 &   15.000 &   15.000 &   15.000 &   15.000 &   15.000 &   15.000 &	15.000\\
4.10 &    1.222 &    0.033 &	1.903 &   11.225 &   15.000 &	15.000 &   15.000 &   15.000 &   15.000 &   15.000 &   15.000 &   15.000 &   15.000 &	15.000\\
4.20 &    1.786 &    0.129 &	0.620 &    7.184 &   15.000 &	15.000 &   15.000 &   15.000 &   15.000 &   15.000 &   15.000 &   15.000 &   15.000 &	15.000\\
4.30 &    2.575 &    0.511 &	0.162 &    4.407 &   13.899 &	15.000 &   15.000 &   15.000 &   15.000 &   15.000 &   15.000 &   15.000 &   15.000 &	15.000\\
4.40 &    3.360 &    0.934 &	0.056 &    2.437 &    9.135 &	15.000 &   15.000 &   15.000 &   15.000 &   15.000 &   15.000 &   15.000 &   15.000 &	15.000\\
4.50 &    4.112 &    1.357 &	0.070 &    0.978 &    5.423 &	14.976 &   15.000 &   15.000 &   15.000 &   15.000 &   15.000 &   15.000 &   15.000 &	15.000\\
4.60 &    5.084 &    2.028 &	0.389 &    0.236 &    2.859 &	 9.674 &   15.000 &   15.000 &   15.000 &   15.000 &   15.000 &   15.000 &   15.000 &	15.000\\
4.70 &    6.196 &    2.864 &	0.910 &    0.085 &    1.265 &	 5.886 &   13.098 &   15.000 &   15.000 &   15.000 &   15.000 &   15.000 &   15.000 &	15.000\\
\enddata
\tablecomments{Same as Table~\protect\ref{tab:Si} but for cobalt.}
\end{deluxetable}

\clearpage

\begin{deluxetable}{cccccccccccccccc}
\setlength{\tabcolsep}{0.05in}
\tabletypesize{\tiny}
\tablecaption{Nickel CIE Fractional Abundances
\label{tab:Ni}}
\tablewidth{0pt}
\tablehead{
\colhead{log(T)} 
& \colhead{Ni{$^{0+}$}}
& \colhead{Ni{$^{1+}$}}
& \colhead{Ni{$^{2+}$}}
& \colhead{Ni{$^{3+}$}}
& \colhead{Ni{$^{4+}$}}
& \colhead{Ni{$^{5+}$}}
& \colhead{Ni{$^{6+}$}}
& \colhead{Ni{$^{7+}$}}
& \colhead{Ni{$^{8+}$}}
& \colhead{Ni{$^{9+}$}}
& \colhead{Ni{$^{10+}$}}
& \colhead{Ni{$^{11+}$}}
& \colhead{Ni{$^{12+}$}}
& \colhead{Ni{$^{13+}$}}}
\startdata
 4.00 &    1.265 &    0.024 &	 4.955 &   15.000 &   15.000 &   15.000 &   15.000 &   15.000 &   15.000 &   15.000 &	15.000 &   15.000 &   15.000 &   15.000\\
 4.10 &    2.154 &    0.004 &	 2.936 &   13.025 &   15.000 &   15.000 &   15.000 &   15.000 &   15.000 &   15.000 &	15.000 &   15.000 &   15.000 &   15.000\\
 4.20 &    2.875 &    0.021 &	 1.343 &    8.551 &   15.000 &   15.000 &   15.000 &   15.000 &   15.000 &   15.000 &	15.000 &   15.000 &   15.000 &   15.000\\
 4.30 &    3.552 &    0.289 &	 0.314 &    5.102 &   15.000 &   15.000 &   15.000 &   15.000 &   15.000 &   15.000 &	15.000 &   15.000 &   15.000 &   15.000\\
 4.40 &    4.404 &    1.022 &	 0.044 &    2.885 &   10.117 &   15.000 &   15.000 &   15.000 &   15.000 &   15.000 &	15.000 &   15.000 &   15.000 &   15.000\\
 4.50 &    5.028 &    1.630 &	 0.033 &    1.310 &    6.154 &   14.805 &   15.000 &   15.000 &   15.000 &   15.000 &	15.000 &   15.000 &   15.000 &   15.000\\
 4.60 &    5.543 &    2.129 &	 0.267 &    0.345 &    3.274 &    9.302 &   15.000 &   15.000 &   15.000 &   15.000 &	15.000 &   15.000 &   15.000 &   15.000\\
 4.70 &    6.208 &    2.753 &	 0.760 &    0.099 &    1.542 &    5.469 &   13.084 &   15.000 &   15.000 &   15.000 &	15.000 &   15.000 &   15.000 &   15.000\\
\enddata
\tablecomments{Same as Table~\protect\ref{tab:Si} but for nickel.}
\end{deluxetable}

\clearpage

\begin{deluxetable}{cccccccccccccccc}
\setlength{\tabcolsep}{0.04in}
\tabletypesize{\tiny}
\tablecaption{Copper CIE Fractional Abundances
\label{tab:Cu}}
\tablewidth{0pt}
\tablehead{
\colhead{log(T)} 
& \colhead{Cu{$^{0+}$}}
& \colhead{Cu{$^{1+}$}}
& \colhead{Cu{$^{2+}$}}
& \colhead{Cu{$^{3+}$}}
& \colhead{Cu{$^{4+}$}}
& \colhead{Cu{$^{5+}$}}
& \colhead{Cu{$^{6+}$}}
& \colhead{Cu{$^{7+}$}}
& \colhead{Cu{$^{8+}$}}
& \colhead{Cu{$^{9+}$}}
& \colhead{Cu{$^{10+}$}}
& \colhead{Cu{$^{11+}$}}
& \colhead{Cu{$^{12+}$}}
& \colhead{Cu{$^{13+}$}}
& \colhead{Cu{$^{14+}$}}}
\startdata
 4.00 &    1.049 &    0.041 &	5.484 &   15.000 &   15.000 &	15.000 &   15.000 &   15.000 &   15.000 &   15.000 &   15.000 &   15.000 &   15.000 &	15.000 &   15.000\\
4.10 &    1.939 &    0.005 &	3.230 &   13.805 &   15.000 &	15.000 &   15.000 &   15.000 &   15.000 &   15.000 &   15.000 &   15.000 &   15.000 &	15.000 &   15.000\\
4.20 &    2.710 &    0.016 &	1.456 &    9.302 &   15.000 &	15.000 &   15.000 &   15.000 &   15.000 &   15.000 &   15.000 &   15.000 &   15.000 &	15.000 &   15.000\\
4.30 &    3.623 &    0.302 &	0.301 &    5.616 &   15.000 &	15.000 &   15.000 &   15.000 &   15.000 &   15.000 &   15.000 &   15.000 &   15.000 &	15.000 &   15.000\\
4.40 &    5.041 &    1.196 &	0.029 &    3.308 &   11.055 &	15.000 &   15.000 &   15.000 &   15.000 &   15.000 &   15.000 &   15.000 &   15.000 &	15.000 &   15.000\\
4.50 &    6.412 &    2.128 &	0.013 &    1.651 &    6.910 &	15.000 &   15.000 &   15.000 &   15.000 &   15.000 &   15.000 &   15.000 &   15.000 &	15.000 &   15.000\\
4.60 &    7.717 &    3.061 &	0.173 &    0.485 &    3.744 &	10.228 &   15.000 &   15.000 &   15.000 &   15.000 &   15.000 &   15.000 &   15.000 &	15.000 &   15.000\\
4.70 &    9.260 &    4.287 &	0.805 &    0.084 &    1.734 &	 6.017 &   12.885 &   15.000 &   15.000 &   15.000 &   15.000 &   15.000 &   15.000 &	15.000 &   15.000\\
\enddata
\tablecomments{Same as Table~\protect\ref{tab:Si} but for copper.}
\end{deluxetable}%

\clearpage

\clearpage
\clearpage

\clearpage

\clearpage
\clearpage

\clearpage

\clearpage

\clearpage

\clearpage

\clearpage
\clearpage

\clearpage

\end{document}